\documentclass[%
 reprint,
 amsmath,amssymb,
 aps,
 prx,
]{revtex4-2}

\usepackage{graphicx}
\usepackage{dcolumn}
\usepackage{bm}
\usepackage{tikz-feynman}

\newcommand{\llrrangle}[1]{
\langle\mkern-4mu \langle #1\rangle \mkern-4mu \rangle}

\tikzfeynmanset{
empty square/.style={
 /tikzfeynman/square dot,
     /tikz/fill=none
  },
}

\begin{document}

\preprint{APS/123-QED}

\title{Field-theoretic approach to compartmental neuronal networks: \\ 
impact of dendritic calcium spike-dependent bursting}

\author{Audrey Teasley}
\affiliation{%
 Graduate Program in Neuroscience, Boston University, Boston MA 02215 USA
}%
\author{Gabriel Koch Ocker}%
 \email{gkocker@bu.edu}
\affiliation{%
Department of Mathematics and Statistics, Boston University, Boston MA 02215 USA
}%




\date{\today}

\begin{abstract}
Neurons are spatially extended cells; different parts of a neuron have specific voltage dynamics.
Important types of neurons even generate different spikes in different parts of the cell. 
Neurons' inputs are also often spatially compartmentalized, with different sources targeting different locations on the cell. 
Classic mean-field theories for neural population activity, however, rely on point-neuron models with at most one type of spike. 
Here, we develop a statistical field-theoretic approach to understanding collective activity in networks of compartmental neurons, including those generating multiple types of spikes. 
We use this to examine simple models of networks with thick-tufted layer 5 pyramidal cells, which generate calcium spikes in their apical dendrite when dendritic depolarization coincides with a back-propagating somatic action potential.
In the weakly-coupled regime, we uncover an exact mean-field limit for these networks that maps them to a marked point process.
We use this mean-field limit to compare the impact of compartmentalized recurrent excitatory and inhibitory connectivity on the equilibrium phase diagram. 
This exposes regions of metastability between various activity states, including activity with silent vs active dendrites, with and without inhibitory activity, and oscillations. 
\end{abstract}

\maketitle



\section{\label{sec:intro1}Introduction}

The collective activity of neuronal networks underlies sensory perception, motor planning and execution, learning and memory, and other cognitive functions.
Mean-field theories for neuronal population activity explain collective activity using a few order parameters. 
Classical mean-field theories exhibit a wide variety of dynamics \cite{coombes_large-scale_2010}.

Individual neurons also exhibit a broad range of nonlinear dynamics. The most famous of these may be the sodium-potassium (Na$^+$/K$^+$) action potential, generated near the cell body, that drives neurotransmitter release to allow fast intercellular communication \cite{eccles_action_1935, hodgkin_quantitative_1952}. 
Furthermore, neurons are spatially extended cells. Cortical neurons have richly branched dendritic trees, on which they receive synaptic inputs.
Neurons can also generate action potentials in their dendrites \cite{larkum_guide_2022}. 

For example, consider thick-tufted pyramidal cells, a main class of cortical long-range projection neurons. 
These neurons have a cell body in layer 5 of the cortex and an apical dendrite that extends up to layer 1 (Fig.~\ref{fig:single-neuron}a). 
The apical dendrites of these neurons generate calcium spikes~\cite{larkum_new_1999}. 
These trigger bursts of somatic action potentials, giving the neuron a parallel output train that can be separately modulated by learning and attention~\cite{naud_sparse_2018, friedenberger_silences_2023}. 
A variety of other important neuron types also generate calcium spikes, including thalamic relay neurons, cerebellar Purkinje cells, and layer 2/3 and hippocampal pyramidal cells~\cite{brumberg_ionic_2000, schwartzkroin_probable_1977, wong_intradendritic_1979, magee_dendritic_1999, larkum_dendritic_2007, jahnsen_ionic_1984, destexhe_dendritic_1998, eccles_excitatory_1966, davie_origin_2008}.



Different types of synaptic input are also spatially localized on pyramidal cells. In the neocortex, long-range feedforward and feedback projections preferentially target specific layers~\cite{felleman_distributed_1991, harris_hierarchical_2019}. 
The major classes of cortical inhibitory neurons also preferentially target specific compartments. Somatostatin-positive (SOM) Martinotti cells preferentially synapse onto pyramidal cells' apical dendrites, while parvalbumin-positive (PV) basket cells synapse onto cell bodies~\cite{rudy_three_2011}. 

Together, these experimental observations beg the question: how does the spatial compartmentalization of synaptic connectivity interact with somatic and dendritic spiking to shape collective activity in cortical networks? 

Due to their complexity, large networks of compartmental neurons are generally studied through careful and laborious simulation, e.g.,~\cite{traub_single-column_2005, markram_blue_2006, markram_reconstruction_2015, billeh_systematic_2020}. 
Using simulation to survey the parameter combinations underlying healthy or pathological network dynamics becomes intractable as the complexity of the model grows.
Classical mean-field theories of neural activity, on the other hand, rely on single-compartment models that either neglect spiking or model only the somatic Na$^+$/K$^+$ spike~\cite{grossberg_learning_1969, amari_characteristics_1971, amari_characteristics_1972, wilson_excitatory_1972, sompolinsky_chaos_1988, ohira_master-equation_1993, ginzburg_theory_1994, vreeswijk_chaotic_1998, brunel_dynamics_2000, gerstner_time_1995, renart_asynchronous_2010, montbrio_macroscopic_2015}.

Here, we propose a statistical field-theoretic approach for networks of compartmental neurons with stochastic spike emission. We construct the joint density functional of the subthreshold voltages of, and action potentials generated by, each compartment of each neuron in a network. 
We apply this to networks with a simple model of dendritic calcium spike-dependent bursting in pyramidal cells.
The large-$N$ mean-field limit of this density functional exposes exact predictions for the rate and fluctuations of somatic and dendritic action potentials.
We use these to study how the compartmentalization of excitatory and inhibitory connectivity controls activity in recurrent networks.

\section{\label{sec:model} Approach and model}
We divide each neuron $i$ into a set $C_i$ of isopotential spatial compartments. 
The membrane voltage of compartment $c$ in neuron $i$ is $v_i^c(t)$.
Rather than describing the nonlinear membrane potential dynamics that generate spikes explicitly, we model them as point processes. Each neuron has a set of point processes associated to it; $\dot{a}_i^d(t)$ is the point process describing spikes of type $d$ in neuron $i$, also called a spike train. 
The intensities of these point processes model the underlying generating mechanisms. 
For example, the classic Na$^+$/K$^+$ action potential generated in the somatic compartment ($c=S$) is typically modeled with rectified power-law voltage-intensity transfer functions:
\begin{equation} \begin{aligned} \label{eq:spike_gen}
\dot{a}_i^S(t) \sim& \mathcal{PP}\big(f(v_i^S(t))\big), \\
f(v_i^S(t)) =& \lfloor v_i^S(t) - \theta \rfloor_+^p,
\end{aligned} \end{equation}
where $\theta$ is a spike threshold and $p$ a power that typically ranges between 1 and 5 in cortical cells~\cite{paliwal_metastability_2025}. 
Other choices for the voltage-intensity transfer function $f$ are admissible. We only require $\lim_{v \rightarrow \infty} f(v) = \infty$ so that if the somatic voltage blows up, the neuron emits an action potential with probability one.

\begin{figure} \includegraphics[scale=0.8]{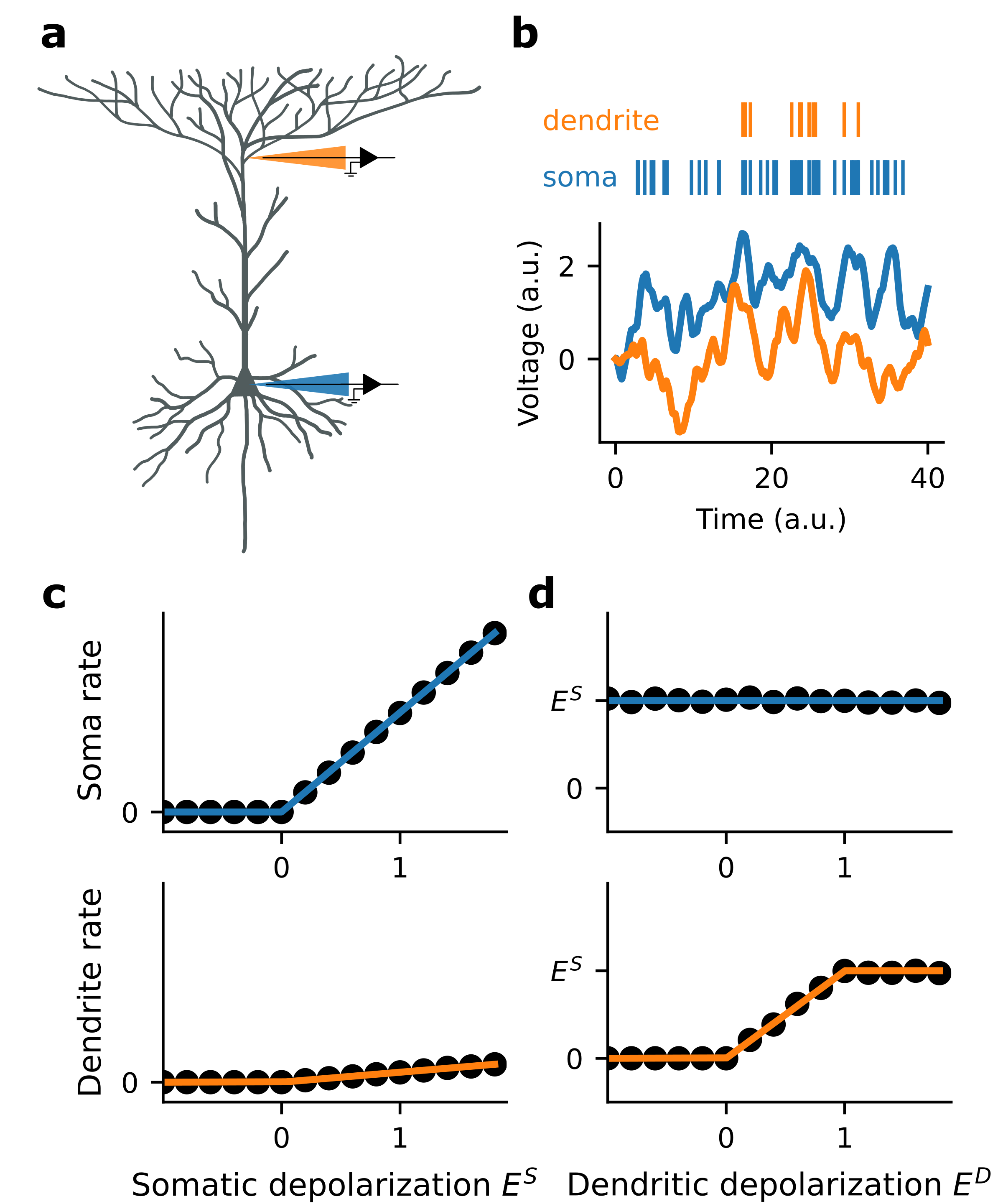}
\caption{A compartmental point process model of thick-tufted layer 5 pyramidal cells. {\bf a)} Rendition of a thick-tufted layer 5 pyramidal cell. {\bf b)} Na$^+$/K$^+$ action potentials at the soma can backpropagate to the apical dendrite and trigger a dendritic calcium action potential. 
{\bf c, d)} Event rates for a neuron with threshold-linear somatic spike intensity and dendritic conditional spike probability functions, as a function of the somatic depolarization ({\bf c}) and dendritic deporalization ({\bf d}). Yellow line: Eq.~\ref{eq:mft_1neuron}. Black dots: simulation.
\label{fig:single-neuron} }
\end{figure}

Here, we focus on a model of thick-tufted layer 5 pyramidal cells with somatic ($S$) and apical dendritic ($D$) compartments (Fig.~\ref{fig:single-neuron}a).  
The somatic compartment generates Na$^+$/K$^+$ action potentials with intensity $f(v_i^S(t))$ (Eq.~\ref{eq:spike_gen}).
In thick-tufted layer 5 pyramidal cells, dendritic calcium spikes are reliably triggered by the backpropagation of a somatic action potential into a depolarized apical dendrite~\cite{larkum_new_1999}. 
Following Naud \& Sprekeler~\cite{naud_sparse_2018}, we model this as
\begin{equation} \begin{aligned} \label{eq:burst_conversion}
\dot{a}_i^D(t) \sim \mathcal{PP}\big(\dot{a}_i^S(t) \, g(v_i^D(t)\big)
\end{aligned}. \end{equation}
where $v_i^D$ is the dendritic voltage and the dendritic transfer function $g(v_i^D(t))$ determines the probability that a somatic Na$^+$/K$^+$ spike triggers a dendritic calcium spike. We require that $0 \leq g \leq 1$; it is a probability.

To focus on the role of compartmentalized action potential types, we here assume linear sub-threshold voltage dynamics within each compartment and linear (current-based) synaptic coupling so that the membrane voltage of compartment $c$ in neuron $i$ obeys
\begin{equation}\label{eq:membrane_voltage_linear}
\left( \partial_t + 1 \right) v_i^c =  E_i^c + \sum_{j=1}^N \sum_{d \in C_j} J_{ij}^{cd} \ast \dot{a}_j^d,
\end{equation}
where 
$E_i^c$ is an effective reversal potential, which may also depend on external inputs, and the synaptic filter $J_{ij}^{cd}(s)$ models the impact of an action potential of type $d$ in neuron $j$ on compartment $c$ of neuron $i$ at time delay $s$. $J_{ij}^{cd} \ast \dot{a}_j^d$ is the convolution of that filter with the presynaptic spike train.
Due to the distance between the two spike-initiating zones in thick-tufted layer 5 pyramidal cells, the two compartments interact only through action potentials (Eq.~\ref{eq:burst_conversion}).
This formulation can be readily extended to passive voltage propagation, as well as nonlinear voltage dynamics and/or spike resets~\cite{chow_path_2015, ocker_republished_2023}.
Without the dendrite, this model is a nonlinear Hawkes process.

In our model, each compartment generates one type of action potential. 
However, one could use this modeling framework for neurons that generate more than one type of action potential in the same compartment by taking their intensities to both depend on the compartment's membrane potential. 

To understand how the location and strength of recurrent connectivity controls collective activity in these networks, we first aim to develop a theory for the expected activity:
\begin{equation} \begin{aligned}
\big \langle \dot{a}^S_i(t) \big\rangle =& \left\langle f \Bigg[G_i^S \ast \Bigg(E_i^S + \sum_{j=1}^N \sum_{d \in C_j} J_{ij}^{Sd} \ast \dot{a}_j^d \Bigg)(t) \Bigg] \right\rangle, \\
\big \langle \dot{a}_i^D (t) \big\rangle =& \left\langle \dot{a}_i^S(t) \, g \Bigg[G_i^D \ast \Bigg(E_i^D + \sum_{j=1}^N \sum_{d \in C_j} J_{ij}^{Dd} \ast \dot{a}_j^d \Bigg)(t) \Bigg] \right\rangle
\end{aligned} \end{equation}
where $G_{i}^c$ are the fundamental solutions associated with the voltage dynamics (Eq.~\ref{eq:membrane_voltage_linear}): $G_i^c \ast x_i^c(t) = v_i^c(t_0)e^{-(t-t_0)} + \int_{t_0}^t ds \, e^{-(t-s)} x_i^c(s) $. Here, a moment closure problem becomes apparent. When $f$ is nonlinear, the mean depends on second or higher-order moments of the presynaptic spike trains. (This can be seen by expanding $f$ in a Taylor series.) This is the same problem faced in similar single-compartment models~\cite{buice_field-theoretic_2007, buice_systematic_2010, ocker_linking_2017}. 

The dendrite faces a more difficult moment closure problem: even if $g$ is linear, the dendritic activity rate depends on the correlation of somatic spikes with the dendritic input. We will address this problem by a mean-field approach in Section~\ref{sec:mft}, and by examining the joint density functional of the network (Section~\ref{sec:path_integral}, Appendices~\ref{app:density}, \ref{app:mft}). We will calculate the mean-field equilibrium phase diagrams in networks with recurrent connectivity targeting the soma vs apical dendrite, and compare the dynamics of excitatory-inhibitory networks with recurrent inhibition targeting either of the two compartments.
First, we briefly consider the case of an uncoupled population.

\section{\label{sec:single-neuron} Somatic and dendritic spike rates in single neurons}
In this case, the expected intensities of the somatic and dendritic spikes can be calculated directly:
\begin{equation} \begin{aligned} \label{eq:mft_1neuron}
\langle \dot{a}^S(t) \rangle =& f\Big[G^S \ast E^S(t)\Big], \\
\langle \dot{a}^D(t) \rangle =& \langle \dot{a}^S(t) \rangle \, g\Big[G^D \ast E^D(t)\Big]
\end{aligned} \end{equation}
As the somatic depolarization $E^S$ increases, the somatic spike rate increases. For simplicity, we will take threshold-linear somatic and dendritic transfer functions: 
\begin{equation} \begin{aligned} \label{eq:relu}
f(x) =& \lfloor x \rfloor_+ \equiv \max(x, 0), \\
g(x) =& [ x]_+^1 \equiv \min(\max(x, 0), 1).
\end{aligned} \end{equation}
The equilibrium dendritic rate also increases with $E^S$, with a proportionality factor given by $g(E^D)$ (Fig.~\ref{fig:single-neuron}c). If the dendritic voltage is below threshold ($E^D \leq 0$), it does not spike. 
As $E^D$ increases, the somatic voltage and therefore also the somatic spike rate are unaffected (Fig.~\ref{fig:single-neuron}d, top) since the two compartments interact only through spikes. The dendritic spike rate, on the other hand, increases with $E^D$ until it reaches an upper bound (Fig.~\ref{fig:single-neuron}d, bottom). This upper bound reflects the saturation of $g$. Once the dendrite converts every somatic spike into a dendritic calcium spike ($E^D \geq 1$), its rate saturates at the somatic rate.

\section{\label{sec:mft} Mean-field theory of recurrent networks}
We next turn to recurrent networks. We assume that each dendritic calcium spike triggers a stereotyped burst of somatic spikes, as in the thick-tufted layer 5 pyramidal cells~\cite{williams_mechanisms_1999, larkum_dendritic_2001, schwindt_mechanisms_1999, kim_apical_1993}. So, for projections from neuron $j$ to $i$, we take
\begin{equation}
J_{ij}^{cD} = \beta J_{ij}^{cS}.
\end{equation}
Here, $\beta$ is the strength of the total postsynaptic potential induced by a presynaptic burst, relative to that from a single presynaptic spike.
From here on, we may also refer to $\dot{a}_i^D(t)$ as the burst process. 
Since dendritic calcium spikes require a somatic spike in this model (Eq.~\ref{eq:burst_conversion}), when a dendritic calcium spike occurs both $\dot{a}^S_i$ and $\dot{a}^D_i$ contain a point. When a somatic spike fails to trigger a dendritic calcium spike, only $\dot{a}_i^S$ contains a point. $\dot{a}_i^S(t)$ contains points marking the first spike in a burst, but not the subsequent within-burst spikes. 
The difference $\dot{a}_i^S - \dot{a}_i^D$ isolates the non-burst singlet spikes.
We will also drop the superscript denoting the presynaptic activity type; $J_{ij}^c \equiv J_{ij}^{cS}$ and $\beta J^c_{ij} \equiv J^{cD}_{ij}$.

To uncover a simple, low-dimensional description of the population activity we scale the synaptic weights as $J_{ij} \sim N^{-1}$ and consider the limit $N \rightarrow \infty$. The net input to each compartment then concentrates around its population mean (Appendix~\ref{app:mft}).
For networks of a single cell type,
\begin{equation} \begin{aligned} \label{eq:mft_single_pop}
\langle \dot{a}^S (t) \rangle =& f \Big[G^S \ast \left(E^S + J^S \ast \langle \dot{a}^S  \rangle + \beta J^S \ast \langle \dot{a}^D \rangle \right)(t)  \Big] \\
\langle \dot{a}^D (t) \rangle =& \langle \dot{a}^S(t) \rangle \\
&\times g \Big[G^D \ast \Big(E^D + J^{D} \ast \langle \dot{a}^S \rangle  + \beta J^{D} \ast \langle \dot{a}^D  \rangle \Big)(t)  \Big].
\end{aligned} \end{equation}
Here, parameters without neuron subscript $i$ refer to the population mean, e.g., $J^{S}$ is the mean of $J^S_{ij}$. This large-$N$ mean-field theory relies on the hypothesis of self-averaging activity: that the behavior of a sufficiently large network follows the average over realizations of the matrices $J$. Otherwise, it involves no approximations. Similar mean-field limits for various versions of the soma-only model, nonlinear Hawkes networks, have recently been rigorously established~\cite{zhu_large_2015, delattre_hawkes_2016, heesen_fluctuation_2021, pfaffelhuber_mean-field_2022}.

A solution of Eq.~\ref{eq:mft_single_pop} describes a joint trajectory of firing rates for each type of action potential. Time-independent solutions describe stationary regimes of the network. These states can be examined in a common type of experiment, namely extracellular electrical or optical physiology experiments in live animals.

Eq.~\ref{eq:mft_single_pop} does not, however, expose information about the stability of those solutions to perturbations or fluctuations in the population activity. To examine stability, we exploit the duality between activity and voltage-based formulations of our model. Averaging Eq.~\ref{eq:membrane_voltage_linear} and again taking the mean-field limit, we find
\begin{equation}\label{eq:mft_single_pop_voltage}
\left( \partial_t + 1 \right) \langle v^c \rangle  =  E^c + J^c \ast f\big(\langle v^S \rangle\big) + \beta J^c \ast \left[ f\big(\langle v^S \rangle \big) \, g\big(\langle v^D \rangle \big) \right]
\end{equation}
where $c \in \{S, D\}$ and $f(\langle v^S \rangle)$ and $f(\langle v^S \rangle) g(\langle v^D \rangle) $ are the somatic and dendritic spike rates. This is a set of two coupled differential equations for the mean voltage in each compartment that exposes standard stability analyses. Solutions of Eq.~\ref{eq:mft_single_pop_voltage} correspond directly to the arguments of $f$ and $g$ in Eq.~\ref{eq:mft_single_pop}. 
We next examine how recurrent connectivity targeting the cell body vs apical dendrites impacts population activity.

\subsection{\label{sec:soma-targeting} Soma-targeting connectivity}
\begin{figure}[ht!] \includegraphics[scale=0.8]{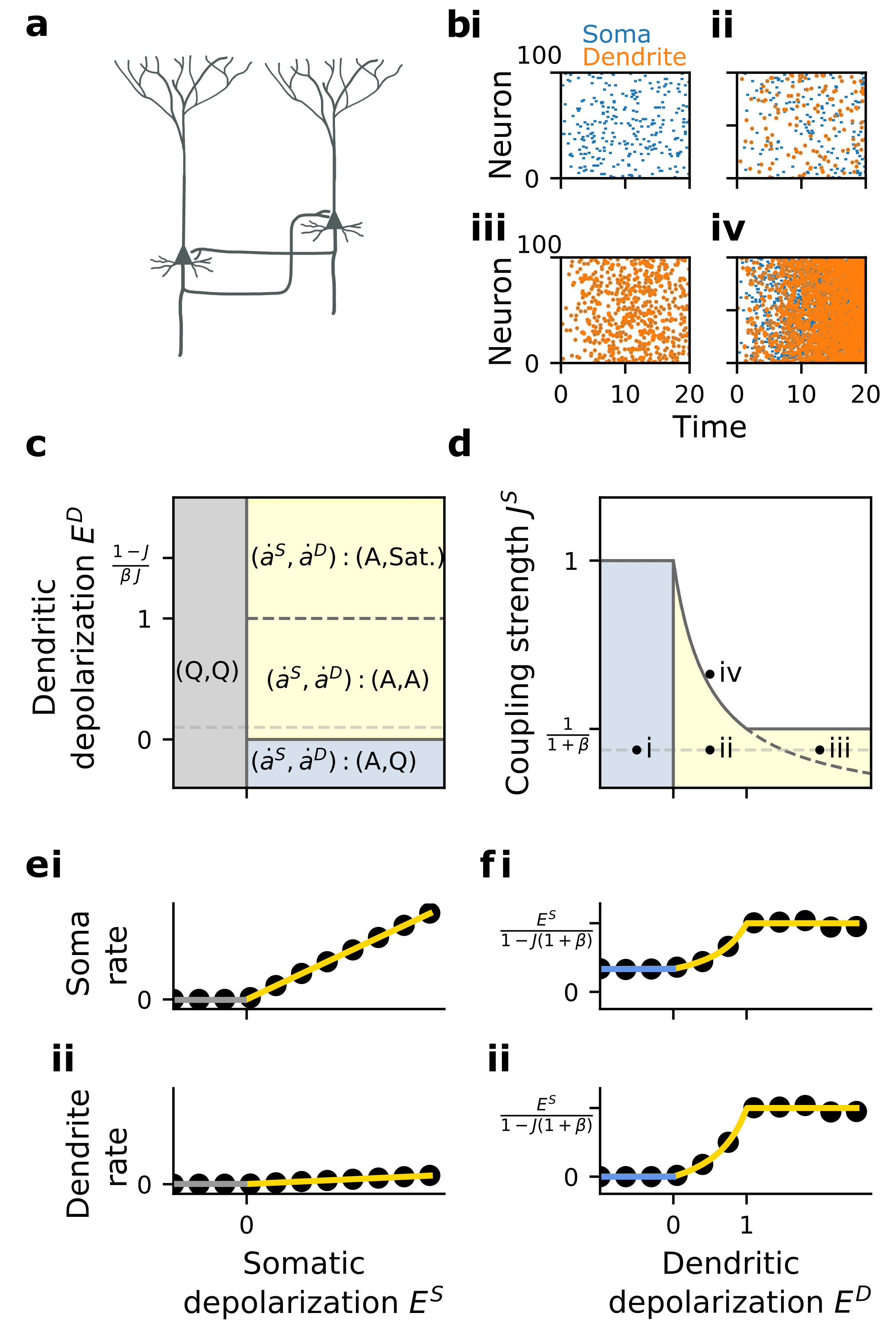}
\caption{Soma-targeting recurrent connectivity. {\bf a)} Rendition of recurrent synapses targeting the somatic compartment. {\bf b)} Raster plots in each of the activity regimes: spiking with no {\bf (i)}, sparse ({\bf ii}), or saturated ({\bf iii}) dendritic activity, and runaway activity {\bf (iv)}. {\bf c)} Mean-field equilibrium phase diagram in the $(E^S, E^D)$ plane. $(\beta, J) = (2, 1/4)$. {\bf d)} Mean-field equilibrium phase diagram in the $(E^D, J^S)$ plane. $(\beta, E^S)= 2, 1/10)$.  {\bf e,f)} Rates along the gray dashed lines in panels {\bf b},{\bf c}. Dots: simulation. Curves: mean-field prediction (Eq.~\ref{eq:mft_soma}; Table~\ref{table:soma}).
\label{fig:soma-syn} }
\end{figure}

We begin by examining networks in which synapses target only the somatic compartment, setting $J^D = 0$ (Fig.~\ref{fig:soma-syn}a).
For simplicity, we again assume threshold-linear intensity functions, Eq.~\ref{eq:relu}. At equilibrium, this yields the mean-field equations
\begin{equation} \begin{aligned} \label{eq:mft_soma}
\langle \dot{a}^S \rangle =& \left\lfloor E^S + J  \langle \dot{a}^S \rangle  + \beta J  \langle \dot{a}^D \rangle \right \rfloor_+ \\
\langle \dot{a}^D \rangle =& \langle \dot{a}^S \rangle \left[ E^D \right]_+^1.
\end{aligned} \end{equation}
Since recurrent connectivity does not target the dendrite, $\langle \dot a^D \rangle $ can be directly substituted into the somatic rate to yield a one-dimensional mean-field theory. We list the solutions to this mean-field equation and their existence and stability conditions in Table~\ref{table:soma} (Appendix~\ref{app:mf_fp}). 

There are three types of mean-field fixed point, each representing a regime of macroscopic population activity (Fig.~\ref{fig:soma-syn}b-d). First, there is a stable silent state ($\langle \dot{a}^S \rangle = \langle \dot{a}^D \rangle = 0$), which exists if the somatic input is below threshold (Fig.~\ref{fig:soma-syn}c, gray). There can be somatic spiking without dendritic calcium spikes if the external drive to the soma is above threshold, the dendrite is below threshold, and recurrent excitation is not too strong (Fig.~\ref{fig:soma-syn}bi; c,d, blue). There can be activity with dendritic calcium spikes if the external drive to both soma and dendrite are above threshold and recurrent excitation is not too strong (Fig.~\ref{fig:soma-syn}bii; c,d yellow). The dendritic rate saturates at $\langle \dot{a}^S \rangle E^D$ if $E^D > 1$; in this regime, every somatic spike triggers a dendritic calcium spike (Fig.~\ref{fig:soma-syn}biii; c,d, black dashed). In each of these regions, the mean-field theory of Eq.~\ref{eq:mft_soma} provides an accurate prediction of the firing rates (Fig.~\ref{fig:soma-syn}e). Finally, if recurrent excitation is strong enough the activity can diverge (Fig.~\ref{fig:soma-syn}biv). This occurs when $J(1 + \beta g(E^D)) > 1$. 
This instability can occur with $E^S > 0$, in which case there is no stable fixed point
(Fig.~\ref{fig:soma-syn}d, white), or with $E^S<0$, in which case there is a stable quiescent state in addition to the unstable active state.

\subsection{\label{sec:dendrite-targeting} Dendrite-targeting connectivity}
\begin{figure}[ht!] \includegraphics[scale=0.8]{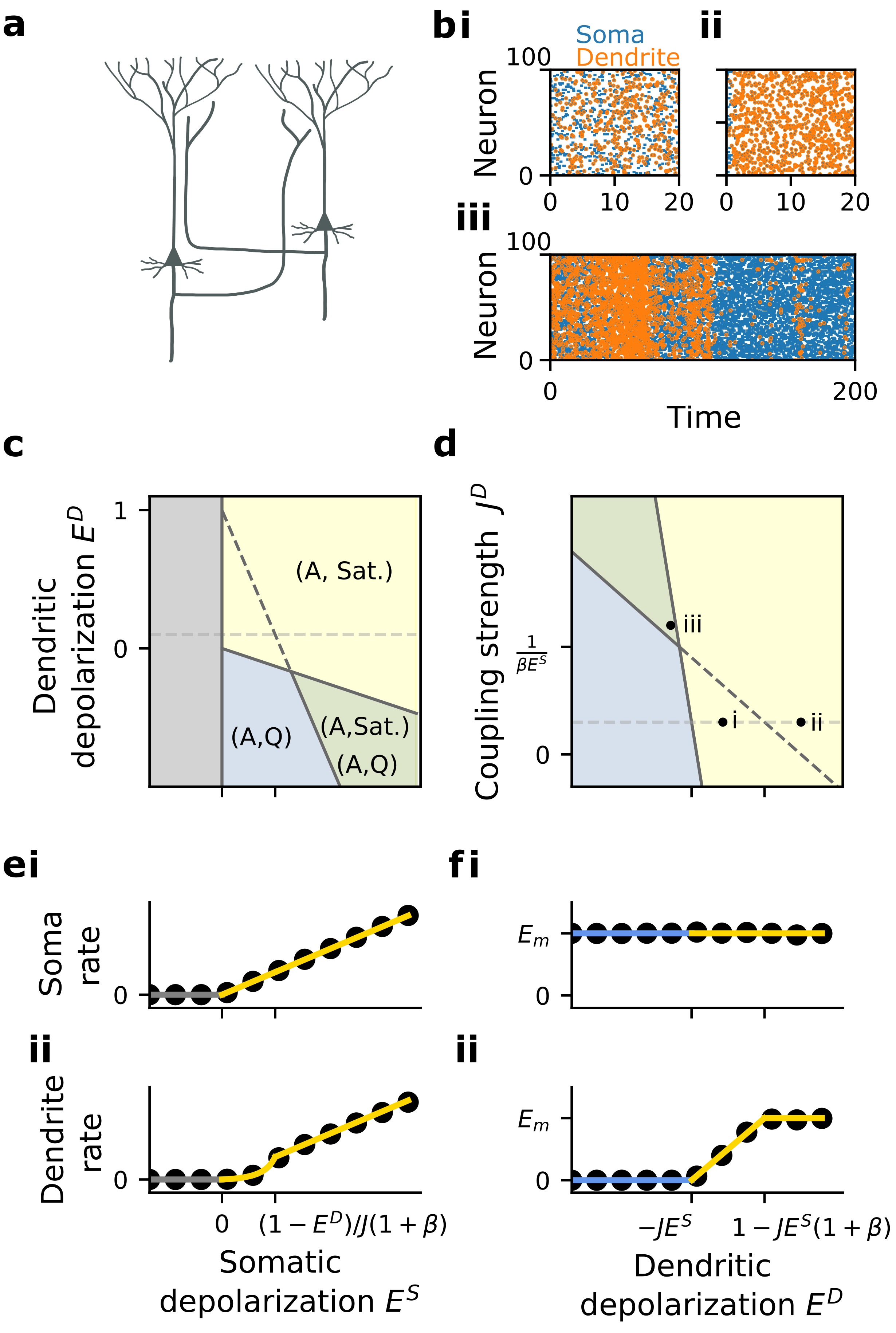}
\caption{ Dendrite-targeting recurrent connectivity. {\bf a)} Rendition of recurrent synapses targeting the apical dendrite compartment. 
{\bf d)} Raster plots in three of the activity regimes: spiking with sparse {\bf (i)} or saturated {\bf (ii)} dendritic activity, and the bistable regime {\bf (iii)} .
{\bf c)} Mean-field equilibrium phase diagram in the $(E^D, E^S)$ plane. $(\beta, J= 6, 1/10)$. 
{\bf d)} Mean-field phase diagram in the $(E^D, J^D)$ plane. $(\beta, E^S= 6, 1/2)$. 
 {\bf e,f)} Rates along the gray dashed lines in panels {\bf b}, {\bf c}. Dots: simulation. Curves: mean-field prediction (Eq.~\ref{eq:mft_dend}; Table~\ref{table:dendrite}).
\label{fig:dend-syn} }
\end{figure}

We next examine networks where recurrent synapses target only the dendrite, setting $J^S = 0$ in Eq.~\ref{eq:mft_single_pop} (Fig.~\ref{fig:dend-syn}a). This yields the equilibrium mean-field equations
\begin{equation} \begin{aligned} \label{eq:mft_dend}
\langle \dot{a}^S \rangle =& \left\lfloor E^S \right\rfloor_+, \\
\langle \dot{a}^D \rangle =& \langle \dot{a}^S \rangle \left[ E^D + J \langle \dot{a}^S \rangle + \beta J \langle \dot{a}^D \rangle \right]_+^1.
\end{aligned} \end{equation}
Since recurrent connectivity does not target the soma, $\langle \dot{a}^S \rangle$ can be directly substituted into the second equation to yield a one-dimensional mean-field theory. The solutions to that mean-field equation and their existence and stability conditions are summarized in Table~\ref{table:dendrite} (Appendix~\ref{app:mf_fp}). Similarly to the network with somatic synapses, there are three types of fixed point: a silent state, spiking without dendritic activity, and spiking with dedritic activity (either saturated or not) (Fig.~\ref{fig:dend-syn}b). The silent state exists so long as the somatic drive is below threshold, and is always stable (Fig.~\ref{fig:dend-syn}c, gray). Spiking with silent dendrites requires the total dendritic input to remain below threshold (Fig.~\ref{fig:dend-syn}c,d, blue). Spiking with dendritic activity requires the total dendritic input to be above threshold (Fig.~\ref{fig:dend-syn}bi; c,d, yellow). When the total dendritic input brings the conditional dendritic spike probability $g$ to 1, the dendritic rate saturates (Fig.~\ref{fig:dend-syn}bii; c,d, dashed black line). Finally, if the dendritic drive is dominated by recurrent excitation rather than external input, $E^D$, there can be bistability between spiking with and without dendritic activity (Fig.~\ref{fig:dend-syn}diii; c,d, green). The mean-field theory of Eq.~\ref{eq:mft_dend} provides a quantitative prediction of the firing rates in each of these states (Fig.~\ref{fig:soma-syn}e,f). The main qualitative difference between the networks with soma-targeting and dendrite-targeting synaptic connectivity is that the former have an instability due to runaway excitation, while in the latter there can be bistability between activity with silent or vocal dendrites.

\section{\label{sec:path_integral} Density functional approach to fluctuations}
Dendritic calcium spikes and somatic bursts can allow neurons to multiplex their coding and are powerful drivers of long-term synaptic plasticity \cite{naud_sparse_2018, friedenberger_silences_2023, bittner_conjunctive_2015, cichon_branch-specific_2015, bittner_behavioral_2017}. The structure of multiplex codes and the dynamics of synaptic plasticity thus depend on correlations between somatic and dendritic spikes. To understand and predict correlations between somatic and dendritic spikes, we study the joint density functional of neurons' spike trains (Appendix~\ref{app:density}). In the response variable path integral formalism~\cite{martin_statistical_1973, dominicis_techniques_1976, janssen_lagrangean_1976, jensen_functional_1981}, that joint density is
\begin{equation} \begin{aligned}
p[\dot{\bf a}] =& \int \mathcal{D} \tilde{\bf a} \; \exp -S[\dot{\bf a}, \tilde{\bf a}], \\
S[\dot{\bf a}, \tilde{\bf a}] =& (\tilde{\bf a}^S)^T \dot{\bf a}^S + (\tilde{\bf a}^D)^T \dot{\bf a}^D \\
-& \left(e^{\tilde{\bf a}^S} -1 \right)^T f\big[ {\bf G}^S \ast \left({\bf E}^S + {\bf J}^S \ast \dot{\bf a}^S + \beta {\bf J}^S \ast \dot{\bf a}^D \right) \big] \\ 
-& (\dot{\bf a}^S)^T \ln \Big\{1 +  \left(e^{\tilde{\bf a}^D} -1 \right) g\big[ {\bf G}^D \ast \big({\bf E}^D + {\bf J}^D \ast \dot{\bf a}^S \\
& \hspace{2in} + \beta {\bf J}^D \ast \dot{\bf a}^D \big) \big] \Big\},
\end{aligned} \end{equation}
where ${\bf x}^T {\bf y} = \int dt \, \sum_i x_i(t) \, y_i(t)$. $S$ is the action functional. 
This density functional can be used to calculate cross-correlations between neurons' somatic and dendritic spikes as functions of the somatic and dendritic drive and coupling matrices, similarly to single-compartment models \cite{ocker_linking_2017}. Here, we focus on the mean-field limit of this density functional. We average over the quenched disorder of the synaptic connectivity (Appendix~\ref{app:mft}) and take the classical mean-field $1/N$ scaling for the synaptic weights. In the limit $N \to \infty$, the density then factorizes over neurons to yield an effective single-neuron system governed by the action
\begin{equation} \begin{aligned} \label{eq:action_mft}
S[\dot{a}, \tilde{a}] =& (\tilde{a}^S)^T \dot{a}^S + (\tilde{a}^D)^T \dot{a}^D \\
-& \left(e^{\tilde{a}^S} -1 \right) f\left[ G^S \ast \left( E^S + J^S \ast \langle \dot{a}^S \rangle + \beta J^S \ast \langle \dot{a}^D \rangle \right) \right] \\
 -& (\dot{a}^S)^T \ln \Big\{1 +  \left(e^{\tilde{a}^D} -1 \right) g\big[G^D \ast \big(E^D + J^D \ast \langle \dot{a}^S \rangle \\
&\hspace{2in} + \beta J^D \ast \langle \dot{a}^D \rangle \big) \big] \Big\}
\end{aligned} \end{equation}
The connectivity matrices have been reduced to their mean elements, $J^S$ and $J^D$. 
Eq.~\ref{eq:action_mft} describes a single neuron receiving self-consistent somatic and dendritic inputs. It can be viewed as a marked point process: the soma generates events as a Poisson process with intensity $f\big[ G \ast \left( E^S + J^S \ast \langle \dot{a}^S \rangle + \beta J^S \ast \langle \dot{a}^D \rangle \right) \big]$. Each of those events is marked as a dendritic calcium spike/burst with conditional probability $g\big[G^D \ast \left(E^D + J^D \ast \langle \dot{a}^S \rangle + \beta J^D \ast \langle \dot{a}^D \rangle \right) \big]$. The corresponding dendritic spike train is a doubly-stochastic point process. The intensities of the somatic and dendritic processes are self-consistently determined by solutions to the mean-field equations. 
The action, Eq.~\ref{eq:action_mft}, exposes the mean-field equations of the previous sections and all joint statistics of the somatic and dendritic spike trains within a typical neuron. In Appendices~\ref{app:mft_expansion}, \ref{app:feynman} we derive a set of graphical rules (Feynman rules) for calculating those statistics. 

\begin{figure}[ht!] \includegraphics[scale=0.8]{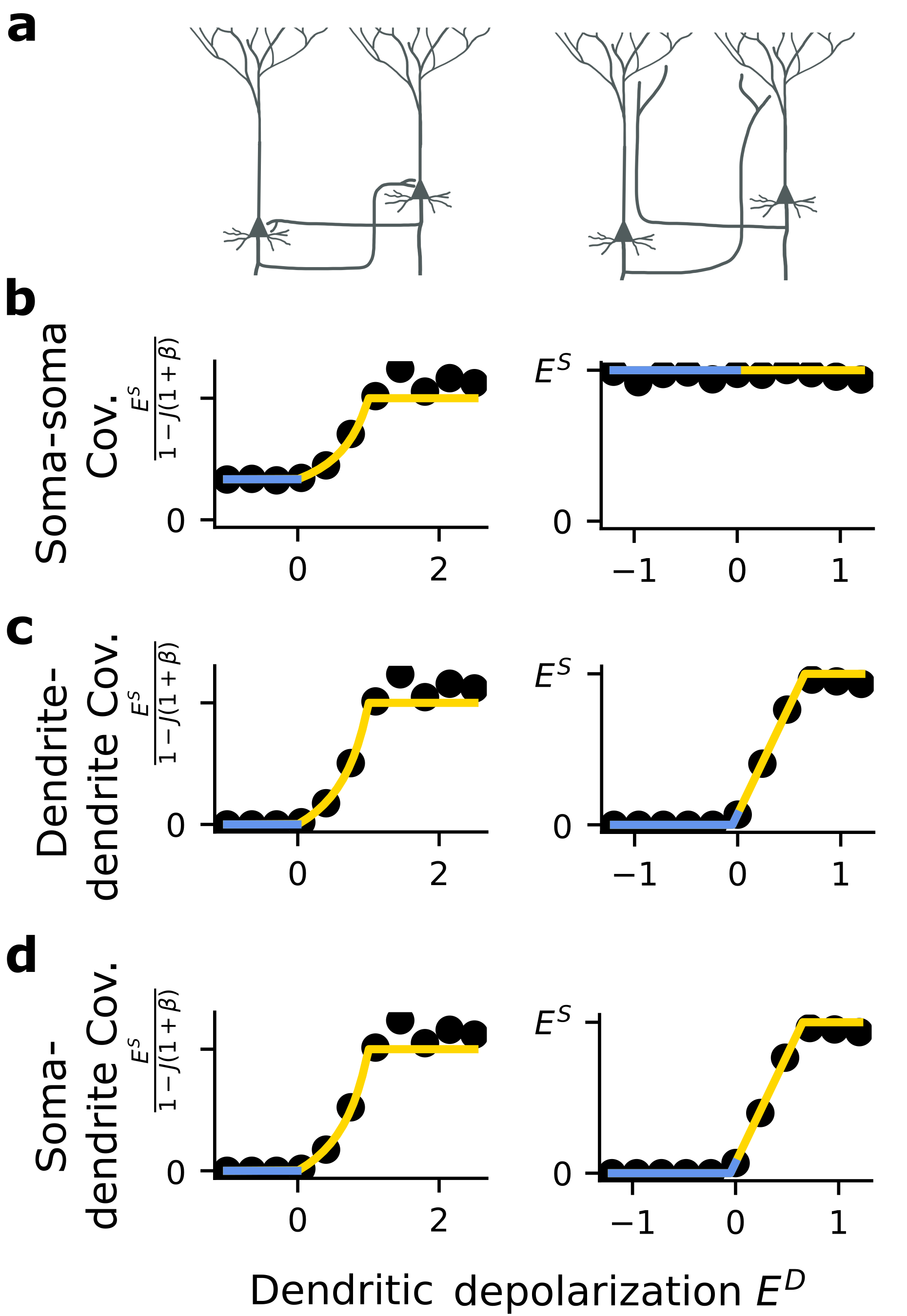}
\caption{Spike and burst covariances in the mean-field limit. {\bf a)} Left column: soma-targeting synapses, parameters $(E^S, J, \beta) = (.1, .25, 2)$. Right column: Dendrite-targeting synapses, parameters $(E^S, J, \beta) = (.5, .1, 6)$.   {\bf b-d)} Soma-soma ({\bf b}), dendrite-dendrite ({\bf c}), and soma-dendrite ({\bf d}) covariances. Black dots: simulations ($N=100$). Solid lines: theoretical predictions, Eqs.~\ref{eq:mft_Cmm}-\ref{eq:mft_Cmb} with the relevant mean-field rates inserted. Colors correspond to those in Figs.~\ref{fig:soma-syn},\ref{fig:dend-syn}.
\label{fig:fluctuations} }
\end{figure}

In the mean-field limit, the somatic event train is a Poisson process and its autocovariance function is
\begin{equation} \label{eq:mft_Cmm}
\llrrangle{\dot{a}^S(t) \, \dot{a}^S(t')} = \delta(t-t') \, \langle \dot{a}^S (t) \rangle.
\end{equation}
Similarly, in the mean-field limit the dendritic spike train is a doubly-stochastic Poisson process and its autocovariance function is
\begin{equation} \label{eq:mft_Cbb}
\llrrangle{\dot{a}^D(t) \, \dot{a}^D(t')} =  \delta(t-t') \, g(t) \langle \dot{a}^S (t) \rangle ,
\end{equation}
with $g(t)\equiv g\big[ G^D \ast \left(E^D + J^D \ast \langle \dot{a}^S \rangle + \beta J^D \ast \langle \dot{a}^D \rangle \right)(t) \big]$.
Finally, the mean-field limit of the cross-covariance of spikes and bursts within a neuron is
\begin{equation} \label{eq:mft_Cmb}
\llrrangle{\dot{a}^S(t) \, \dot{a}^D(t')} = \delta(t-t') g(t') \langle \dot{a}^S (t) \rangle.
\end{equation}
These predictions match well simulations of sparsely connected networks of 100 neurons (Fig.~\ref{fig:fluctuations}). 

 \section{\label{sec:ei} Excitatory-inhibitory networks}
 Inhibitory neurons in the cerebral cortex can be divided into three main types, based on their expression of the proteins parvalbumin (PV), somatostatin (SOM) and 5-HT3A receptor~\cite{rudy_three_2011}. SOM and PV cells provide most of the inhibition onto pyramidal cells.
 Most SOM interneurons are anatomically classified as Martinotti cells, which preferentially target the apical dendrites of pyramidal neurons. Most PV cells are anatomically classified as basket cells, which preferentially target the cell body and proximal dendrites of pyramidal cells. 
 We next investigate how these different types of inhibition regulate collective somatic and dendritic activity in recurrent excitatory-inhibitory networks. We model the inhibitory neurons with a single isopotential compartment, without dendritic calcium spikes.

 \subsection{\label{sec:som} Dendrite-targeting inhibition}
We first examine recurrent networks with SOM inhibitory cells, where both excitatory and inhibitory synapses on pyramidal cells occur on their apical dendrite compartment (Fig.~\ref{fig:som}a). We expect that the inhibitory population may exert direct control over the dendritic activity. To determine how, we study the population rates of somatic, dendritic, and inhibitory spikes. 

 We take symmetric output connectivity: $J_{EE} = J_{IE} \equiv J_E$ and $J_{EI} = J_{II} \equiv J_I$. We take $J_E$ and $J_I$ to be both non-negative.
 For threshold-linear $f_E, f_I, g$ (Eq.~\ref{eq:relu}), the equilibrium mean-field equations are
 \begin{equation} \begin{aligned}
 \langle \dot{a}^S_E \rangle &= \lfloor E^S \rfloor_+ \\
 \langle \dot{a}^D_E \rangle &= \langle \dot{a}^S_E \rangle [ E^D + J_E \langle \dot{a}^S_E \rangle + \beta J_E \langle \dot{a}^D_E \rangle - J_I \langle \dot{a}_I \rangle ]_+^1 \\
 \langle \dot{a}_I \rangle &= \lfloor E_I + J_E \langle \dot{a}^S_E \rangle + \beta J_E \langle \dot{a}^S_E \rangle - J_I \langle \dot{a}_I \rangle \rfloor_+,
 \end{aligned} \end{equation}
a two-dimensional piecewise-polynomial system for the dendritic and inhibitory rates. We summarize the equilibrium solutions to these mean-field equations for threshold-linear $f,g,h$, and their existence and stability conditions, in Table~\ref{table:SOM} (Appendix~\ref{app:mf_fp}). We observe eight types of equilibrium: 
\begin{enumerate} \label{list:ei_fp}
\item a silent state,
\item inhibitory activity alone,
\item[3-5.] pyramidal activity alone with no (3), sparse (4), or saturated (5) dendritic activity,
\item[6-8.] pyramidal and inhibitory activity with no (6), sparse (7), or saturated (8) pyramidal dendritic activity
\end{enumerate}
We observe metastability between activity with silent or active dendrites, which can occur either with or without SOM activity (Fig.~\ref{fig:som}bi vs bii). This metastability is predicted by bistable regions in the mean-field phase diagram, which we will now describe.

If $E^S$ and $E_I$ are both non-positive there is a quiescent state (1), and if $E^S \leq 0$ but $E_I > 0$ there is SOM activity alone (2). When $E^S$ is positive, our analysis of the mean-field fixed points reveals three characteristic types of phase diagram in the $E_I, E^D$ plane, depending on the relative strengths of bursting excitation $\beta J_E$ and inhibition $J_I$ (Fig.~\ref{fig:som}c-e). 

 \begin{widetext}
\begin{center}
 \begin{figure}[ht!] \includegraphics[scale=0.8]{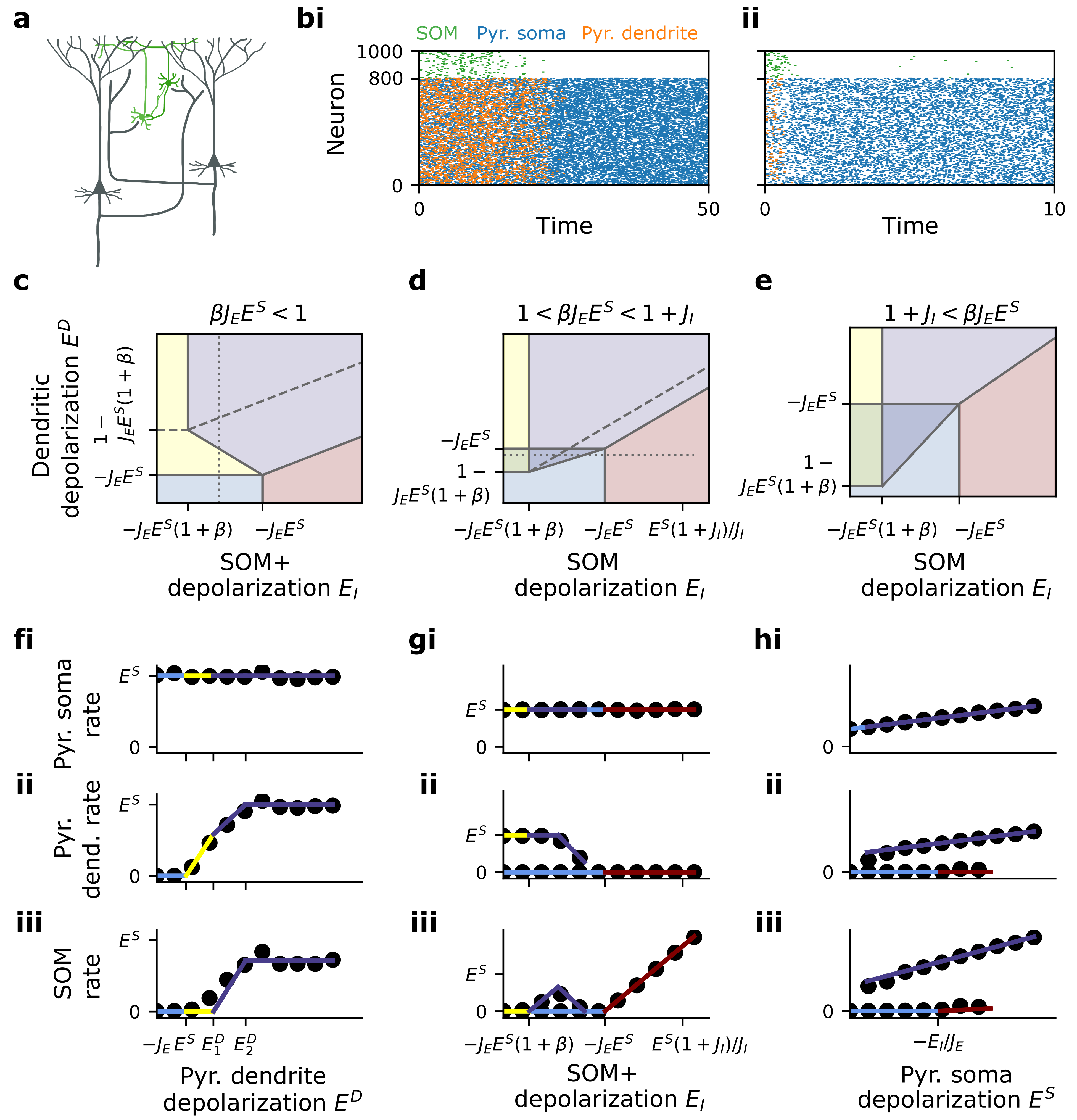}
 \caption{  {\bf a)} Rendition of thick-tufted layer 5 pyramidal cells and SOM inhibitory cells. {\bf b)} Raster plots in two activity regimes: {\bf i)} bistability between full activity and pyramidal soma-only activity, and {\bf ii)} bistability between activity with or without pyramidal dendritic spikes. {\bf c-e)} Slices of the mean-field equilibrium phase diagram in the $(E^D, E_I)$ plane. Yellow: somatic and dendritic activity in pyramidal cells, quiescent SOM cells. Blue: somatic activity in pyramidal cells only. Purple: somatic and dendritic activity in pyramidal cells with SOM activity. Red: Somatic activity in pyramidal cells and SOM activity, no pyramidal dendritic activity. Dashed line: saturation of the dendritic rate above the dashed line. Dotted lines: $(E_I, E^S)$ values for panels f (c) and g (d). {\bf c)} $E^S = 0.2$. Dotted line: $(E_I, E^S)$ values for panel g. {\bf d)} $E^S = 0.43$.  {\bf e)} $E^S = 1$.
 {\bf f)} Rates versus pyramidal dendritic depolarization, $E^D$. $(E^S, E_I) = (0.2, -0.5)$. The x ticks are $E^D_1=-(E_I + J_E E^S)(1-\beta J_E  E^S)/(\beta J_E E^S) - J_E E^S$ and $E^D_2 = 1 + (J_I  E_I - J_E E^S (1+\beta))/(1+J_I)$. {\bf g)} Rates vs SOM depolarization. $(E^S, E^D) = (0.43, -0.4)$. {\bf h)} Rates vs pyramidal somatic depolarization, $E^S$. $(E^D, E_I)=(-0.9, -0.8)$. In panels {\bf c-h}, $(J_E, J_I, \beta) = (3/4, 3/4, 4)$.
 \label{fig:som} }
 \end{figure}
 \end{center}
 \end{widetext}

First, if burst-driven excitation is weak ($\beta J_E < 1/E^S$, or equivalently $\beta J_E E^S < 1$) the phase diagram includes pyramidal activity with no, sparse, or saturated dendritic activity and no inhibitory activity (3-5; Fig.~\ref{fig:som}c,f,  blue, yellow) and full network activity with no, sparse, or saturated dendritic activity (6-8; Fig.~\ref{fig:som}c,f, purple, red). In this case, there is no bistability.

If burst-driven excitation is strong enough ($1 < \beta J_E E^S$), two regions of bistability emerge in the phase diagram (Fig.~\ref{fig:som}d,e). The first region has bistability between pyramidal-only activity with silent or active dendrites (with quiescent inhibitory cells in both states; Fig.~\ref{fig:som}d, green; g, yellow/blue). This region requires sufficiently negative input to the inhibitory cells $E_I$; it corresponds to the bistable region observed in the single-population network (Fig.~\ref{fig:dend-syn}).

As the inhibitory drive $E_I$ increases, we move into a second region of bistability between pyramidal-only activity with silent dendrites and full network activity (Fig.~\ref{fig:som}d,e, dark purple). The pyramidal dendrites' rate is initially saturated, with every somatic event triggering a dendritic calcium spike, and stimulating the inhibitory cells does not affect the pyramidal rates (Fig.~\ref{fig:som}g, purple). As the inhibitory drive increases further, the network moves into a regime with sparse dendritic activity. In this region, stimulation of the inhibitory cells paradoxically reduces their firing rate (Fig.~\ref{fig:som}giii, purple). Here, the paradoxical reduction of inhibitory rates comes with a decrease in the pyramidal cells' dendritic rate but does not affect their somatic event rate (Fig.~\ref{fig:som}gi, ii).
If the inhibitory drive is sufficiently strong, inhibition silences the pyramidal dendrites (Fig.~\ref{fig:som}d,g, red). 

When burst-driven excitation is intermediate ($1 < \beta J_E E^S < 1 + J_I$), the second bistable region may feature sparse or saturated dendritic activity. As burst-driven excitation becomes stronger, the dendritic saturation boundary (Fig.~\ref{fig:som}d, dashed line) meets the boundary between the regions with and without dendritic activity (Fig.~\ref{fig:som}d, boundary between dark purple and blue). After this, the second bistable region always features saturated dendritic activity (Fig.~\ref{fig:som}e,h, purple). 

In summary, dendrite-targeting recurrent inhibition allows for a variety of stationary activity regimes, including bistability between states with silent or active dendrites. The inhibitory cells can control the strength of recurrent dendrite-dependent burst-driven excitation, including silencing the dendrite. 
The bistable regions we uncover in the mean-field theory describe metastable active states in finite-size networks (Fig.~\ref{fig:som}b).
Finally, we found a paradoxical response to inhibitory stimulation that appears in the dendritic rate without affecting the somatic event rate (because all recurrent inputs target the dendrite).

 \subsection{\label{sec:pv} Soma-targeting inhibition}
We next examine PV inhibition, targeting the pyramidal cells' somatic compartment (Fig.~\ref{fig:pv}a). 
Similarly as for the network with SOM interneurons, we assume symmetric mean output synaptic weights for each population and threshold-linear intensity functions to obtain the equilibrium mean-field equations
 \begin{equation} \begin{aligned} \label{eq:mf_equil_pv}
 \langle \dot{a}^S_E \rangle =& \lfloor E^S - J_I \langle \dot{a}_I \rangle  \rfloor_+ \\
 \langle \dot{a}^D_E \rangle =& \langle \dot{a}^S_E \rangle [ E^D + J_E \langle \dot{a}^S_E \rangle + \beta J_E \langle \dot{a}^D_E \rangle  ]_+^1  \\
 \langle \dot{a}_I \rangle =& \lfloor E_I + J_E \langle \dot{a}^S_E \rangle + \beta J_E \langle \dot{a}^D_E \rangle - J_I \langle \dot{a}_I \rangle \rfloor_+
 \end{aligned} \end{equation}
 We list the solutions to this piecewise polynomial system and their existence and stability conditions in Table~\ref{table:PV} (Appendix~\ref{app:mf_fp}). 
 There are the same eight types of equilibria as listed previously for the pyramidal-SOM network, describing stationary regimes of the spiking network.
We also observe metastability between these states, including between full network activity and pyramidal-only activity without bursting (Fig.~\ref{fig:pv}bi) and between full network activity with and without bursting (Fig.~\ref{fig:pv}bii). We next describe the various regions of the mean-field phase diagram.
 
There is a quiescent regime if $E^S \leq 0 $ and $E_I \leq 0$, and PV-only activity if $E^S \leq 0$ but $E_I > 0$. When both $E^S$ and $E_I$ are above threshold, our analysis of the mean-field equations again reveals three characteristic types of phase diagram in the $E_I, E^D$ plane, depending on the relative strengths of burst-related excitation and inhibitory feedback (Fig.~\ref{fig:pv}c-e).

First, if dendrite-dependent burst-related excitation is weak ($\beta J_E E^S < 1$), we can observe all types of active fixed point: PV-only activity (2; Fig.~\ref{fig:pv}c, orange), pyramidal-only activity with no, sparse, or saturated dendritic activity (3-5; Fig.~\ref{fig:pv}c,f blue, yellow), and pyramidal and PV activity with no, sparse, or saturated pyramidal dendritic activity (6-8; Fig.~\ref{fig:pv}c,f, purple).

When burst-driven excitation is intermediate ($1 < \beta J_E E^S < \frac{1 + J_I(1+J_I)}{1+J_I}$), two regions of bistability emerge (Fig.~\ref{fig:pv}d,e). The first is bistability between pyramidal-only activity with silent or active dendrites (Fig.~\ref{fig:pv}d, green; g, yellow/blue). This region requires sufficiently negative input to the inhibitory cells to shut them off, as in the network with SOM inhibition.

As the inhibitory drive increases, that bistable region transitions into bistability between pyramidal-only activity with silent dendrites and full network activity (Fig.~\ref{fig:pv}d, dark purple; g, blue/purple). As in the network with SOM cells, the pyramidal dendritic rate is initially saturated and increasing inhibitory drive suppresses pyramidal activity. As the inhibitory drive increases further, the network moves into a regime with sparse dendritic activity. Here, a small increase in the inhibitory drive leads to a paradoxical suppression of the inhibitory rate (Fig.~\ref{fig:pv}giii, purple). This paradoxical reduction of the inhibitory rate comes with a decrease in the pyramidal cells' dendritic rate but an increase in the somatic event rate (Fig.~\ref{fig:pv}gi,ii, purple). 

From that region of the phase diagram, a further increase in the inhibitory drive reduces the pyramidal activity enough that their dendritic input falls below threshold (Fig.~\ref{fig:pv}d,g, red). In contrast to the SOM network, further increasing the inhibitory drive can also silence the pyramidal cells' somatic activity (Fig.~\ref{fig:pv}d,g, orange).

 \begin{widetext}
\begin{center}
 \begin{figure}[ht!] \includegraphics[scale=0.8]{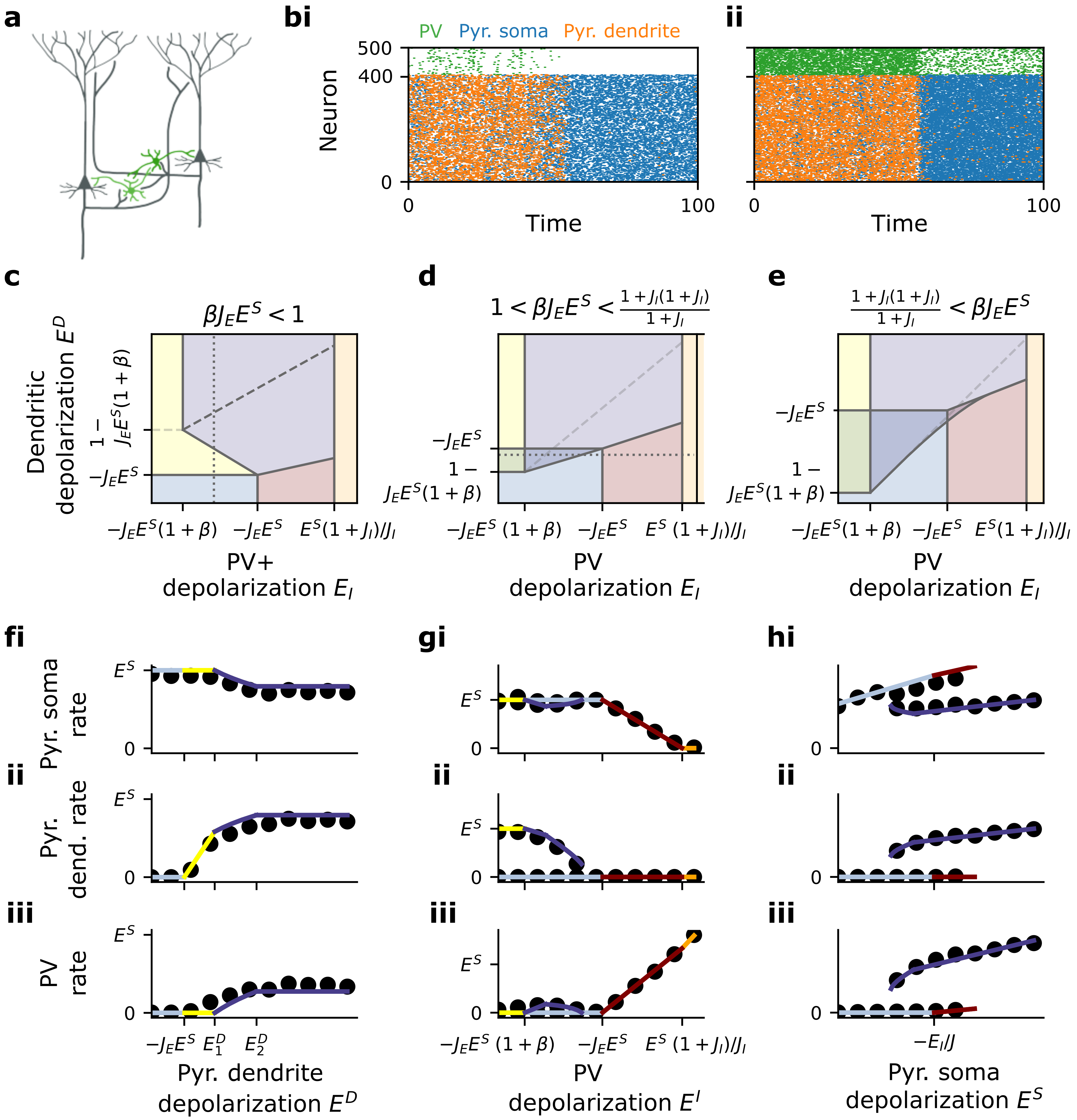}
 \caption{  {\bf a)} Rendition of thick-tufted layer 5 pyramidal cells and PV inhibitory cells. {\bf b)} Raster plots in two activity regimes: {\bf i)} bistability between 1) sparse pyramidal dendrite and PV activity and 2) silent pyramidal dendrites and PV cells, and {\bf ii)} bistability between 1) sparse pyramidal dendrite and PV activity and 2) silent pyramidal dendrites with PV activity. {\bf c-e)} Slices of the mean-field equilibrium phase diagram in the $(E^D, E_I)$ plane. Yellow: somatic and dendritic activity in pyramidal cells, quiescent PV cells. Blue: somatic activity in pyramidal cells only. Purple: somatic and dendritic activity in pyramidal cells with PV activity. Red: Somatic activity in pyramidal cells and PV activity, no pyramidal dendritic activity. Orange: PV activity alone, no pyramidal activity. Dashed line: saturation of the dendritic rate above the dashed line. Dotted line: $(E_I, E^S)$ values for panel f. {\bf c)} $E^S = 0.2$. Dotted line: $(E_I, E^S)$ values for panels f ({\bf c}) and g ({\bf d}). {\bf d)} $E^S = 0.43$.  {\bf e)} $E^S = 1$.
 {\bf f)} Rates versus pyramidal dendritic depolarization, $E^D$. $(E^S, E_I) = (0.2, -0.5)$. The x ticks are $E^D_1=-[E_I/(J_E E^S) + 1](1 - \beta J_E E^S)/\beta - J_E E^S$ and 
 $E^D_2 = 1 - J_E(1+\beta)\{E^S - J_I[E_I + J_E E^S(1+\beta)]\} / \{1 + J_I[1+J_E(1+\beta)]\}$. 
 {\bf g)} Rates vs PV depolarization. $(E^S, E^D) = (0.43, -0.4)$. {\bf h)} Rates vs pyramidal somatic depolarization, $E^S$. $(E^D, E_I)=(-0.9, -0.8)$. In panels {\bf c-h}, $(J_E, J_I, \beta) = (3/4, 3/4, 4)$.
 \label{fig:pv} }
 \end{figure}
\end{center}
\end{widetext}

When burst-driven excitation is strong ($\frac{1 + J_I(1+J_I)}{1+J_I} < \beta J_E E^S $), a third bistable region emerges (Fig.~\ref{fig:pv}e,h, purple/red). This region features bistability between pyramidal somatic and PV activity without dendritic activity (Fig.~\ref{fig:pv}e,h, red) and full network activity (Fig.~\ref{fig:pv}e,h, purple). 

In summary, the network with PV inhibition has the same regions of fixed-point bistability as the SOM network does and an additional region with bistability between pyramidal and PV activity with silent or active pyramidal dendrites. It also has a region of paradoxical response to inhibitory stimulation. In the pyramidal-PV network the paradoxical response to inhibitory stimulation features increased pyramidal somatic activity, while in the pyramidal-SOM network the pyramidal somatic activity is unaffected. In both networks, the paradoxical response features reduced pyramidal dendritic activity.

\begin{figure}[ht!] \includegraphics[scale=.8]{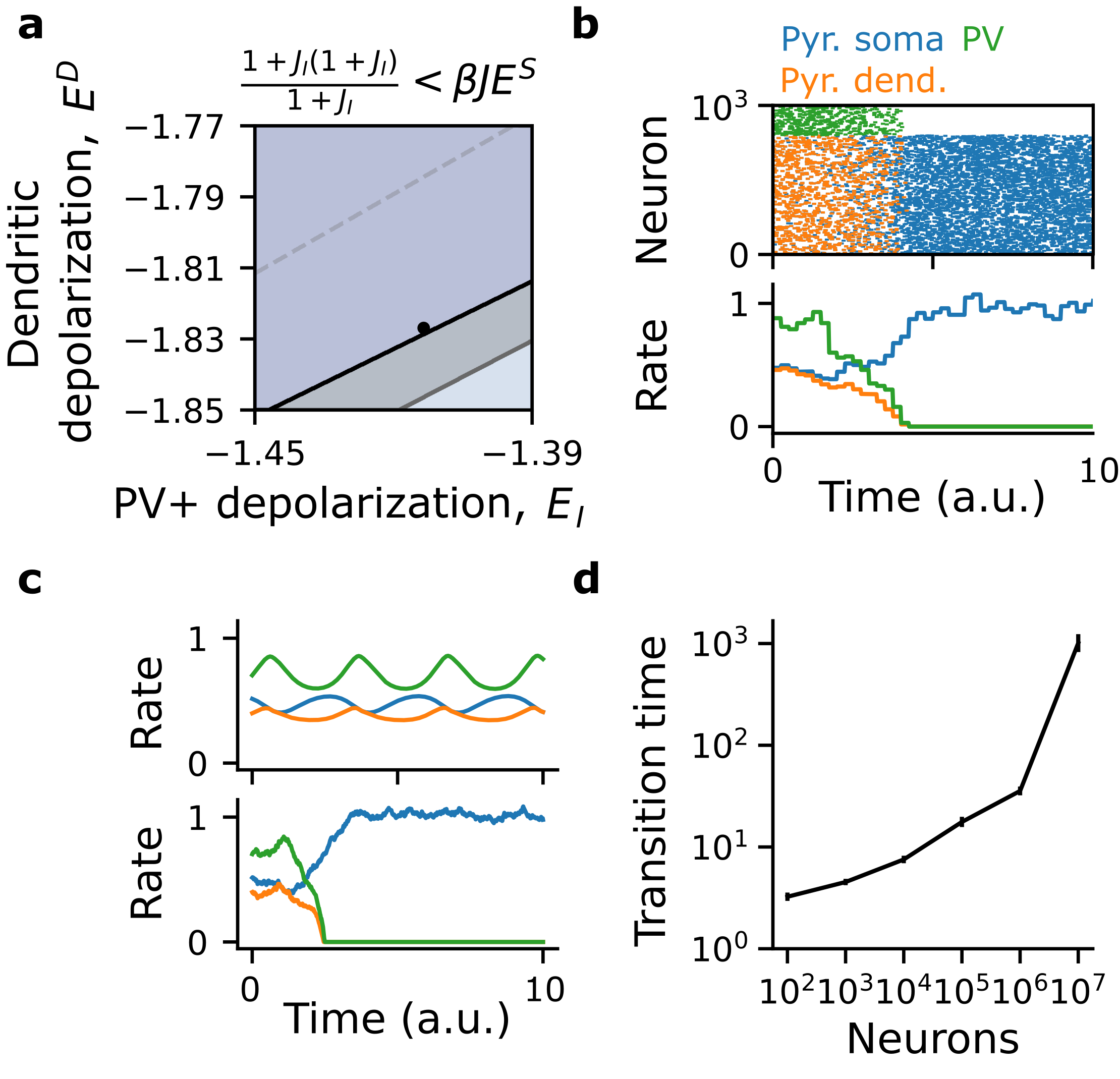}
 \caption{  {\bf a)} Phase diagram in the $E_I, E^D$ plane for $(J_E, J_I, \beta, E^S) = (3/4, 3/4, 8, 1)$, zoomed in around the location of the Hopf bifurcation. Here we show boundaries that involve the appearance of unstable fixed points. 
 Light blue: Pyramidal-only activity with silent dendrites is the only equilibrium and it is stable. gray: stable pyramidal activity with silent dendrites, and unstable full network activity. Purple: Stable full network activity and stable pyramidal-only activity with silent dendrites. Dashed gray line: the pyramidal dendritic rate saturates above the dashed line. Black line: Hopf bifurcation of the full network activity state. Gray line: appearance of the full network activity state. Black dot: parameter for panel b. {\bf b)} Example simulation of the stochastic network. Top: raster plot. Blue: pyramidal spikes. Orange: pyramidal bursts. Green: PV spikes. {\bf c)} Simulation of the mean-field dynamics (top) and the stochastic mean-field dynamics, Eq.~\ref{eq:pv_finite_size} (bottom). {\bf d)} Transition times from the limit cycle to the fixed point in the stochastic mean-field dynamics. Error bars: standard error of the mean. 
 Example raster plot at the parameter value marked with a black dot in panel a. 
 \label{fig:hopf} }
 \end{figure}

In the PV network, we also find a subcritical Hopf bifurcation of a full network-activity state. 
From the region where the only stable fixed point is pyramidal somatic-only activity (Fig.~\ref{fig:hopf}a, light blue), increasing the dendritic drive $E^D$ leads to the appearance of an unstable state with full network activity (Fig.~\ref{fig:hopf}a, gray) which then stabilizes through a subcritical Hopf bifurcation (Fig.~\ref{fig:hopf}a, black line). Numerically continuing the unstable limit cycle from the bifurcation, we saw that it undergoes a fold to generate a region of tristability between two fixed points and a limit cycle (not shown).
Simulating of the stochastic spiking network in this region, we observed that initial conditions on the stable limit cycle quickly transitioned to the stationary state with pyramidal somatic activity (Fig.~\ref{fig:hopf}b). 

Integrating the mean-field voltage dynamics confirmed the presence of a stable limit cycle (Fig.~\ref{fig:hopf}c, top). Since the mean-field theory is exact in the limit $N \to \infty$, the transitions are driven by finite-size fluctuations in the synaptic field. 
To characterize these transitions, we examine a heuristic finite-size mean-field theory for the PV network:
\begin{equation} \begin{aligned} \label{eq:pv_finite_size}
\dot{v}^S =& -v^S + E^S - J_I \lfloor v_I \rfloor_+ + \frac{\xi^S}{\sqrt{N}} \\
\dot{v}^D =& -v^D + E^D + J_E \lfloor v_E \rfloor_+ + \beta J_E [v^D]_+^1 + \frac{\xi^D}{\sqrt{N}} \\
\dot{v}_I =& -v_E + E_I + J_E \lfloor v_E \rfloor_+ + \beta J_E [v^D]_+^1 - J_I \lfloor v_I \rfloor_+ + \frac{\xi_I}{\sqrt{N}},
\end{aligned} \end{equation}
where $\boldsymbol{\xi}$ are zero-mean, uncorrelated, white Gaussian processes. The $1/\sqrt{N}$ scaling of the finite-size fluctuations is inherited from the $1/N$ scaling of the synaptic weights. These dynamics neglect temporal correlation of the fluctuations and their potential correlation across pyramidal compartments or pyramidal and PV cells. Despite this simplification, these dynamics display similar transitions from the stable limit cycle to the pyramidal soma-only fixed point as in the spiking network (Fig.~\ref{fig:hopf}c, bottom). We then calculated the lifetime of the limit cycle in Eq.~\ref{eq:pv_finite_size} as a function of $N$ (Fig.~\ref{fig:hopf}d). The mean lifetime of the limit cycle was less than 10 time units for $N \leq 10^4$. Long persistence in the limit cycle required very large networks of $10^5$ or more neurons. This suggests that in the $(v^S, v^D, v_I)$ phase space, the limit cycle lies very near a transition path to that one stable fixed point so that even small fluctuations can destabilize the orbit. 

 \section{Discussion}
 Here, we developed a statistical field-theoretic approach to multi-compartmental neuron models with stochastic spike emission. We applied it to recurrent networks with dendritic calcium spike-dependent bursting, using a single-neuron model developed in \cite{naud_sparse_2018}
 We examined the impact of the compartmental targeting of excitatory and inhibitory synapses on the networks' mean-field equilibrium phase diagrams.  

In our recurrent model networks, the interaction between the soma and dendrite allows network activity with no, sparse, or saturated dendritic calcium spikes and bistability between activity with silent or active dendrites. With recurrent SOM inhibition targeting pyramidal dendrites, there was also bistability between full network activity and pyramidal soma-only activity. Recurrent PV inhibition targeting pyramidal somata could silence the pyramidal cells fully, while the SOM inhibition could only silence their dendrites. Recurrent PV inhibition also generated a subcritical Hopf bifurcation of the fully active state and a corresponding region of tristability between a fully active stationary state, stationary pyramidal soma-only activity, and a limit cycle involving the pyramidal somatic and dendritic and inhibitory activity. However, for the parameter values we explored that limit cycle had short lifetimes due to finite-size fluctuations in the net synaptic input. A rigorous, rather than heuristic, finite-size theory could expose explicit predictions for those transition rates~\cite{schmutz_finite-size_2021, vinci_self-consistent_2023, fleurantin_dynamical_2023}.

In both types of excitatory-inhibitory network, we observed a regime of paradoxical reduction of inhibitory rates after stimulation of inhibitory cells~\cite{tsodyks_paradoxical_1997}. These paradoxical responses have been widely observed in the mouse cortex~\cite{sanzeni_inhibition_2020}. In the classic excitatory/inhibitory model, that paradoxical reduction of inhibitory rates after inhibitory stabilization is due to a loss of recurrent excitation; excitatory rates are also reduced. Here, we saw that paradoxical reductions in inhibitory activity after inhibition stimulation can come with either unchanged (SOM network) or increased (PV network) pyramidal firing rates. In both cases, the paradoxical response still involved a reduction of pyramidal dendritic activity. In single-compartment networks with PV and SOM cells, the paradoxical response involves a reduction in the net inhibitory input to pyramidal cells~\cite{litwin-kumar_inhibitory_2016}. 
The dynamics of inhibitory stabilization in compartmental networks with multiple inhibitory and excitatory cell types remain to be analyzed.

A variety of other important neuron types also generate somatic or dendritic calcium spikes. 
These include thalamic relay neurons, cerebellar Purkinje cells, and layer 2/3 and hippocampal pyramidal cells~\cite{brumberg_ionic_2000, schwartzkroin_probable_1977, wong_intradendritic_1979, magee_dendritic_1999, larkum_dendritic_2007, jahnsen_ionic_1984, destexhe_dendritic_1998, eccles_excitatory_1966, davie_origin_2008}.
The methods we developed here may thus be useful in analyzing the dynamics of thalamocortical sensorimotor processing, cerebellar motor control, and hippocampal spatial and relational processing and memory dynamics. 

\subsection*{Limitations and future directions}
We analyzed a highly simplified model of thick-tufted layer 5 pyramidal cells. 
This model assumed linear subthreshold voltage dynamics, electrotonically isolated compartments, current-based synaptic interactions, and a piecewise-linear maps from the somatic and dendritic voltages to spike probabilities. 
Nonlinear voltage-rate intensity functions can generate additional qualitative dynamics \cite{paliwal_metastability_2025}. This model could also be extended by including current or conductance-based interactions between the somatic and dendritic compartments, as increases in dendritic calcium concentrations can shape somatic responses even without a dendritic calcium spike \cite{palmer_cellular_2012}). 
The model we studied assumes that dendritic calcium spikes are only generated by backpropagating action potentials (Eq.~\ref{eq:burst_conversion}). 
In more detailed biophysical models, sufficiently strong dendritic stimulation can also trigger calcium spikes, although it is not clear whether this occurs in vivo \cite{hay_models_2011}. 
Despite the model's simplicity, our mean-field theory uncovered a variety of macroscopic dynamics. It remains to be seen whether these predictions are accurate with more detailed single-cell models.

Here, we examined excitatory-inhibitory networks with excitatory synapses targeting the apical dendrite. 
Long-range cortical projections of thick-tufted layer 5 pyramidal cells  target both layer 5 and the superficial layers, and thus could potentially provide both forms of input. 
Synapses between nearby thick-tufted layer 5 pyramidal cells in the juvenile rat target both their basal and apical dendrites, with a bias towards the basal dendrites \cite{markram_physiology_1997, markram_network_1997}. The majority of their local synapses are, however, onto inhibitory neurons~\cite{bodor_synaptic_2025}. The excitatory-inhibitory network models studied here could thus serve a preliminary examination of the homogenous equilibria of networks with local inhibitory connectivity and long-range excitation.
The methods we developed here can be directly applied to networks including multiple pyramidal and inhibitory cell types; how the relative spatial profiles of their projections to, and inputs from, different inhibitory and excitatory cell types and different compartments affects pattern formation remains an interesting open question.

Thick-tufted layer 5 pyramidal neurons can also generate other types of action potential in their dendrites, including NMDA spikes in the basal and apical tuft dendrites \cite{schiller_nmda_2000, nevian_properties_2007, larkum_synaptic_2009, larkum_guide_2022}. The methods developed here can be applied to models with finer dendritic compartmentalizations to study the interaction of different types of dendritic spike.

Finally, calcium is an important signal for long-term synaptic plasticity~\cite{rubin_calcium_2005, shouval_spike_2010, graupner_calcium-based_2012}. Dendritic calcium spikes drive synaptic plasticity in pyramidal cells during learning~\cite{bittner_conjunctive_2015, cichon_branch-specific_2015, bittner_behavioral_2017}.
Here, we analyzed the mean within-neuron somatic/dendritic spike correlations in the large-$N$ limit. These methods could also be used to calculate the correlation between different neurons in finite-size networks, and thus to predict synaptic plasticity driven by dendritic calcium spikes.

\section*{Acknowledgments}
The authors were supported by a grant from the Allen Institute Mindscope Phase 4 program. 

\section*{Data and code availability}
Related code and data can be found at \url{https://github.com/gkocker/burst_mft.git}

\appendix

\section{Model} 
We describe our model in discrete time first, before taking a continuous-time limit. We consider a network of $N$ neurons, the first $N_E$ of which are pyramidal cells and the remainder inhibitory cells.
Each neuron $i$ has a set of compartment, $C_i$.
$v_{i,t}^c$ is the membrane voltage of compartment $c$ in neuron $i$ at time $t$. 
The neuron may generate different types of spike.
So, each neuron has an activity vector: $da^d_{j,t} \in \{0, 1 \}$ is spike type $d$ of neuron $j$ at time $t$.
We assume that each compartment generates one type of action potential.
Here, we assume the membrane voltages evolve according to:
\begin{equation} \label{equ 1: voltage_discrete}
v^{c}_{i, t+1} = v^{c}_{i, t} + \frac{dt}{\tau_i^{c}} \left(E^{c}_i-v^c_{i,t} +  \sum_{j=1}^{N} \sum_{d \in C_j} \sum_{s=1}^{T-1} J_{ij, s}^{cd} \; da^d_{j, t-s} \right)
\end{equation}
In the above equations, $E^c$  are the leak reversal potentials (and/or external drives) of compartment $c$ in neuron $i$. The last term in Eq.~\ref{equ 1: voltage_discrete} models the input to compartment $c$ due to presynaptic activity of type $d$. $J_{ij}^{cd}$ contain the weighted synaptic filters for compartment $c$ of neuron $i$, in response to inputs of type $d$ in neuron $j$. 
The spike increments $da_{i,t}^c$ are generated as conditionally Bernoulli random variables. 

Here, we study pyramidal neurons with somatic and apical dendrite compartments ($C_i = \{S, D\}$) and inhibitory neurons with a somatic compartment ($C=\{S\}$). 
We introduce the model for the pyramidal neurons; the inhibitory neurons follow the same construction without the dendritic compartment. 

The pyramidal cells generate somatic events $da^S_{i,t}$ with event probability $f(v^S_{i,t}) \; dt$, where $v^S_{i,t}$ is the somatic membrane voltage. Each somatic event models a Na$^+$/K$^+$ action potential. One of these somatic events may trigger a dendritic calcium spike. These are also generated as conditionally Bernoulli random variables with success probability $da^S_{i,t} \; g(v^D_{i,t})$, where $v^D_{i,t}$ is the dendritic membrane voltage. The function $g(v^D_{i,t})$ determines the probability that a somatic action potential will trigger a dendritic calcium spike.

As described in the main text, we assume that each dendritic calcium spike triggers a stereotyped burst of somatic Na$^+$/K$^+$ action potentials. This is why we call $da^S$ the somatic events, rather than somatic spikes. A dendritic calcium spike/burst has $(da^S_{i,t}, \; da^D_{i,t})=(1,1)$ while an isolated somatic spike has $(da^S_{i,t}, \; da^D_{i,t})=(1,0)$.
We will employ pulse coupling such that $J_{ij,s} = J_{ij} \delta_{s,1}$, but present our derivations with general synaptic filters. 

\begin{table}[ht]
\begin{tabular}{ | m{3 cm} | m{5 cm} | } \hline
Variable/Term & Interpretation \\
  \hline
  $v^c_{i,t}$ & membrane potential of compartment $c$ of neuron $i$ at time step $t$ \\ 
  \hline
  $\tau^c_i$ & membrane potential time constant of compartment $c$ of neuron $i$ \\ 
  \hline
  $E^c_i$ & external input / depolarization / excitability / leak/reversal potential of compartment $c$ of neuron $i$ \\ 
  \hline
  $N$ & total number of neurons \\ 
  \hline
  $T$ & total number of time steps \\ 
  \hline
  $J^{cd}_{ij, s}$ & impact of activity of type $d$ in neuron $j$ on the membrane voltage in compartment $c$ of neuron $i$ at time lag $s \cdot dt$ \\ 
  \hline
  $da^{d}_{i, t} \in \{0,1\}$ & action potential of type $d$ generated by neuron $i$ at time step $t$ \\
  \hline
  $f (v^S_{i,t}) $ & intensity function for somatic spike generation of cell $i$ at time step $t$ \\
  \hline
  $g (v^D_{i,t}) $ & probability that a somatic spike in neuron $i$ at time $t$ triggers a dendritic calcium spike \\
  \hline
\end{tabular}
\caption{Model variables.}
\end{table}

\onecolumngrid
\section{Joint probability density functional of voltage and activity} \label{app:density}
We write the conditional joint probability density function of the membrane potentials and activity as
\begin{equation} \label{equ 2: JPDF} \begin{aligned}
p ({\bf v}, d{\bf a} \; \vert \; {\bf J}, \boldsymbol{\chi})  =& \prod_{t=0}^{T-1} \prod_{i=1}^{i=N} \prod_{c \in C_i} \delta \left(da^c_{i,t}-\chi^c_{i,t} \right) \delta \left( v^{c}_{i, t+1} - \left[ v^c_{i, t} + \frac{dt}{\tau^c} \left(E^c_i-v^c_{i,t}  + \sum_{j=1}^{N} \sum_{d \in C_j} \sum_{s=1}^{T-1} J_{ij, s}^{cd} \; da^d_{j, t-s} \right) \right] \right),
\end{aligned} 
\end{equation}    
where $C_i$ is the set of compartment indices associated with neuron $i$: $C_i = \{S, D\}$ for $i =1, \ldots, N_E$ and $C_i = \{S\}$ for $i=N_E+1, \ldots, N$.
$\chi^S_{i,t}$ and $\chi^D_{i,t} $ are Bernoulli random variables with success probabilities $f(v^S_{i,t}) \; dt$ and $da^S_{i,t} \; g(v^D_{i,t})$ respectively. Introducing the Fourier representation of those delta functions yields
\begin{align} \label{equ 3: FT JPDF} 
p ({\bf v}, d{\bf a} \; \vert \; {\bf J}, \boldsymbol{\chi}) =& \int D \tilde{\bf v} \int D \tilde{\bf a} \; \exp \left[ -S({\bf v}, d{\bf a}, \tilde{\bf v}, \tilde{\bf a}, \boldsymbol{\chi}) \right], \\ \notag
S({\bf v}, d{\bf a}, \tilde{\bf v}, \tilde{\bf a}, \boldsymbol{\chi}) =& 
\sum_{t=0}^{t=T} \sum_{i=1}^{N}  \sum_{c \in C_i} \tilde{a}^c_{i,t} \left( da^c_{i,t} - \chi^c_{i,t} \right) + \tilde{v}^c_{i, t} \Biggl ( v^{c}_{i, t+1} - \left[ v^c_{i, t} + \frac{dt}{\tau_i^c} \left(E^c_{i} -v^c_{i,t}  + \sum_{j=1}^{N} \sum_{d \in C_j} \sum_{s=1}^{T-1} J_{ij, s}^{cd} \; da^d_{j, t-s} \right) \right] \Biggr )
\end{align}  
where $D \tilde{\bf v}= \prod_{t=0}^{T-1}  \prod_{i=1}^{i=N}  \prod_{c \in C_i} \frac{d\tilde{v}_{i,t}}{2 \pi i}$ and similarly for $D\tilde{\bf a}$, and the negative exponent, $S$, is called the action. The integrals over the response variables, $\tilde{v}_{i,t}^c$ and $\tilde{a}_{i,t}^c$, are along the imaginary axis. We marginalize over the auxiliary noise variables $\boldsymbol{\chi}$. Since the action is linear in $\boldsymbol{\chi}$, this introduces the cumulant generating functions of $\boldsymbol{\chi}$ into the action:
\begin{align} \label{equ 4: FTJPDF with CGF, marginalized noise}
p ({\bf v}, d{\bf a} \; \vert \; J) =& \int D \tilde{\bf v} \int D \tilde{\bf a} \; \exp \left[ -S({\bf v}, d{\bf a}, \tilde{\bf v}, \tilde{\bf a}) \right], \\ 
\notag S({\bf v}, d{\bf a}, \tilde{\bf v}, \tilde{\bf a}) =& 
\sum_{t=0}^{t=T} \sum_{i=1}^{N}  \sum_{c \in C_i} \tilde{a}^c_{i,t} da^c_{i,t} - W_{c}(\tilde{a}_{i,t}^c) + \tilde{v}^c_{i, t} \Bigg[ v^{c}_{i, t+1} - \Bigg( v^c_{i, t} + \frac{dt}{\tau_i^c} \left( E^c_{i}-v^c_{i,t} +  \sum_{j=1}^{N} \sum_{d \in C_j} \sum_{s=1}^{T-1} J_{ij, s}^{cd} \; da^d_{j, t-s} \right) \Bigg) \Bigg] 
\end{align}
where $W_{c}(\tilde{a}_{i,t}^c)  = \ln \langle \exp \tilde{a}_{i,t}^c \rangle_{\chi^c}$. Since the spikes and bursts are Bernoulli,
\begin{align} \label{equ 5: CGF m}
W_{S}(\tilde{a}^S_{i,t}) =& \ln \left[1 + f(v_{i,t}^S) \; dt \; \left(e^{\tilde{a}^S_{i,t}} -1 \right) \right] \\ 
\label{equ 6: CGF b} W_{D}(\tilde{a}^D_{i,t}) =& \ln \left[1 + da^S_{i,t} \; g(v^D_{i,t}) \; \left(e^{\tilde{a}^D_{i,t}} -1 \right) \right] \\
\notag =& da^S_{i, t} \ln \left[1 + g(v^D_{i,t}) \; \left(e^{\tilde{a}^D_{i,t}} -1 \right) \right].
\end{align} 
We take the continuous time limit, $dt \rightarrow 0$ and $T \rightarrow \infty$ with their product fixed, and expand $W_S$ around one to obtain
\begin{align} \label{eq:msrdj_density}
p[{\bf v}, \dot{\bf a} \vert {\bf J}] =& \int \mathcal{D}\tilde{\bf v} \int \mathcal{D}\tilde{\bf a} \; \exp \left(-S[{\bf v}, \dot{\bf a}, \tilde{\bf v}, \tilde{\bf a}, {\bf J}] \right), \\
\notag S[{\bf v}, \dot{\bf a}, \tilde{\bf v}, \tilde{\bf a}, {\bf J}] =& \int dt \; \sum_{i=1}^{N}  \sum_{c \in C_i} \tilde{a}^c_{i}(t) \dot{a}^c_{i}(t) - \delta_{c S} \,  \bigl( e^{\tilde{a}_i^S(t)}-1 \bigr) f(v_i^S(t)) -  \delta_{cD} \, \dot{a}_i^S(t) \ln \left[1 + g(v^D_i(t)) \bigl( e^{\tilde{a}^D_i(t)} - 1 \bigr)  \right] \\
\notag &\hspace{2cm} + \tilde{v}^c_{i}(t) \Bigg[\dot{v}_{i}^c(t) - \frac{1}{\tau^c} \left( E^c_{i}-v^c_{i}(t)  + \sum_{j=1}^{N} \sum_{d \in C_j} J_{ij}^{cd} \ast \; \dot{a}^d_{j}(t) \right) \Bigg]
\end{align} 
where $\mathcal{D}\tilde{\bf v} = \lim_{T \rightarrow \infty} D\tilde{\bf v}$, similarly for $\mathcal{D} \tilde{\bf a}$, and $\dot{x} = dx/dt$.
With the functional inner product ${\bf x}^T {\bf y} = \int dt \; \sum_i \sum_{c \in C_i} x_i^c(t) y_i^c(t)$, and similarly for the inner product of vector-valued functions of time, we write this in condensed notation as
\begin{equation} \begin{aligned} \label{equ 8: condensed action}
S[{\bf v}, \dot{\bf a}, \tilde{\bf v}, \tilde{\bf a}, {\bf J}] =&  \tilde{\bf a}^T \dot{\bf a} -  W[\tilde{\bf a}; \boldsymbol{\mu}({\bf v})] + \tilde{\bf v}^T \left(\boldsymbol{\tau} \dot{\bf v} + {\bf v} - {\bf E} - {\bf J} \ast \dot{\bf a}\right), \\
W[\tilde{\bf a}; \boldsymbol{\mu}({\bf v})] =& \left(e^{\tilde{\bf a}^S}-1 \right)^T {\bf f}({\bf v}^S) + (\dot{\bf a}^S)^T \ln \left[1 + {\bf g}({\bf v}^D) \; \left(e^{\tilde{\bf a}^D} -1 \right) \right],
\end{aligned} \end{equation}
where $\mu_i^S(t) = f(v_i^S(t))$, and $\mu^D_i(t) = g(v^D_i(t))$.

\section{Large-$N$ limit and average over the quenched disorder} \label{app:mft}
To obtain a reduced, lower-dimensional description of the collective activity in large networks, we marginalize out the synaptic weights from the density functional.
This assumes that the system is self-averaging with respect to the connectivity: that the average over realizations of J will give us an accurate description of a single large system. 

\begin{equation} \begin{aligned} 
p [\bf{v}, \dot{\bf a}] &= \int  \mathcal{D} \bf{J} \; p(\bf{J}) \, p [\bf{v} ,\dot{\bf a} | \bf{ J} ] \\
&= \int \mathcal{D} \tilde{\bf v} \int \mathcal{D} \tilde{\bf{a}} \, \exp \left(-\left \{ \tilde{\bf a}^T \dot{\bf a} -  W[\tilde{\bf a}; \boldsymbol{\mu}({\bf v})] + \tilde{\bf v}^T \left(\dot{\bf v} + {\bf v} - {\bf E} \right) \right \} \right)
\int \bf{\mathcal{D} J} \; p(\bf{J}) \; e^{ \tilde{\bf v}^T \bf{J} \ast \dot{\bf a} }  \\ 
&= \int \mathcal{D} \tilde{\bf v} \int \mathcal{D} \tilde{\bf{a}} \, \exp \left(-\left\{\tilde{\bf a}^T \dot{\bf a} -  W[\tilde{\bf a}; \boldsymbol{\mu}({\bf v})] + \tilde{\bf v}^T \left(\dot{\bf v} + {\bf v} - {\bf E} \right) - W_J(\tilde{\bf v} \dot{\bf a}) \right \} \right)
\label{eq:density_marginalize_weights} 
\end{aligned} \end{equation}
\twocolumngrid
so that a cumulant generating function of the synaptic weights appears in the action. We then consider the limit of a large network, taking $N \rightarrow \infty$.
We assume that the elements of ${\bf J}^{cd}$ are drawn from a distribution with mean of order $1/N$ and negligible higher cumulants.  (For example, if the connection probability is order 1 and ${\bf J} \sim 1/N$, the $n$th cumulant of $J^{cd}_{ij}$ scales as $N^{-n}$, so we can truncate $W_J$ at first order when taking the large network limit.) The synaptic connectivity matrix is then effectively replaced with its first cumulant.

Here, we endow the mean connectivity with a block structure reflecting the connectivity between neuron types.
We assume that the neurons are divided into $T$ types, with $N_t$ neurons of type $t$. Each neuron type $t$ has a common set of compartments $C_t$, so that the net synaptic field onto compartment $c$ of any neuron $i$ can be organized as 
\begin{equation}
\label{eq:net_input}
\sum_{t=1}^T \sum_{d \in C_t}  \frac{J^{cd}_{st}}{N_t} \sum_{j=1}^{N_t} \dot{a}_j^d
\end{equation}
where $J^{cd}_{st}/N$ is the mean strength of activity type $d$ in neuron type $t$ onto compartment $c$ of neurons of type $s$.

In the limit of a large network, with the number of cell types fixed, the sample mean in Eq.~\ref{eq:net_input} concentrates around the population mean of each activity type $d$, $\langle \dot{a} \rangle$, and the density factorizes into that of $N$ independent neurons each receiving the same mean-field input. This can be seen through a standard auxiliary-field calculation~\cite{helias_statistical_2020}. This concentration, which justifies the self-averaging assumption above, yields the reduced joint density of the typical neuron's voltage in the somatic and dendritic compartments and event trains:
\begin{equation} \begin{aligned} \label{eq:reduced_action}
S[{\bf v}, \dot{\bf a}, \tilde{\bf v}, \tilde{\bf a}] =&  \tilde{\bf a}^T \dot{\bf a} -  W[\tilde{\bf a}; \boldsymbol{\mu}({\bf v})] + \tilde{\bf v}^T \left(\dot{\bf v} + {\bf v} - {\bf E} - {\bf J} \ast \langle \dot{\bf a} \rangle \right)
\end{aligned} \end{equation}
where now the inner products involve sums over neuron types instead of neurons. If there are $T$ types, each with $\lvert C_t \rvert$ compartments, the mean-field variables have dimensionality $\sum_{t=1}^T \lvert C_t \rvert$.

\subsection{Mean-field expansion for the activity} \label{app:mft_expansion}
We marginalize out the membrane potentials to obtain the mean-field limit of the density functional of the activity:
\begin{equation} \begin{aligned} \label{eq:mf_density_activity}
p[\dot{\bf a}] =& \int \mathcal{D}\tilde{\bf a} \; \exp \left(-S[\dot{\bf a}, \tilde{\bf a}] \right), \\
S[\dot{\bf a}, \tilde{\bf a}] =& \tilde{\bf a}^T \dot{\bf a} -  {\bf W}[\tilde{\bf a}; \boldsymbol{\mu}\big({\bf G} \ast ({\bf E} + \bf J \ast \langle \dot{\bf a} \rangle) \big) ].
\end{aligned} \end{equation}
We expand the activity around a background field: $\bf{\dot{a}=\dot{a}^*+\delta \dot{a}}$.
Here, we give the expansion for a network with one neuron type.
The free and interacting actions for the expansion around the mean-field theory are
\begin{equation} \begin{aligned} \label{eq:mf_action_separated}
S_0 =& \; \tilde{\bf a}^T \delta\dot{\bf a} - (\tilde{a}^D)^T g \, \delta \dot{a}^S  \\
S_V =& - \sum_{p=2}^{\infty} \left( \frac{(\tilde{a}^S)^p}{p!} \right) ^T f \\
&- \left( (\dot{a}^S)^* + \delta\dot{a}^S \right) ^T \left[ \sum_{q=1}^{\infty} \frac{-1^{q+1}}{q} \left(\sum_{\substack{p=1 \\ \backslash p=q=1}}^{\infty} \frac{ (\tilde{a}^D)^p}{p!}g  \right)^q \right]
\end{aligned} \end{equation}
where $f:= f \left( \left( {\bf G} \ast \left( {\bf E} + {\bf J} \ast \langle \dot{\bf a} \rangle \right) \right)^S \right) $ and $g := g \left( \left( {\bf G} \ast \left( {\bf E} + {\bf J} \ast \langle \dot{\bf a} \rangle \right) \right)^D \right)$. Note that, unlike the derivation of the free and interacting components of the action in section 4.4, the derivatives of the intensity functions $\bf f$ and $\bf g$ with respect to the activity variables in the population-averaged action vanish as the intensity functions are functions of the population-averaged activities $\langle \dot{a} \rangle$.

The mean-field approximation $\dot{\bf a}^* \approx \langle \dot{\bf a} \rangle$ is exact in the large-$N$ limit since all vertices in the action have greater out-degree than in-degree (Table~\ref{table:vertices}). 
So without loss of generality, we here take $\dot{\bf a}^* = \langle \dot{\bf a} \rangle$. 
(If the mean-field approximation were not the exact, then to enforce the constraint $\dot{\bf a}^* = \langle \dot{\bf a} \rangle $ one would make a Legendre transformation from the action $S$ to an effective action for the mean, yielding a hierarchy of equations for $\langle \dot{\bf a} \rangle$~\cite{zinn-justin_quantum_2002, berges_introduction_2004, ocker_republished_2023}.) 

\subsection{Feynman rules} \label{app:feynman}
Here we give the Feynman rules for an expansion around the mean-field theory. These define a graphical algorithm for computing arbitrary joint cumulants of the activity variables $\dot{a}^S, \dot{a}^D$.
The Feynman rules for this model are derived in the standard way~\cite{chow_path_2015, ocker_linking_2017}. The main difference with our previous works~\cite{ocker_linking_2017, ocker_republished_2023} is that there are two stochastic activity variables in this model.
The propagator determined by the free action of Eq.~\ref{eq:mf_action_separated} is
\begin{equation} \label{eq:propagator}
\boldsymbol{\Delta}(t, s)
= \begin{pmatrix}
\Delta_{SS}(t, s), & \Delta_{SD}(t, s) \\
\Delta_{DS}(t, s), & \Delta_{DD} (t, s)
\end{pmatrix}  
= \begin{pmatrix}
1, & 0 \\
g, & 1
\end{pmatrix} \delta(t-s)
\end{equation}
We denote its non-zero elements with the edges in Table~\ref{table:edges}. The source and internal vertices corresponding to the interacting action are in Table~\ref{table:vertices}. To calculate the joint cumulant density $\llrrangle{\prod_{i=1}^c \dot{a}^S(t_i) \, \prod_{j=1}^d \dot{a}^D(t_j)}$:
\begin{enumerate}
\item Place an external point for each of the $c+d$ factors of $\dot{a}^S$ and $\dot{a}^D$. Each external point carries the corresponding time variable.
\item Using the vertices and edges in Tables~\ref{table:vertices} and~\ref{table:edges}, construct all connected graphs such that each external point has one incoming propagator edge. Each of these vertices carries an `internal' time variable. The time arguments of each propagator correspond to the vertices it links.
\item To evaluate a connected diagram, multiply the factors of every edge and vertex together and integrate over all internal time variables. 
\item Sum the contributions of all the connected diagrams constructed in step 2.
\end{enumerate}
The vertices due to the dendritic spike/burst generation can be thought of as having multiplicity. For the same total out-degree, there are multiple possible versions of those vertices that may exist due to the nested expansion of the logarithm and exponential in $S_V$. For example, both $(p,q)=(1,2)$ and $(p,q)=(2,1)$ give rise to source vertices with out-degree two. Rather than drawing these vertices separately, we sum their contributions in the vertex factors of the burst vertices. 

{\renewcommand{\arraystretch}{2} 
\begin{table}[ht!]
\begin{tabular}{| c | c | c | c |} \hline
Vertex & Factor &  {\shortstack{ In-degree \\ ($\delta m, \; \delta b$)}} & {\shortstack{Out-degree \\ ($\tilde{a}^S, \; \tilde{a}^D$)}} \\ \hline
$\vcenter{\hbox{\feynmandiagram{a[dot]};}} $ & $f$ & (0, 0) & $(p \geq 2, \; 0)$  \\ \hline
$\vcenter{\hbox{\feynmandiagram{a[square dot]};}} $ & $(\dot{a}^S)^* \displaystyle\sum_{\substack{p, q =1 \\ pq = r}}^r (-1)^{q+1} \frac{(pq)!}{q (p!)^q} g^q$ & (0, 0) & $(0, \; r \geq 2)$  \\ \hline
$\vcenter{\hbox{\feynmandiagram[]{a[empty square]};}} $ & $  \displaystyle\sum_{\substack{p, q =1 \\ pq = r }}^r (-1)^{q+1} \frac{(pq)!}{q (p!)^q} g^q $ & $ (1, \; 0)$ & $(0, \; r \geq 2)$ \\ \hline 
\end{tabular} 
\caption{\label{table:vertices}Vertices corresponding to the interacting action, $S_V$, in Eq.~\ref{eq:mf_action_separated}. All vertices have a greater out-degree than in-degree.
 Filled vertices are source vertices, and the open vertex is internal. $p$ and $q$ are both positive integers. $g$ is the dendritic nonlinearity evaluated at the mean-field dendritic input as below Eq.~\ref{eq:mf_action_separated}.}
\end{table}
}

{ \renewcommand{\arraystretch}{2}
\begin{table}[ht!]
\begin{tabular}{| c | c | c |} \hline 
Edge & Propagator & Factor \\ \hline
$\vcenter{\hbox{\feynmandiagram[horizontal=a to b, inline=(a.base)]{a --[] b};}}$ &  $\Delta_{SS}(t, s)$ & $\delta(t-s)$ \\[2\baselineskip] \hline
$\vcenter{\hbox{\feynmandiagram[horizontal=a to b, inline=(a.base)]{a --[photon] b};}}$ & $\Delta_{ DS}(t, s)$ & $ g \, \delta(t-s) $  \\ \hline
$\vcenter{\hbox{\feynmandiagram[horizontal=a to b, inline=(a.base)]{a --[gluon] b};}}$ & $\Delta_{ DD}(t, s) $ &  $ \delta(t-s) $  \\ \hline
\end{tabular}
\caption{\label{table:edges} Edges corresponding to the components of the propagator from $S_0$ in Eq.~\ref{eq:mf_action_separated}. Each measures the linear response of one configuration variable to a perturbation of another. For example, $\bar{\Delta}_{ DS}$ measures the linear response of the dendritic rate to a somatic fluctuation. $g$ is the dendritic nonlinearity evaluated at the mean-field dendritic input as below Eq.~\ref{eq:mf_action_separated}.
}
\end{table}
}
As an example, below is the calculation for the typical dendritic spike train autocovariance in a stationary regime.
\begin{equation} \begin{aligned}
\llrrangle{\dot{a}^D(t) \dot{a}^D(t')} =&
\vcenter{ \hbox{\begin{tikzpicture}
\begin{feynman}[small]
\vertex  (a) at (0, 1) {};
\vertex (b) at (0, 0) {};
\vertex[dot] (c) at (.5, .5) {};
\diagram*{(a)  -- [photon] (c)  -- [photon] (b)};
\end{feynman}\end{tikzpicture} } }
+
\vcenter{ \hbox{
\begin{tikzpicture}
\begin{feynman}[small]
\vertex (a) at (2.5, 1) {};
\vertex (b) at (2.5, 0) {};
\vertex[square dot] (c) at (3, .5) {};
\diagram*{
(a) -- [gluon] (c) -- [gluon] (b);
};
\end{feynman}\end{tikzpicture} } } \\
=& \int \Delta_{DS}(t, s) \, \Delta_{DS}(t', s) \, f(s) \, ds \\
&+ \int \Delta_{DD}(t, s) \, \Delta_{DD}(t', s) \, (\dot{a}^S)^*(s) \\
&\hspace{.2in} \times \sum_{\substack{p, q =1 \\ pq = 2}}^2 (-1)^{q+1} \frac{(pq)!}{q (p!)^q} g^q(s)\, ds \\
=& g^2 \delta(t-t') + \hspace{-12pt} \sum_{\substack{(p,q) \in \\ \{ (1,2),(2,1)\}}} \hspace{-12pt} (-1)^{q+1} \frac{(pq)!}{q (p!)^q}  g^q (\dot{a}^*) \delta(t-t') \\
=& g^2 f \delta(t-t') - g^2 (\dot{a}^S)^* \delta(t-t') + g (\dot{a}^S)^* \delta(t-t').
\end{aligned} \end{equation}
Using the fact that $f = (\dot{a}^*)^S$ then yields Eq.~\ref{eq:mft_Cbb}.

\subsection{Stability}
The mean-field limit exposes a very simple description of the network's activity: neurons spike as uncorrelated Poisson processes with self-consistently determined rates (Eq.~\ref{eq:mf_density_activity}).
With recurrent coupling, the nonlinear interaction between soma and dendrite can lead to multiple co-existing self-consistent activity states.
This description, however, does not describe the stability of those states.
To analyze the stability of the mean-field fixed points, we turn to the equivalent mean-field voltage dynamics. These are
\begin{equation} \begin{aligned} \label{eq:mf_voltage}
\langle \dot{v}_\alpha^c \rangle =& E_\alpha^c - \langle v_\alpha^c \rangle + \sum_t \sum_{d \in C_t} J_{\alpha t}^{c d} \ast \langle \dot{a}_t^d \rangle,
\end{aligned} \end{equation}
which can be derived by marginalizing out the activity variables from Eq.~\ref{eq:reduced_action}. Stationary regimes of the network correspond equivalently to equilibria of Eq.~\ref{eq:mf_voltage} or solutions of the equilibrium mean-field rate equations described in the main text. The maximum real part of the eigenvalues of the Jacobian of Eq.~\ref{eq:mf_voltage} around its equilibria describe stability of those stationary regimes in the large-$N$ limit. Due to the piecewise-linear nonlinearities used here, those can be evaluated at the rate solutions described in the main text.

\section{Mean-field fixed points} \label{app:mf_fp}
Here, we collect the mean-field fixed points and their existence and stability conditions.
\renewcommand{\arraystretch}{1.5}
\begin{table}[h!]
\begin{center}
\begin{tabular}{ |m{3cm} |m{2.5cm}|m{2.5cm}| } 
\hline
$\langle \dot{a}^S \rangle, \langle \dot{a}^D \rangle$ & Existence & Stability \\ \hline
$(0,0)$  & $E^S<0$ & stable \\ \hline
$\left( \displaystyle\frac{E^S}{1-J}, 0\right)$ &
$E^D \leq 0$ and
\newline  $E^S<0$, $J>1$ or
\newline  $E^S>0, J < 1$ & $J<1$  
\\ \hline
$\Bigg( \displaystyle\frac{E^S}{1-J(1+\beta E^D)}, \newline \displaystyle\frac{E^S E^D}{1-J(1+\beta E^D)} \Bigg)$ & 
$0<E^D \leq 1$
\newline $E^S<0$
\newline $J \left( 1 + \beta \; E^D \right)>1$
& $J \left( 1 + \beta \; E^D \right)<1$  \\ \cline{2-3} &
 $0<E^D \leq 1$
\newline $E^S>0$
\newline $J \left( 1 + \beta \; E^D \right)<1$
& $J \left( 1 + \beta \; E^D \right)<1$  \\ \hline
$\Bigg( \displaystyle\frac{E^S}{1-J(1+\beta)}, \newline \frac{E^S}{1-J(1+\beta)}\Bigg)$ &  
$E^D>1$
\newline $E^S<0$
\newline $J \left( 1 + \beta \right)>1$
& $J \left( 1 + \beta \right)<1$ \\ \cline{2-3} &
 $E^D \geq 1$
\newline $E^S>0$
\newline $J \left( 1 + \beta \right)<1$
 & $J \left( 1 + \beta \right)<1$ \\ \hline
\end{tabular}
\caption{mean-field fixed points and their existence and stability conditions in the soma-targeting network (Fig.~\ref{fig:soma-syn}). Inequalities in the stability column are the stability conditions (stable if the inequality is satisfied). }
\label{table:soma}
\end{center}
\end{table}

\renewcommand{\arraystretch}{1.5}
\begin{table}[h!]
\begin{center}
\begin{tabular}{ |m{3cm}| m{3.7cm}|m{1.5cm}| } 
\hline
$\langle \dot{a}^S \rangle, \langle \dot{a}^D \rangle$ & Existence & Stability \\ \hline
$(0,0)$ & $E^S \leq 0$ & stable \\ \hline
$\left( E^S, 0 \right)$ & 
$E^S>0$ 
\newline $E^D \leq -JE^S$
& stable  \\ \hline
$\Bigg( E^S, \newline \displaystyle\frac{E^S \left( E^D +J E^S \right)}{1-\beta J E^S } \Bigg)$ & 
$E^S>0$ 
\newline $ 1 - \beta J E^S \leq E^D + JE^S < 0 $
&  $ \beta J E^S < 1 $  \\ \cline{2-3} & 
 $E^S>0$
\newline $ 0 < E^D + JE^S \leq 1 - \beta J E^S $
& $ \beta J E^S < 1 $  \\ \hline
$\left( E^S, E^S \right)$ &
$E^S>0$
\newline $ E^D + JE^S \left( 1+ \beta \right) \geq 1 $
& stable \\ \hline
\end{tabular}
\caption{Conditions for existence and stability of fixed points in the dendrite-targeting network. Inequalities in the stability column are the stability conditions (stable if the inequality is satisfied).}
\label{table:dendrite}
\end{center}
\end{table}

\clearpage
\begin{widetext}
\renewcommand{\arraystretch}{2}
\begin{table}[h!]
\begin{center}
\begin{tabular}{ |m{4cm}| m{10cm}|m{2cm}| } 
\hline
$\langle \dot{a}^S_E \rangle, \langle \dot{a}^D_E \rangle, \langle \dot{a}^S_I \rangle$ & Existence & Stability \\ \hline
$(0,0,0)$ & 
$E^S \leq 0$ 
\newline $E_I \leq 0$ & stable \\ \hline

$(0,0,\displaystyle\frac{E_I}{1+gJ})$ &  
$E^S \leq 0$ 
\newline $ \displaystyle\frac{E_I}{1+gJ} > 0$ & $1+gJ>0$ \\ \hline

$\left( E^S, 0, 0 \right)$ & 
$E^S>0$ 
\newline $E^D \leq -JE^S$
\newline $E_I \leq -JE^S$
& stable  \\ \hline

$\left( E^S, 0, \displaystyle\frac{E_I+J_E E^S}{1+J_I} \right)$ &
$E^S > 0$ 
\newline $ E^D \leq -\left(J_E E^S -J_I\left(\displaystyle\frac{E_I+J_E E^S}{1+J_I} \right)\right)$
\newline $ \displaystyle\frac{E_I+J_E E^S}{1+J_I}>0 $
& $1+gJ>0$ \\ \hline

$\left( E^S, \frac{E^S \left( E^D +J_E E^S \right)}{1-\beta J_E E^S }, 0 \right)$ &
$E^S>0$ 
\newline $ 1 - \beta J_E E^S \leq E^D + J_E E^S < 0 $
\newline $E_I \leq - J_E E^S\left(1 + \beta \left( \frac{ E^D +J_E E^S }{1-\beta J_E E^S } \right) \right) $
& $ \beta J_E E^S < 1 $  \\ \cline{2-3} & 
 $E^S>0$
\newline $ 0 < E^D + J_E E^S \leq 1 - \beta J_E E^S $
\newline $E_I \leq - J_E E^S\left(1 + \beta \left( \frac{  E^D +J_E E^S }{1-\beta J_E E^S } \right) \right) $
& $\beta J_E E^S < 1 $  \\ \hline

$\left( E^S, E^S, 0 \right)$ & $E^S>0$
\newline $E^D+J_E E^S \left( 1+ \beta \right) \geq 1$
\newline $E_I \leq - J_E E^S\left(1 + \beta \right) $
& stable \\ \hline

$\Biggl ( E^S, \newline E^S \frac{E^D \left( 1+J_I \right) + J_E E^S - J_I E_I}{1 +J_I -\beta J_E E^S }, \newline \frac{E_I \left(1-\beta J_E E^S \right) +J_E E^S \left( 1+ \beta E^D \right)}{1+J_I-\beta J_E E^S} \Biggr )$ & 
$E^S>0$ 
\newline $1+J_I- \beta J_E E^S  \leq E^D \left( 1+J_I \right) +J_E E^S - J_I E_I<0 $
\newline $E_I \left(\beta J_E E^S - 1 \right)>J_E E^S \left( 1+ \beta E^D \right) $
\newline $ 1 + J_I < \beta J_E E^S  $
& $ \beta J_E E^S < 1 + J_I $  \\ \cline{2-3} & 
 $E^S>0$
\newline $0 < E^D \left( 1+J_I \right) +J_E E^S - J_I E_I \leq 1+J_I - \beta J_E E^S$
\newline $E_I \left(\beta J_E E^S - 1 \right)<J_E E^S \left( 1+ \beta E^D \right) $
\newline $ 1 + J_I > \beta J_E E^S  $
&  $ \beta J_E E^S < 1+ J_I $  \\ \hline

$\left( E^S, E^S, \frac{E_I +J_E E^S \left( 1+ \beta \right)}{1+J_I} \right)$ &
$E^S>0$
\newline $ E^D + J_E E^S \left(1+\beta \right) \geq 1+J_I \left( \frac{E_I+J_E E^S \left(1+\beta \right)}{1+J_I} \right)$
\newline $\frac{E_I+J_E E^S \left(1+\beta \right)}{1+J_I}>0$
& $1+J_I>0$ \\ \hline
\end{tabular}
\caption{Conditions for existence and stability of fixed points in the dendrite-targeting excitation and inhibition network}
\label{table:SOM}
\end{center}
\end{table}

\renewcommand{\arraystretch}{2}
\begin{table}[h!]
\begin{center}
\begin{tabular}{ |m{4.5cm}|m{9cm}|m{2cm}| } 
\hline
$\langle \dot{a}_E^S \rangle, \langle \dot{a}_E^D \rangle, \langle \dot{a}_I \rangle$ & Existence & Stability \\ \hline
$(0,0,0)$ & 
$E^S, \, E_I \leq 0$  & stable \\ \hline
$(0,0,\frac{E_I}{1+J_I})$ & 
$E^S \leq J_I \frac{E_I}{1+J_I} $ \newline
$E_I>0$  & $1+J_I>0$ \\ \hline

$\left( E^S, 0, 0 \right)$ &
$E^S>0$ 
\newline $E^D, \, E_I \leq -J_E E^S$
& stable  \\ \hline

$\Bigg( E^S-\frac{E_I+J_E E^S}{1+J_I(1+J_E)}, 0, \newline \frac{E_I+J_E E^S}{1+J_I(1+J_E)} \Bigg)$ &
$E^S > J_I \frac{E_I+J_E E^S}{1+J_I(1+J_E)}$ 
\newline $E^D \leq -\left(J_E E^S - J_I J_E \left( \frac{E_I+J_E E^S}{1+J_I(1+J_E)}\right) \right)$
\newline $\frac{E_I+J_E E^S}{1+J_I(1+J_E)}>0$
& stable$^1$ \\ \hline

$\left( E^S, \frac{E^S \left( E^D +J_E E^S \right)}{1-\beta J_E E^S }, 0 \right)$ & 
$E^S>0$ 
\newline $ 1 - \beta J_E E^S \leq E^D + J_E E^S < 0 $
\newline $E_I \leq - J_E E^S\left(1 + \beta \left( \frac{ \left( E^D +J_E E^S \right) }{1-\beta J_E E^S } \right) \right) $
& $ \beta J_E E^S < 1 $  \\ \cline{2-3} & 
 $E^S>0$
\newline $ 0 < E^D + J_E E^S \leq 1 - \beta J_E E^S $
\newline $ E_I \leq - J_E E^S\left(1 + \beta \left( \frac{ \left( E^D +J_E E^S \right)}{1-\beta J_E E^S } \right) \right) $
& $ \beta J_E E^S < 1 $  \\ \hline

$\left( E^S, E^S, 0 \right)$ & $E^S>0$
\newline $ E^D+J_E E^S \left( 1+ \beta \right) \geq 1 $
\newline $ E_I \leq - J_E E^S\left(1 + \beta \right) $
& stable \\ \hline

$(\dot{a}^S_E)_{\pm}^{\ast}, (\dot{a}^D_E)_{\pm}^{\ast}, (\dot{a}_I)_{\pm}^{\ast}$ \; (Eq.~\ref{PV_unsat_means}) & 
$(\dot{a}^S_E)_{\pm}^{\ast}, \, (\dot{a}^D_E)_{\pm}^{\ast}, \,(\dot{a}_I)_{\pm}^{\ast} >0$ 
\newline $ \left( 1 - \beta J_E \phi + \psi \left( E^D +2J_E \phi \right) \right)^2 \geq 4 \phi \left( E^D+J_E \phi \right) \left( \psi \left( J_E \psi - \beta J_E \right) \right) $
\newline $(\dot{a}^D_E)_{\pm}^{\ast} \leq (\dot{a}^S_E)_{\pm}^{\ast}$
& (un)stable$^1$ \\ \hline

$\Biggl ( E^S-J_I\frac{E_I+J_E E^S \left( 1+ \beta \right)}{1+J_I(1+J_E) \left( 1+ \beta \right)},$ \newline $E^S-J_I \frac{E_I+J_E E^S\left( 1+ \beta \right)}{1+J_I(1+J_E) \left( 1+ \beta \right)},$ \newline 
$ \frac{E_I+J_E E^S \left( 1+ \beta \right)}{1+J_I(1+J_E) \left( 1+ \beta \right)} \Biggr )$ & 
$E^S-J_I\frac{E_I+J_E E^S \left( 1+ \beta \right)}{1+J_I(1+J_E) \left( 1+ \beta \right)} > 0$
\newline $ E^D + J_E \left(1+\beta \right) \left( E^S - J_I \left( \frac{E_I+J_E E^S \left(1+\beta \right)}{1+J_I(1+J_E) \left( 1+ \beta \right)} \right) \right) \geq 1 $
\newline $\frac{E_I+J_E E^S \left(1+\beta \right)}{1+J_I(1+J_E) \left( 1+ \beta \right)}>0$
& stable$^2$ \\ \hline
\end{tabular}
\caption{\scriptsize Conditions for existence and stability of fixed points in the dendrite-targeting excitation and soma-targeting inhibition network. $^1$ unstable (stable) for  $\text{Re} \bigg \{ -2-J_I+\beta J_E \langle v^s \rangle \pm  \sqrt{\left( J_I+\beta J_E \langle v^s \rangle \right)^2 -4 \left( J_I + \beta J_E \langle v^d \rangle \right)} \bigg \} < 0 $; the $+$ fixed point is always unstable due to the conditions for the existence of the fixed point and sign constraints on parameter values, while the $-$ fixed point is unstable for some sets of parameter values and stable for others.
$^2$ stable for $\text{Re} \bigg \{ -2-J_I \pm \sqrt{\left( 2+J_I \right)^2 -4 \left( 1+ J_I(1+J_E) \left( 1+ \beta \right) \right)} \bigg \} < 0 $ which is always true since $\beta, J_E, J_I$ are non-negative. The constants $\phi, \psi, \gamma$ are defined below.
} 
\label{table:PV}
\end{center}
\end{table}

\subsection{PV inhibition}
The network with dendritic excitation and somatic inhibition has a three-dimensional piecewise-polynomial mean-field equilibrium system (Eq.~\ref{eq:mf_equil_pv}).
When this network is fully active and the dendrite is not saturated the system is quadratic, and else linear. In the quadratic case, the roots are
\begin{equation} \begin{aligned}\label{PV_unsat_means}
(\dot{a}^S_E)_{\pm}^{*} =& E^S-J_I (\dot{a}_I)_{\pm}^*, \\
(\dot{a}^D_E)_{\pm}^{*} =& \frac{1 - \beta J_E \phi + \psi \left( E^D +2J_E \phi \right)  \pm \sqrt{\left[ 1 - \beta J_E \phi + \psi \left( E^D +2J_E \phi \right) \right]^2 - 4 \phi \left( E^D+J_E \phi \right) \left[ \psi \left( J_E \psi  \beta J_E \right) \right]} }{2 \left( J_E \psi - \beta J_E \right)}, \\ 
(\dot{a}_I)_{\pm}^{*} =& \frac{E_I+J_E E^S+\beta J_E (\dot{a}^D_{-})^{\ast}}{\gamma}, \; \mathrm{where} \\
\psi =& \frac{\beta J_E J_I }{\gamma}, \; \gamma = 1+J_I(1+J_E), \; \phi = E^S - J_I \left( \frac{E_I+J_E E^S}{\gamma} \right).
\end{aligned} \end{equation}
\end{widetext}

\bibliography{MyLibrary}

\begin{thebibliography}{76}%
\makeatletter
\providecommand \@ifxundefined [1]{%
 \@ifx{#1\undefined}
}%
\providecommand \@ifnum [1]{%
 \ifnum #1\expandafter \@firstoftwo
 \else \expandafter \@secondoftwo
 \fi
}%
\providecommand \@ifx [1]{%
 \ifx #1\expandafter \@firstoftwo
 \else \expandafter \@secondoftwo
 \fi
}%
\providecommand \natexlab [1]{#1}%
\providecommand \enquote  [1]{``#1''}%
\providecommand \bibnamefont  [1]{#1}%
\providecommand \bibfnamefont [1]{#1}%
\providecommand \citenamefont [1]{#1}%
\providecommand \href@noop [0]{\@secondoftwo}%
\providecommand \href [0]{\begingroup \@sanitize@url \@href}%
\providecommand \@href[1]{\@@startlink{#1}\@@href}%
\providecommand \@@href[1]{\endgroup#1\@@endlink}%
\providecommand \@sanitize@url [0]{\catcode `\\12\catcode `\$12\catcode
  `\&12\catcode `\#12\catcode `\^12\catcode `\_12\catcode `\%12\relax}%
\providecommand \@@startlink[1]{}%
\providecommand \@@endlink[0]{}%
\providecommand \url  [0]{\begingroup\@sanitize@url \@url }%
\providecommand \@url [1]{\endgroup\@href {#1}{\urlprefix }}%
\providecommand \urlprefix  [0]{URL }%
\providecommand \Eprint [0]{\href }%
\providecommand \doibase [0]{https://doi.org/}%
\providecommand \selectlanguage [0]{\@gobble}%
\providecommand \bibinfo  [0]{\@secondoftwo}%
\providecommand \bibfield  [0]{\@secondoftwo}%
\providecommand \translation [1]{[#1]}%
\providecommand \BibitemOpen [0]{}%
\providecommand \bibitemStop [0]{}%
\providecommand \bibitemNoStop [0]{.\EOS\space}%
\providecommand \EOS [0]{\spacefactor3000\relax}%
\providecommand \BibitemShut  [1]{\csname bibitem#1\endcsname}%
\let\auto@bib@innerbib\@empty
\bibitem [{\citenamefont {Coombes}(2010)}]{coombes_large-scale_2010}%
  \BibitemOpen
  \bibfield  {author} {\bibinfo {author} {\bibfnamefont {S.}~\bibnamefont
  {Coombes}},\ }\bibfield  {title} {\bibinfo {title} {Large-scale neural
  dynamics: {{Simple}} and complex},\ }\href
  {https://doi.org/10.1016/j.neuroimage.2010.01.045} {\bibfield  {journal}
  {\bibinfo  {journal} {NeuroImage}\ }\bibinfo {series} {Computational
  {{Models}} of the {{Brain}}},\ \textbf {\bibinfo {volume} {52}},\ \bibinfo
  {pages} {731} (\bibinfo {year} {2010})}\BibitemShut {NoStop}%
\bibitem [{\citenamefont {Eccles}(1935)}]{eccles_action_1935}%
  \BibitemOpen
  \bibfield  {author} {\bibinfo {author} {\bibfnamefont {J.~C.}\ \bibnamefont
  {Eccles}},\ }\bibfield  {title} {\bibinfo {title} {The action potential of
  the superior cervical ganglion},\ }\href
  {https://doi.org/10.1113/jphysiol.1935.sp003313} {\bibfield  {journal}
  {\bibinfo  {journal} {The Journal of Physiology}\ }\textbf {\bibinfo {volume}
  {85}},\ \bibinfo {pages} {179} (\bibinfo {year} {1935})}\BibitemShut
  {NoStop}%
\bibitem [{\citenamefont {Hodgkin}\ and\ \citenamefont
  {Huxley}(1952)}]{hodgkin_quantitative_1952}%
  \BibitemOpen
  \bibfield  {author} {\bibinfo {author} {\bibfnamefont {A.~L.}\ \bibnamefont
  {Hodgkin}}\ and\ \bibinfo {author} {\bibfnamefont {A.~F.}\ \bibnamefont
  {Huxley}},\ }\bibfield  {title} {\bibinfo {title} {A quantitative description
  of membrane current and its application to conduction and excitation in
  nerve},\ }\href {https://doi.org/10.1113/jphysiol.1952.sp004764} {\bibfield
  {journal} {\bibinfo  {journal} {The Journal of Physiology}\ }\textbf
  {\bibinfo {volume} {117}},\ \bibinfo {pages} {500} (\bibinfo {year}
  {1952})}\BibitemShut {NoStop}%
\bibitem [{\citenamefont {Larkum}\ \emph {et~al.}(2022)\citenamefont {Larkum},
  \citenamefont {Wu}, \citenamefont {Duverdin},\ and\ \citenamefont
  {Gidon}}]{larkum_guide_2022}%
  \BibitemOpen
  \bibfield  {author} {\bibinfo {author} {\bibfnamefont {M.~E.}\ \bibnamefont
  {Larkum}}, \bibinfo {author} {\bibfnamefont {J.}~\bibnamefont {Wu}}, \bibinfo
  {author} {\bibfnamefont {S.~A.}\ \bibnamefont {Duverdin}},\ and\ \bibinfo
  {author} {\bibfnamefont {A.}~\bibnamefont {Gidon}},\ }\bibfield  {title}
  {\bibinfo {title} {The {{Guide}} to {{Dendritic Spikes}} of the {{Mammalian
  Cortex In Vitro}} and {{In Vivo}}},\ }\href
  {https://doi.org/10.1016/j.neuroscience.2022.02.009} {\bibfield  {journal}
  {\bibinfo  {journal} {Neuroscience}\ }\bibinfo {series} {Dendritic
  Contributions to Biological and Artificial Computations},\ \textbf {\bibinfo
  {volume} {489}},\ \bibinfo {pages} {15} (\bibinfo {year} {2022})}\BibitemShut
  {NoStop}%
\bibitem [{\citenamefont {Larkum}\ \emph {et~al.}(1999)\citenamefont {Larkum},
  \citenamefont {Zhu},\ and\ \citenamefont {Sakmann}}]{larkum_new_1999}%
  \BibitemOpen
  \bibfield  {author} {\bibinfo {author} {\bibfnamefont {M.~E.}\ \bibnamefont
  {Larkum}}, \bibinfo {author} {\bibfnamefont {J.~J.}\ \bibnamefont {Zhu}},\
  and\ \bibinfo {author} {\bibfnamefont {B.}~\bibnamefont {Sakmann}},\
  }\bibfield  {title} {\bibinfo {title} {A new cellular mechanism for coupling
  inputs arriving at different cortical layers},\ }\href
  {https://doi.org/10.1038/18686} {\bibfield  {journal} {\bibinfo  {journal}
  {Nature}\ }\textbf {\bibinfo {volume} {398}},\ \bibinfo {pages} {338}
  (\bibinfo {year} {1999})}\BibitemShut {NoStop}%
\bibitem [{\citenamefont {Naud}\ and\ \citenamefont
  {Sprekeler}(2018)}]{naud_sparse_2018}%
  \BibitemOpen
  \bibfield  {author} {\bibinfo {author} {\bibfnamefont {R.}~\bibnamefont
  {Naud}}\ and\ \bibinfo {author} {\bibfnamefont {H.}~\bibnamefont
  {Sprekeler}},\ }\bibfield  {title} {\bibinfo {title} {Sparse bursts optimize
  information transmission in a multiplexed neural code},\ }\href
  {https://doi.org/10.1073/pnas.1720995115} {\bibfield  {journal} {\bibinfo
  {journal} {Proceedings of the National Academy of Sciences}\ ,\ \bibinfo
  {pages} {201720995}} (\bibinfo {year} {2018})}\BibitemShut {NoStop}%
\bibitem [{\citenamefont {Friedenberger}\ \emph {et~al.}(2023)\citenamefont
  {Friedenberger}, \citenamefont {Harkin}, \citenamefont {T{\'o}th},\ and\
  \citenamefont {Naud}}]{friedenberger_silences_2023}%
  \BibitemOpen
  \bibfield  {author} {\bibinfo {author} {\bibfnamefont {Z.}~\bibnamefont
  {Friedenberger}}, \bibinfo {author} {\bibfnamefont {E.}~\bibnamefont
  {Harkin}}, \bibinfo {author} {\bibfnamefont {K.}~\bibnamefont {T{\'o}th}},\
  and\ \bibinfo {author} {\bibfnamefont {R.}~\bibnamefont {Naud}},\ }\bibfield
  {title} {\bibinfo {title} {Silences, spikes and bursts: {{Three-part}} knot
  of the neural code},\ }\href {https://doi.org/10.1113/JP281510} {\bibfield
  {journal} {\bibinfo  {journal} {The Journal of Physiology}\ }\textbf
  {\bibinfo {volume} {601}},\ \bibinfo {pages} {5165} (\bibinfo {year}
  {2023})}\BibitemShut {NoStop}%
\bibitem [{\citenamefont {Brumberg}\ \emph {et~al.}(2000)\citenamefont
  {Brumberg}, \citenamefont {Nowak},\ and\ \citenamefont
  {McCormick}}]{brumberg_ionic_2000}%
  \BibitemOpen
  \bibfield  {author} {\bibinfo {author} {\bibfnamefont {J.~C.}\ \bibnamefont
  {Brumberg}}, \bibinfo {author} {\bibfnamefont {L.~G.}\ \bibnamefont
  {Nowak}},\ and\ \bibinfo {author} {\bibfnamefont {D.~A.}\ \bibnamefont
  {McCormick}},\ }\bibfield  {title} {\bibinfo {title} {Ionic mechanisms
  underlying repetitive high-frequency burst firing in supragranular cortical
  neurons},\ }\href@noop {} {\bibfield  {journal} {\bibinfo  {journal} {The
  Journal of Neuroscience: The Official Journal of the Society for
  Neuroscience}\ }\textbf {\bibinfo {volume} {20}},\ \bibinfo {pages} {4829}
  (\bibinfo {year} {2000})}\BibitemShut {NoStop}%
\bibitem [{\citenamefont {Schwartzkroin}\ and\ \citenamefont
  {Slawsky}(1977)}]{schwartzkroin_probable_1977}%
  \BibitemOpen
  \bibfield  {author} {\bibinfo {author} {\bibfnamefont {P.~A.}\ \bibnamefont
  {Schwartzkroin}}\ and\ \bibinfo {author} {\bibfnamefont {M.}~\bibnamefont
  {Slawsky}},\ }\bibfield  {title} {\bibinfo {title} {Probable calcium spikes
  in hippocampal neurons},\ }\href
  {https://doi.org/10.1016/0006-8993(77)91060-5} {\bibfield  {journal}
  {\bibinfo  {journal} {Brain Research}\ }\textbf {\bibinfo {volume} {135}},\
  \bibinfo {pages} {157} (\bibinfo {year} {1977})}\BibitemShut {NoStop}%
\bibitem [{\citenamefont {Wong}\ \emph {et~al.}(1979)\citenamefont {Wong},
  \citenamefont {Prince},\ and\ \citenamefont
  {Basbaum}}]{wong_intradendritic_1979}%
  \BibitemOpen
  \bibfield  {author} {\bibinfo {author} {\bibfnamefont {R.~K.}\ \bibnamefont
  {Wong}}, \bibinfo {author} {\bibfnamefont {D.~A.}\ \bibnamefont {Prince}},\
  and\ \bibinfo {author} {\bibfnamefont {A.~I.}\ \bibnamefont {Basbaum}},\
  }\bibfield  {title} {\bibinfo {title} {Intradendritic recordings from
  hippocampal neurons.},\ }\href {https://doi.org/10.1073/pnas.76.2.986}
  {\bibfield  {journal} {\bibinfo  {journal} {Proceedings of the National
  Academy of Sciences}\ }\textbf {\bibinfo {volume} {76}},\ \bibinfo {pages}
  {986} (\bibinfo {year} {1979})}\BibitemShut {NoStop}%
\bibitem [{\citenamefont {Magee}\ and\ \citenamefont
  {Carruth}(1999)}]{magee_dendritic_1999}%
  \BibitemOpen
  \bibfield  {author} {\bibinfo {author} {\bibfnamefont {J.~C.}\ \bibnamefont
  {Magee}}\ and\ \bibinfo {author} {\bibfnamefont {M.}~\bibnamefont
  {Carruth}},\ }\bibfield  {title} {\bibinfo {title} {Dendritic voltage-gated
  ion channels regulate the action potential firing mode of hippocampal {{CA1}}
  pyramidal neurons},\ }\href {https://doi.org/10.1152/jn.1999.82.4.1895}
  {\bibfield  {journal} {\bibinfo  {journal} {Journal of Neurophysiology}\
  }\textbf {\bibinfo {volume} {82}},\ \bibinfo {pages} {1895} (\bibinfo {year}
  {1999})}\BibitemShut {NoStop}%
\bibitem [{\citenamefont {Larkum}\ \emph {et~al.}(2007)\citenamefont {Larkum},
  \citenamefont {Waters}, \citenamefont {Sakmann},\ and\ \citenamefont
  {Helmchen}}]{larkum_dendritic_2007}%
  \BibitemOpen
  \bibfield  {author} {\bibinfo {author} {\bibfnamefont {M.~E.}\ \bibnamefont
  {Larkum}}, \bibinfo {author} {\bibfnamefont {J.}~\bibnamefont {Waters}},
  \bibinfo {author} {\bibfnamefont {B.}~\bibnamefont {Sakmann}},\ and\ \bibinfo
  {author} {\bibfnamefont {F.}~\bibnamefont {Helmchen}},\ }\bibfield  {title}
  {\bibinfo {title} {Dendritic {{Spikes}} in {{Apical Dendrites}} of
  {{Neocortical Layer}} 2/3 {{Pyramidal Neurons}}},\ }\href
  {https://doi.org/10.1523/JNEUROSCI.1717-07.2007} {\bibfield  {journal}
  {\bibinfo  {journal} {Journal of Neuroscience}\ }\textbf {\bibinfo {volume}
  {27}},\ \bibinfo {pages} {8999} (\bibinfo {year} {2007})}\BibitemShut
  {NoStop}%
\bibitem [{\citenamefont {Jahnsen}\ and\ \citenamefont
  {Llin{\'a}s}(1984)}]{jahnsen_ionic_1984}%
  \BibitemOpen
  \bibfield  {author} {\bibinfo {author} {\bibfnamefont {H.}~\bibnamefont
  {Jahnsen}}\ and\ \bibinfo {author} {\bibfnamefont {R.}~\bibnamefont
  {Llin{\'a}s}},\ }\bibfield  {title} {\bibinfo {title} {Ionic basis for the
  electro-responsiveness and oscillatory properties of guinea-pig thalamic
  neurones in vitro.},\ }\href {https://doi.org/10.1113/jphysiol.1984.sp015154}
  {\bibfield  {journal} {\bibinfo  {journal} {The Journal of Physiology}\
  }\textbf {\bibinfo {volume} {349}},\ \bibinfo {pages} {227} (\bibinfo {year}
  {1984})}\BibitemShut {NoStop}%
\bibitem [{\citenamefont {Destexhe}\ \emph {et~al.}(1998)\citenamefont
  {Destexhe}, \citenamefont {Neubig}, \citenamefont {Ulrich},\ and\
  \citenamefont {Huguenard}}]{destexhe_dendritic_1998}%
  \BibitemOpen
  \bibfield  {author} {\bibinfo {author} {\bibfnamefont {A.}~\bibnamefont
  {Destexhe}}, \bibinfo {author} {\bibfnamefont {M.}~\bibnamefont {Neubig}},
  \bibinfo {author} {\bibfnamefont {D.}~\bibnamefont {Ulrich}},\ and\ \bibinfo
  {author} {\bibfnamefont {J.}~\bibnamefont {Huguenard}},\ }\bibfield  {title}
  {\bibinfo {title} {Dendritic {{Low-Threshold Calcium Currents}} in {{Thalamic
  Relay Cells}}},\ }\href {https://doi.org/10.1523/JNEUROSCI.18-10-03574.1998}
  {\bibfield  {journal} {\bibinfo  {journal} {Journal of Neuroscience}\
  }\textbf {\bibinfo {volume} {18}},\ \bibinfo {pages} {3574} (\bibinfo {year}
  {1998})}\BibitemShut {NoStop}%
\bibitem [{\citenamefont {Eccles}\ \emph {et~al.}(1966)\citenamefont {Eccles},
  \citenamefont {Llin{\'a}s},\ and\ \citenamefont
  {Sasaki}}]{eccles_excitatory_1966}%
  \BibitemOpen
  \bibfield  {author} {\bibinfo {author} {\bibfnamefont {J.~C.}\ \bibnamefont
  {Eccles}}, \bibinfo {author} {\bibfnamefont {R.}~\bibnamefont {Llin{\'a}s}},\
  and\ \bibinfo {author} {\bibfnamefont {K.}~\bibnamefont {Sasaki}},\
  }\bibfield  {title} {\bibinfo {title} {The excitatory synaptic action of
  climbing fibres on the {{Purkinje}} cells of the cerebellum},\ }\href
  {https://doi.org/10.1113/jphysiol.1966.sp007824} {\bibfield  {journal}
  {\bibinfo  {journal} {The Journal of Physiology}\ }\textbf {\bibinfo {volume}
  {182}},\ \bibinfo {pages} {268} (\bibinfo {year} {1966})}\BibitemShut
  {NoStop}%
\bibitem [{\citenamefont {Davie}\ \emph {et~al.}(2008)\citenamefont {Davie},
  \citenamefont {Clark},\ and\ \citenamefont
  {H{\"a}usser}}]{davie_origin_2008}%
  \BibitemOpen
  \bibfield  {author} {\bibinfo {author} {\bibfnamefont {J.~T.}\ \bibnamefont
  {Davie}}, \bibinfo {author} {\bibfnamefont {B.~A.}\ \bibnamefont {Clark}},\
  and\ \bibinfo {author} {\bibfnamefont {M.}~\bibnamefont {H{\"a}usser}},\
  }\bibfield  {title} {\bibinfo {title} {The {{Origin}} of the {{Complex
  Spike}} in {{Cerebellar Purkinje Cells}}},\ }\href
  {https://doi.org/10.1523/JNEUROSCI.0559-08.2008} {\bibfield  {journal}
  {\bibinfo  {journal} {The Journal of Neuroscience}\ }\textbf {\bibinfo
  {volume} {28}},\ \bibinfo {pages} {7599} (\bibinfo {year}
  {2008})}\BibitemShut {NoStop}%
\bibitem [{\citenamefont {Felleman}\ and\ \citenamefont {Van~Essen}(
  Feb)}]{felleman_distributed_1991}%
  \BibitemOpen
  \bibfield  {author} {\bibinfo {author} {\bibfnamefont {D.~J.}\ \bibnamefont
  {Felleman}}\ and\ \bibinfo {author} {\bibfnamefont {D.~C.}\ \bibnamefont
  {Van~Essen}},\ }\bibfield  {title} {\bibinfo {title} {Distributed
  hierarchical processing in the primate cerebral cortex},\ }\href@noop {}
  {\bibfield  {journal} {\bibinfo  {journal} {Cerebral Cortex (New York, N.Y.:
  1991)}\ }\textbf {\bibinfo {volume} {1}},\ \bibinfo {pages} {1} (\bibinfo
  {year} {1991 Jan-Feb})}\BibitemShut {NoStop}%
\bibitem [{\citenamefont {Harris}\ \emph {et~al.}(2019)\citenamefont {Harris},
  \citenamefont {Mihalas}, \citenamefont {Hirokawa}, \citenamefont {Whitesell},
  \citenamefont {Choi}, \citenamefont {Bernard}, \citenamefont {Bohn},
  \citenamefont {Caldejon}, \citenamefont {Casal}, \citenamefont {Cho},
  \citenamefont {Feiner}, \citenamefont {Feng}, \citenamefont {Gaudreault},
  \citenamefont {Gerfen}, \citenamefont {Graddis}, \citenamefont {Groblewski},
  \citenamefont {Henry}, \citenamefont {Ho}, \citenamefont {Howard},
  \citenamefont {Knox}, \citenamefont {Kuan}, \citenamefont {Kuang},
  \citenamefont {Lecoq}, \citenamefont {Lesnar}, \citenamefont {Li},
  \citenamefont {Luviano}, \citenamefont {McConoughey}, \citenamefont
  {Mortrud}, \citenamefont {Naeemi}, \citenamefont {Ng}, \citenamefont {Oh},
  \citenamefont {Ouellette}, \citenamefont {Shen}, \citenamefont {Sorensen},
  \citenamefont {Wakeman}, \citenamefont {Wang}, \citenamefont {Wang},
  \citenamefont {Williford}, \citenamefont {Phillips}, \citenamefont {Jones},
  \citenamefont {Koch},\ and\ \citenamefont {Zeng}}]{harris_hierarchical_2019}%
  \BibitemOpen
  \bibfield  {author} {\bibinfo {author} {\bibfnamefont {J.~A.}\ \bibnamefont
  {Harris}}, \bibinfo {author} {\bibfnamefont {S.}~\bibnamefont {Mihalas}},
  \bibinfo {author} {\bibfnamefont {K.~E.}\ \bibnamefont {Hirokawa}}, \bibinfo
  {author} {\bibfnamefont {J.~D.}\ \bibnamefont {Whitesell}}, \bibinfo {author}
  {\bibfnamefont {H.}~\bibnamefont {Choi}}, \bibinfo {author} {\bibfnamefont
  {A.}~\bibnamefont {Bernard}}, \bibinfo {author} {\bibfnamefont
  {P.}~\bibnamefont {Bohn}}, \bibinfo {author} {\bibfnamefont {S.}~\bibnamefont
  {Caldejon}}, \bibinfo {author} {\bibfnamefont {L.}~\bibnamefont {Casal}},
  \bibinfo {author} {\bibfnamefont {A.}~\bibnamefont {Cho}}, \bibinfo {author}
  {\bibfnamefont {A.}~\bibnamefont {Feiner}}, \bibinfo {author} {\bibfnamefont
  {D.}~\bibnamefont {Feng}}, \bibinfo {author} {\bibfnamefont {N.}~\bibnamefont
  {Gaudreault}}, \bibinfo {author} {\bibfnamefont {C.~R.}\ \bibnamefont
  {Gerfen}}, \bibinfo {author} {\bibfnamefont {N.}~\bibnamefont {Graddis}},
  \bibinfo {author} {\bibfnamefont {P.~A.}\ \bibnamefont {Groblewski}},
  \bibinfo {author} {\bibfnamefont {A.~M.}\ \bibnamefont {Henry}}, \bibinfo
  {author} {\bibfnamefont {A.}~\bibnamefont {Ho}}, \bibinfo {author}
  {\bibfnamefont {R.}~\bibnamefont {Howard}}, \bibinfo {author} {\bibfnamefont
  {J.~E.}\ \bibnamefont {Knox}}, \bibinfo {author} {\bibfnamefont
  {L.}~\bibnamefont {Kuan}}, \bibinfo {author} {\bibfnamefont {X.}~\bibnamefont
  {Kuang}}, \bibinfo {author} {\bibfnamefont {J.}~\bibnamefont {Lecoq}},
  \bibinfo {author} {\bibfnamefont {P.}~\bibnamefont {Lesnar}}, \bibinfo
  {author} {\bibfnamefont {Y.}~\bibnamefont {Li}}, \bibinfo {author}
  {\bibfnamefont {J.}~\bibnamefont {Luviano}}, \bibinfo {author} {\bibfnamefont
  {S.}~\bibnamefont {McConoughey}}, \bibinfo {author} {\bibfnamefont {M.~T.}\
  \bibnamefont {Mortrud}}, \bibinfo {author} {\bibfnamefont {M.}~\bibnamefont
  {Naeemi}}, \bibinfo {author} {\bibfnamefont {L.}~\bibnamefont {Ng}}, \bibinfo
  {author} {\bibfnamefont {S.~W.}\ \bibnamefont {Oh}}, \bibinfo {author}
  {\bibfnamefont {B.}~\bibnamefont {Ouellette}}, \bibinfo {author}
  {\bibfnamefont {E.}~\bibnamefont {Shen}}, \bibinfo {author} {\bibfnamefont
  {S.~A.}\ \bibnamefont {Sorensen}}, \bibinfo {author} {\bibfnamefont
  {W.}~\bibnamefont {Wakeman}}, \bibinfo {author} {\bibfnamefont
  {Q.}~\bibnamefont {Wang}}, \bibinfo {author} {\bibfnamefont {Y.}~\bibnamefont
  {Wang}}, \bibinfo {author} {\bibfnamefont {A.}~\bibnamefont {Williford}},
  \bibinfo {author} {\bibfnamefont {J.~W.}\ \bibnamefont {Phillips}}, \bibinfo
  {author} {\bibfnamefont {A.~R.}\ \bibnamefont {Jones}}, \bibinfo {author}
  {\bibfnamefont {C.}~\bibnamefont {Koch}},\ and\ \bibinfo {author}
  {\bibfnamefont {H.}~\bibnamefont {Zeng}},\ }\bibfield  {title} {\bibinfo
  {title} {Hierarchical organization of cortical and thalamic connectivity},\
  }\href {https://doi.org/10.1038/s41586-019-1716-z} {\bibfield  {journal}
  {\bibinfo  {journal} {Nature}\ ,\ \bibinfo {pages} {1}} (\bibinfo {year}
  {2019})}\BibitemShut {NoStop}%
\bibitem [{\citenamefont {Rudy}\ \emph {et~al.}(2011)\citenamefont {Rudy},
  \citenamefont {Fishell}, \citenamefont {Lee},\ and\ \citenamefont
  {{Hjerling-Leffler}}}]{rudy_three_2011}%
  \BibitemOpen
  \bibfield  {author} {\bibinfo {author} {\bibfnamefont {B.}~\bibnamefont
  {Rudy}}, \bibinfo {author} {\bibfnamefont {G.}~\bibnamefont {Fishell}},
  \bibinfo {author} {\bibfnamefont {S.}~\bibnamefont {Lee}},\ and\ \bibinfo
  {author} {\bibfnamefont {J.}~\bibnamefont {{Hjerling-Leffler}}},\ }\bibfield
  {title} {\bibinfo {title} {Three groups of interneurons account for nearly
  100\% of neocortical {{GABAergic}} neurons},\ }\href
  {https://doi.org/10.1002/dneu.20853} {\bibfield  {journal} {\bibinfo
  {journal} {Developmental Neurobiology}\ }\textbf {\bibinfo {volume} {71}},\
  \bibinfo {pages} {45} (\bibinfo {year} {2011})}\BibitemShut {NoStop}%
\bibitem [{\citenamefont {Traub}\ \emph {et~al.}(2005)\citenamefont {Traub},
  \citenamefont {Contreras}, \citenamefont {Cunningham}, \citenamefont
  {Murray}, \citenamefont {LeBeau}, \citenamefont {Roopun}, \citenamefont
  {Bibbig}, \citenamefont {Wilent}, \citenamefont {Higley},\ and\ \citenamefont
  {Whittington}}]{traub_single-column_2005}%
  \BibitemOpen
  \bibfield  {author} {\bibinfo {author} {\bibfnamefont {R.~D.}\ \bibnamefont
  {Traub}}, \bibinfo {author} {\bibfnamefont {D.}~\bibnamefont {Contreras}},
  \bibinfo {author} {\bibfnamefont {M.~O.}\ \bibnamefont {Cunningham}},
  \bibinfo {author} {\bibfnamefont {H.}~\bibnamefont {Murray}}, \bibinfo
  {author} {\bibfnamefont {F.~E.~N.}\ \bibnamefont {LeBeau}}, \bibinfo {author}
  {\bibfnamefont {A.}~\bibnamefont {Roopun}}, \bibinfo {author} {\bibfnamefont
  {A.}~\bibnamefont {Bibbig}}, \bibinfo {author} {\bibfnamefont {W.~B.}\
  \bibnamefont {Wilent}}, \bibinfo {author} {\bibfnamefont {M.~J.}\
  \bibnamefont {Higley}},\ and\ \bibinfo {author} {\bibfnamefont {M.~A.}\
  \bibnamefont {Whittington}},\ }\bibfield  {title} {\bibinfo {title}
  {Single-{{Column Thalamocortical Network Model Exhibiting Gamma
  Oscillations}}, {{Sleep Spindles}}, and {{Epileptogenic Bursts}}},\ }\href
  {https://doi.org/10.1152/jn.00983.2004} {\bibfield  {journal} {\bibinfo
  {journal} {Journal of Neurophysiology}\ }\textbf {\bibinfo {volume} {93}},\
  \bibinfo {pages} {2194} (\bibinfo {year} {2005})}\BibitemShut {NoStop}%
\bibitem [{\citenamefont {Markram}(2006)}]{markram_blue_2006}%
  \BibitemOpen
  \bibfield  {author} {\bibinfo {author} {\bibfnamefont {H.}~\bibnamefont
  {Markram}},\ }\bibfield  {title} {\bibinfo {title} {The {{Blue Brain
  Project}}},\ }\href {https://doi.org/10.1038/nrn1848} {\bibfield  {journal}
  {\bibinfo  {journal} {Nature Reviews Neuroscience}\ }\textbf {\bibinfo
  {volume} {7}},\ \bibinfo {pages} {153} (\bibinfo {year} {2006})}\BibitemShut
  {NoStop}%
\bibitem [{\citenamefont {Markram}\ \emph {et~al.}(2015)\citenamefont
  {Markram}, \citenamefont {Muller}, \citenamefont {Ramaswamy}, \citenamefont
  {Reimann}, \citenamefont {Abdellah}, \citenamefont {Sanchez}, \citenamefont
  {Ailamaki}, \citenamefont {{Alonso-Nanclares}}, \citenamefont {Antille},
  \citenamefont {Arsever}, \citenamefont {Kahou}, \citenamefont {Berger},
  \citenamefont {Bilgili}, \citenamefont {Buncic}, \citenamefont {Chalimourda},
  \citenamefont {Chindemi}, \citenamefont {Courcol}, \citenamefont
  {Delalondre}, \citenamefont {Delattre}, \citenamefont {Druckmann},
  \citenamefont {Dumusc}, \citenamefont {Dynes}, \citenamefont {Eilemann},
  \citenamefont {Gal}, \citenamefont {Gevaert}, \citenamefont {Ghobril},
  \citenamefont {Gidon}, \citenamefont {Graham}, \citenamefont {Gupta},
  \citenamefont {Haenel}, \citenamefont {Hay}, \citenamefont {Heinis},
  \citenamefont {Hernando}, \citenamefont {Hines}, \citenamefont {Kanari},
  \citenamefont {Keller}, \citenamefont {Kenyon}, \citenamefont {Khazen},
  \citenamefont {Kim}, \citenamefont {King}, \citenamefont {Kisvarday},
  \citenamefont {Kumbhar}, \citenamefont {Lasserre}, \citenamefont {Le~B{\'e}},
  \citenamefont {Magalh{\~a}es}, \citenamefont {{Merch{\'a}n-P{\'e}rez}},
  \citenamefont {Meystre}, \citenamefont {Morrice}, \citenamefont {Muller},
  \citenamefont {{Mu{\~n}oz-C{\'e}spedes}}, \citenamefont {Muralidhar},
  \citenamefont {Muthurasa}, \citenamefont {Nachbaur}, \citenamefont {Newton},
  \citenamefont {Nolte}, \citenamefont {Ovcharenko}, \citenamefont {Palacios},
  \citenamefont {Pastor}, \citenamefont {Perin}, \citenamefont {Ranjan},
  \citenamefont {Riachi}, \citenamefont {Rodr{\'i}guez}, \citenamefont
  {Riquelme}, \citenamefont {R{\"o}ssert}, \citenamefont {Sfyrakis},
  \citenamefont {Shi}, \citenamefont {Shillcock}, \citenamefont {Silberberg},
  \citenamefont {Silva}, \citenamefont {Tauheed}, \citenamefont {Telefont},
  \citenamefont {{Toledo-Rodriguez}}, \citenamefont {Tr{\"a}nkler},
  \citenamefont {Van~Geit}, \citenamefont {D{\'i}az}, \citenamefont {Walker},
  \citenamefont {Wang}, \citenamefont {Zaninetta}, \citenamefont {DeFelipe},
  \citenamefont {Hill}, \citenamefont {Segev},\ and\ \citenamefont
  {Sch{\"u}rmann}}]{markram_reconstruction_2015}%
  \BibitemOpen
  \bibfield  {author} {\bibinfo {author} {\bibfnamefont {H.}~\bibnamefont
  {Markram}}, \bibinfo {author} {\bibfnamefont {E.}~\bibnamefont {Muller}},
  \bibinfo {author} {\bibfnamefont {S.}~\bibnamefont {Ramaswamy}}, \bibinfo
  {author} {\bibfnamefont {M.~W.}\ \bibnamefont {Reimann}}, \bibinfo {author}
  {\bibfnamefont {M.}~\bibnamefont {Abdellah}}, \bibinfo {author}
  {\bibfnamefont {C.~A.}\ \bibnamefont {Sanchez}}, \bibinfo {author}
  {\bibfnamefont {A.}~\bibnamefont {Ailamaki}}, \bibinfo {author}
  {\bibfnamefont {L.}~\bibnamefont {{Alonso-Nanclares}}}, \bibinfo {author}
  {\bibfnamefont {N.}~\bibnamefont {Antille}}, \bibinfo {author} {\bibfnamefont
  {S.}~\bibnamefont {Arsever}}, \bibinfo {author} {\bibfnamefont {G.~A.~A.}\
  \bibnamefont {Kahou}}, \bibinfo {author} {\bibfnamefont {T.~K.}\ \bibnamefont
  {Berger}}, \bibinfo {author} {\bibfnamefont {A.}~\bibnamefont {Bilgili}},
  \bibinfo {author} {\bibfnamefont {N.}~\bibnamefont {Buncic}}, \bibinfo
  {author} {\bibfnamefont {A.}~\bibnamefont {Chalimourda}}, \bibinfo {author}
  {\bibfnamefont {G.}~\bibnamefont {Chindemi}}, \bibinfo {author}
  {\bibfnamefont {J.-D.}\ \bibnamefont {Courcol}}, \bibinfo {author}
  {\bibfnamefont {F.}~\bibnamefont {Delalondre}}, \bibinfo {author}
  {\bibfnamefont {V.}~\bibnamefont {Delattre}}, \bibinfo {author}
  {\bibfnamefont {S.}~\bibnamefont {Druckmann}}, \bibinfo {author}
  {\bibfnamefont {R.}~\bibnamefont {Dumusc}}, \bibinfo {author} {\bibfnamefont
  {J.}~\bibnamefont {Dynes}}, \bibinfo {author} {\bibfnamefont
  {S.}~\bibnamefont {Eilemann}}, \bibinfo {author} {\bibfnamefont
  {E.}~\bibnamefont {Gal}}, \bibinfo {author} {\bibfnamefont {M.~E.}\
  \bibnamefont {Gevaert}}, \bibinfo {author} {\bibfnamefont {J.-P.}\
  \bibnamefont {Ghobril}}, \bibinfo {author} {\bibfnamefont {A.}~\bibnamefont
  {Gidon}}, \bibinfo {author} {\bibfnamefont {J.~W.}\ \bibnamefont {Graham}},
  \bibinfo {author} {\bibfnamefont {A.}~\bibnamefont {Gupta}}, \bibinfo
  {author} {\bibfnamefont {V.}~\bibnamefont {Haenel}}, \bibinfo {author}
  {\bibfnamefont {E.}~\bibnamefont {Hay}}, \bibinfo {author} {\bibfnamefont
  {T.}~\bibnamefont {Heinis}}, \bibinfo {author} {\bibfnamefont {J.~B.}\
  \bibnamefont {Hernando}}, \bibinfo {author} {\bibfnamefont {M.}~\bibnamefont
  {Hines}}, \bibinfo {author} {\bibfnamefont {L.}~\bibnamefont {Kanari}},
  \bibinfo {author} {\bibfnamefont {D.}~\bibnamefont {Keller}}, \bibinfo
  {author} {\bibfnamefont {J.}~\bibnamefont {Kenyon}}, \bibinfo {author}
  {\bibfnamefont {G.}~\bibnamefont {Khazen}}, \bibinfo {author} {\bibfnamefont
  {Y.}~\bibnamefont {Kim}}, \bibinfo {author} {\bibfnamefont {J.~G.}\
  \bibnamefont {King}}, \bibinfo {author} {\bibfnamefont {Z.}~\bibnamefont
  {Kisvarday}}, \bibinfo {author} {\bibfnamefont {P.}~\bibnamefont {Kumbhar}},
  \bibinfo {author} {\bibfnamefont {S.}~\bibnamefont {Lasserre}}, \bibinfo
  {author} {\bibfnamefont {J.-V.}\ \bibnamefont {Le~B{\'e}}}, \bibinfo {author}
  {\bibfnamefont {B.~R.~C.}\ \bibnamefont {Magalh{\~a}es}}, \bibinfo {author}
  {\bibfnamefont {A.}~\bibnamefont {{Merch{\'a}n-P{\'e}rez}}}, \bibinfo
  {author} {\bibfnamefont {J.}~\bibnamefont {Meystre}}, \bibinfo {author}
  {\bibfnamefont {B.~R.}\ \bibnamefont {Morrice}}, \bibinfo {author}
  {\bibfnamefont {J.}~\bibnamefont {Muller}}, \bibinfo {author} {\bibfnamefont
  {A.}~\bibnamefont {{Mu{\~n}oz-C{\'e}spedes}}}, \bibinfo {author}
  {\bibfnamefont {S.}~\bibnamefont {Muralidhar}}, \bibinfo {author}
  {\bibfnamefont {K.}~\bibnamefont {Muthurasa}}, \bibinfo {author}
  {\bibfnamefont {D.}~\bibnamefont {Nachbaur}}, \bibinfo {author}
  {\bibfnamefont {T.~H.}\ \bibnamefont {Newton}}, \bibinfo {author}
  {\bibfnamefont {M.}~\bibnamefont {Nolte}}, \bibinfo {author} {\bibfnamefont
  {A.}~\bibnamefont {Ovcharenko}}, \bibinfo {author} {\bibfnamefont
  {J.}~\bibnamefont {Palacios}}, \bibinfo {author} {\bibfnamefont
  {L.}~\bibnamefont {Pastor}}, \bibinfo {author} {\bibfnamefont
  {R.}~\bibnamefont {Perin}}, \bibinfo {author} {\bibfnamefont
  {R.}~\bibnamefont {Ranjan}}, \bibinfo {author} {\bibfnamefont
  {I.}~\bibnamefont {Riachi}}, \bibinfo {author} {\bibfnamefont {J.-R.}\
  \bibnamefont {Rodr{\'i}guez}}, \bibinfo {author} {\bibfnamefont {J.~L.}\
  \bibnamefont {Riquelme}}, \bibinfo {author} {\bibfnamefont {C.}~\bibnamefont
  {R{\"o}ssert}}, \bibinfo {author} {\bibfnamefont {K.}~\bibnamefont
  {Sfyrakis}}, \bibinfo {author} {\bibfnamefont {Y.}~\bibnamefont {Shi}},
  \bibinfo {author} {\bibfnamefont {J.~C.}\ \bibnamefont {Shillcock}}, \bibinfo
  {author} {\bibfnamefont {G.}~\bibnamefont {Silberberg}}, \bibinfo {author}
  {\bibfnamefont {R.}~\bibnamefont {Silva}}, \bibinfo {author} {\bibfnamefont
  {F.}~\bibnamefont {Tauheed}}, \bibinfo {author} {\bibfnamefont
  {M.}~\bibnamefont {Telefont}}, \bibinfo {author} {\bibfnamefont
  {M.}~\bibnamefont {{Toledo-Rodriguez}}}, \bibinfo {author} {\bibfnamefont
  {T.}~\bibnamefont {Tr{\"a}nkler}}, \bibinfo {author} {\bibfnamefont
  {W.}~\bibnamefont {Van~Geit}}, \bibinfo {author} {\bibfnamefont {J.~V.}\
  \bibnamefont {D{\'i}az}}, \bibinfo {author} {\bibfnamefont {R.}~\bibnamefont
  {Walker}}, \bibinfo {author} {\bibfnamefont {Y.}~\bibnamefont {Wang}},
  \bibinfo {author} {\bibfnamefont {S.~M.}\ \bibnamefont {Zaninetta}}, \bibinfo
  {author} {\bibfnamefont {J.}~\bibnamefont {DeFelipe}}, \bibinfo {author}
  {\bibfnamefont {S.~L.}\ \bibnamefont {Hill}}, \bibinfo {author}
  {\bibfnamefont {I.}~\bibnamefont {Segev}},\ and\ \bibinfo {author}
  {\bibfnamefont {F.}~\bibnamefont {Sch{\"u}rmann}},\ }\bibfield  {title}
  {\bibinfo {title} {Reconstruction and {{Simulation}} of {{Neocortical
  Microcircuitry}}},\ }\href {https://doi.org/10.1016/j.cell.2015.09.029}
  {\bibfield  {journal} {\bibinfo  {journal} {Cell}\ }\textbf {\bibinfo
  {volume} {163}},\ \bibinfo {pages} {456} (\bibinfo {year}
  {2015})}\BibitemShut {NoStop}%
\bibitem [{\citenamefont {Billeh}\ \emph {et~al.}(2020)\citenamefont {Billeh},
  \citenamefont {Cai}, \citenamefont {Gratiy}, \citenamefont {Dai},
  \citenamefont {Iyer}, \citenamefont {Gouwens}, \citenamefont {{Abbasi-Asl}},
  \citenamefont {Jia}, \citenamefont {Siegle}, \citenamefont {Olsen},
  \citenamefont {Koch}, \citenamefont {Mihalas},\ and\ \citenamefont
  {Arkhipov}}]{billeh_systematic_2020}%
  \BibitemOpen
  \bibfield  {author} {\bibinfo {author} {\bibfnamefont {Y.~N.}\ \bibnamefont
  {Billeh}}, \bibinfo {author} {\bibfnamefont {B.}~\bibnamefont {Cai}},
  \bibinfo {author} {\bibfnamefont {S.~L.}\ \bibnamefont {Gratiy}}, \bibinfo
  {author} {\bibfnamefont {K.}~\bibnamefont {Dai}}, \bibinfo {author}
  {\bibfnamefont {R.}~\bibnamefont {Iyer}}, \bibinfo {author} {\bibfnamefont
  {N.~W.}\ \bibnamefont {Gouwens}}, \bibinfo {author} {\bibfnamefont
  {R.}~\bibnamefont {{Abbasi-Asl}}}, \bibinfo {author} {\bibfnamefont
  {X.}~\bibnamefont {Jia}}, \bibinfo {author} {\bibfnamefont {J.~H.}\
  \bibnamefont {Siegle}}, \bibinfo {author} {\bibfnamefont {S.~R.}\
  \bibnamefont {Olsen}}, \bibinfo {author} {\bibfnamefont {C.}~\bibnamefont
  {Koch}}, \bibinfo {author} {\bibfnamefont {S.}~\bibnamefont {Mihalas}},\ and\
  \bibinfo {author} {\bibfnamefont {A.}~\bibnamefont {Arkhipov}},\ }\bibfield
  {title} {\bibinfo {title} {Systematic {{Integration}} of {{Structural}} and
  {{Functional Data}} into {{Multi-scale Models}} of {{Mouse Primary Visual
  Cortex}}},\ }\href {https://doi.org/10.1016/j.neuron.2020.01.040} {\bibfield
  {journal} {\bibinfo  {journal} {Neuron}\ }\textbf {\bibinfo {volume} {106}},\
  \bibinfo {pages} {388} (\bibinfo {year} {2020})}\BibitemShut {NoStop}%
\bibitem [{\citenamefont {Grossberg}(1969)}]{grossberg_learning_1969}%
  \BibitemOpen
  \bibfield  {author} {\bibinfo {author} {\bibfnamefont {S.}~\bibnamefont
  {Grossberg}},\ }\bibfield  {title} {\bibinfo {title} {On learning and
  energy-entropy dependence in recurrent and nonrecurrent signed networks},\
  }\href {https://doi.org/10.1007/BF01007484} {\bibfield  {journal} {\bibinfo
  {journal} {Journal of Statistical Physics}\ }\textbf {\bibinfo {volume}
  {1}},\ \bibinfo {pages} {319} (\bibinfo {year} {1969})}\BibitemShut {NoStop}%
\bibitem [{\citenamefont {Amari}(1971)}]{amari_characteristics_1971}%
  \BibitemOpen
  \bibfield  {author} {\bibinfo {author} {\bibfnamefont {S.-I.}\ \bibnamefont
  {Amari}},\ }\bibfield  {title} {\bibinfo {title} {Characteristics of randomly
  connected threshold-element networks and network systems},\ }\href
  {https://doi.org/10.1109/PROC.1971.8087} {\bibfield  {journal} {\bibinfo
  {journal} {Proceedings of the IEEE}\ }\textbf {\bibinfo {volume} {59}},\
  \bibinfo {pages} {35} (\bibinfo {year} {1971})}\BibitemShut {NoStop}%
\bibitem [{\citenamefont {Amari}(1972)}]{amari_characteristics_1972}%
  \BibitemOpen
  \bibfield  {author} {\bibinfo {author} {\bibfnamefont {S.-I.}\ \bibnamefont
  {Amari}},\ }\bibfield  {title} {\bibinfo {title} {Characteristics of {{Random
  Nets}} of {{Analog Neuron-Like Elements}}},\ }\href
  {https://doi.org/10.1109/TSMC.1972.4309193} {\bibfield  {journal} {\bibinfo
  {journal} {IEEE Transactions on Systems, Man, and Cybernetics}\ }\textbf
  {\bibinfo {volume} {SMC-2}},\ \bibinfo {pages} {643} (\bibinfo {year}
  {1972})}\BibitemShut {NoStop}%
\bibitem [{\citenamefont {Wilson}\ and\ \citenamefont
  {Cowan}(1972)}]{wilson_excitatory_1972}%
  \BibitemOpen
  \bibfield  {author} {\bibinfo {author} {\bibfnamefont {H.~R.}\ \bibnamefont
  {Wilson}}\ and\ \bibinfo {author} {\bibfnamefont {J.~D.}\ \bibnamefont
  {Cowan}},\ }\bibfield  {title} {\bibinfo {title} {Excitatory and {{Inhibitory
  Interactions}} in {{Localized Populations}} of {{Model Neurons}}},\ }\href
  {https://doi.org/10.1016/S0006-3495(72)86068-5} {\bibfield  {journal}
  {\bibinfo  {journal} {Biophysical Journal}\ }\textbf {\bibinfo {volume}
  {12}},\ \bibinfo {pages} {1} (\bibinfo {year} {1972})}\BibitemShut {NoStop}%
\bibitem [{\citenamefont {Sompolinsky}\ \emph {et~al.}(1988)\citenamefont
  {Sompolinsky}, \citenamefont {Crisanti},\ and\ \citenamefont
  {Sommers}}]{sompolinsky_chaos_1988}%
  \BibitemOpen
  \bibfield  {author} {\bibinfo {author} {\bibfnamefont {H.}~\bibnamefont
  {Sompolinsky}}, \bibinfo {author} {\bibfnamefont {A.}~\bibnamefont
  {Crisanti}},\ and\ \bibinfo {author} {\bibfnamefont {H.~J.}\ \bibnamefont
  {Sommers}},\ }\bibfield  {title} {\bibinfo {title} {Chaos in {{Random Neural
  Networks}}},\ }\href {https://doi.org/10.1103/PhysRevLett.61.259} {\bibfield
  {journal} {\bibinfo  {journal} {Physical Review Letters}\ }\textbf {\bibinfo
  {volume} {61}},\ \bibinfo {pages} {259} (\bibinfo {year} {1988})}\BibitemShut
  {NoStop}%
\bibitem [{\citenamefont {Ohira}\ and\ \citenamefont
  {Cowan}(1993)}]{ohira_master-equation_1993}%
  \BibitemOpen
  \bibfield  {author} {\bibinfo {author} {\bibfnamefont {T.}~\bibnamefont
  {Ohira}}\ and\ \bibinfo {author} {\bibfnamefont {J.~D.}\ \bibnamefont
  {Cowan}},\ }\bibfield  {title} {\bibinfo {title} {Master-equation approach to
  stochastic neurodynamics},\ }\href {https://doi.org/10.1103/PhysRevE.48.2259}
  {\bibfield  {journal} {\bibinfo  {journal} {Physical Review E}\ }\textbf
  {\bibinfo {volume} {48}},\ \bibinfo {pages} {2259} (\bibinfo {year}
  {1993})}\BibitemShut {NoStop}%
\bibitem [{\citenamefont {Ginzburg}\ and\ \citenamefont
  {Sompolinsky}(1994)}]{ginzburg_theory_1994}%
  \BibitemOpen
  \bibfield  {author} {\bibinfo {author} {\bibfnamefont {I.}~\bibnamefont
  {Ginzburg}}\ and\ \bibinfo {author} {\bibfnamefont {H.}~\bibnamefont
  {Sompolinsky}},\ }\bibfield  {title} {\bibinfo {title} {Theory of
  correlations in stochastic neural networks},\ }\href
  {https://doi.org/10.1103/PhysRevE.50.3171} {\bibfield  {journal} {\bibinfo
  {journal} {Physical Review E}\ }\textbf {\bibinfo {volume} {50}},\ \bibinfo
  {pages} {3171} (\bibinfo {year} {1994})}\BibitemShut {NoStop}%
\bibitem [{\citenamefont {van Vreeswijk}\ and\ \citenamefont
  {Sompolinsky}(1998)}]{vreeswijk_chaotic_1998}%
  \BibitemOpen
  \bibfield  {author} {\bibinfo {author} {\bibfnamefont {C.}~\bibnamefont {van
  Vreeswijk}}\ and\ \bibinfo {author} {\bibfnamefont {H.}~\bibnamefont
  {Sompolinsky}},\ }\bibfield  {title} {\bibinfo {title} {Chaotic {{Balanced
  State}} in a {{Model}} of {{Cortical Circuits}}},\ }\href
  {https://doi.org/10.1162/089976698300017214} {\bibfield  {journal} {\bibinfo
  {journal} {Neural Computation}\ }\textbf {\bibinfo {volume} {10}},\ \bibinfo
  {pages} {1321} (\bibinfo {year} {1998})}\BibitemShut {NoStop}%
\bibitem [{\citenamefont {Brunel}(2000)}]{brunel_dynamics_2000}%
  \BibitemOpen
  \bibfield  {author} {\bibinfo {author} {\bibfnamefont {N.}~\bibnamefont
  {Brunel}},\ }\bibfield  {title} {\bibinfo {title} {Dynamics of {{Sparsely
  Connected Networks}} of {{Excitatory}} and {{Inhibitory Spiking Neurons}}},\
  }\href {https://doi.org/10.1023/A:1008925309027} {\bibfield  {journal}
  {\bibinfo  {journal} {Journal of Computational Neuroscience}\ }\textbf
  {\bibinfo {volume} {8}},\ \bibinfo {pages} {183} (\bibinfo {year}
  {2000})}\BibitemShut {NoStop}%
\bibitem [{\citenamefont {Gerstner}(1995)}]{gerstner_time_1995}%
  \BibitemOpen
  \bibfield  {author} {\bibinfo {author} {\bibfnamefont {W.}~\bibnamefont
  {Gerstner}},\ }\bibfield  {title} {\bibinfo {title} {Time structure of the
  activity in neural network models},\ }\href
  {https://doi.org/10.1103/PhysRevE.51.738} {\bibfield  {journal} {\bibinfo
  {journal} {Physical Review E}\ }\textbf {\bibinfo {volume} {51}},\ \bibinfo
  {pages} {738} (\bibinfo {year} {1995})}\BibitemShut {NoStop}%
\bibitem [{\citenamefont {Renart}\ \emph {et~al.}(2010)\citenamefont {Renart},
  \citenamefont {de~la Rocha}, \citenamefont {Bartho}, \citenamefont
  {Hollender}, \citenamefont {Parga}, \citenamefont {Reyes},\ and\
  \citenamefont {Harris}}]{renart_asynchronous_2010}%
  \BibitemOpen
  \bibfield  {author} {\bibinfo {author} {\bibfnamefont {A.}~\bibnamefont
  {Renart}}, \bibinfo {author} {\bibfnamefont {J.}~\bibnamefont {de~la Rocha}},
  \bibinfo {author} {\bibfnamefont {P.}~\bibnamefont {Bartho}}, \bibinfo
  {author} {\bibfnamefont {L.}~\bibnamefont {Hollender}}, \bibinfo {author}
  {\bibfnamefont {N.}~\bibnamefont {Parga}}, \bibinfo {author} {\bibfnamefont
  {A.}~\bibnamefont {Reyes}},\ and\ \bibinfo {author} {\bibfnamefont {K.~D.}\
  \bibnamefont {Harris}},\ }\bibfield  {title} {\bibinfo {title} {The
  {{Asynchronous State}} in {{Cortical Circuits}}},\ }\href
  {https://doi.org/10.1126/science.1179850} {\bibfield  {journal} {\bibinfo
  {journal} {Science}\ }\textbf {\bibinfo {volume} {327}},\ \bibinfo {pages}
  {587} (\bibinfo {year} {2010})}\BibitemShut {NoStop}%
\bibitem [{\citenamefont {Montbri{\'o}}\ \emph {et~al.}(2015)\citenamefont
  {Montbri{\'o}}, \citenamefont {Paz{\'o}},\ and\ \citenamefont
  {Roxin}}]{montbrio_macroscopic_2015}%
  \BibitemOpen
  \bibfield  {author} {\bibinfo {author} {\bibfnamefont {E.}~\bibnamefont
  {Montbri{\'o}}}, \bibinfo {author} {\bibfnamefont {D.}~\bibnamefont
  {Paz{\'o}}},\ and\ \bibinfo {author} {\bibfnamefont {A.}~\bibnamefont
  {Roxin}},\ }\bibfield  {title} {\bibinfo {title} {Macroscopic {{Description}}
  for {{Networks}} of {{Spiking Neurons}}},\ }\href
  {https://doi.org/10.1103/PhysRevX.5.021028} {\bibfield  {journal} {\bibinfo
  {journal} {Physical Review X}\ }\textbf {\bibinfo {volume} {5}},\ \bibinfo
  {pages} {021028} (\bibinfo {year} {2015})}\BibitemShut {NoStop}%
\bibitem [{\citenamefont {Paliwal}\ \emph {et~al.}(2025)\citenamefont
  {Paliwal}, \citenamefont {Ocker},\ and\ \citenamefont
  {Brinkman}}]{paliwal_metastability_2025}%
  \BibitemOpen
  \bibfield  {author} {\bibinfo {author} {\bibfnamefont {S.}~\bibnamefont
  {Paliwal}}, \bibinfo {author} {\bibfnamefont {G.~K.}\ \bibnamefont {Ocker}},\
  and\ \bibinfo {author} {\bibfnamefont {B.~A.~W.}\ \bibnamefont {Brinkman}},\
  }\bibfield  {title} {\bibinfo {title} {Metastability in networks of
  stochastic integrate-and-fire neurons},\ }\href
  {https://doi.org/10.1103/PhysRevE.111.064402} {\bibfield  {journal} {\bibinfo
   {journal} {Physical Review E}\ }\textbf {\bibinfo {volume} {111}},\ \bibinfo
  {pages} {064402} (\bibinfo {year} {2025})}\BibitemShut {NoStop}%
\bibitem [{\citenamefont {Chow}\ and\ \citenamefont
  {Buice}(2015)}]{chow_path_2015}%
  \BibitemOpen
  \bibfield  {author} {\bibinfo {author} {\bibfnamefont {C.~C.}\ \bibnamefont
  {Chow}}\ and\ \bibinfo {author} {\bibfnamefont {M.~A.}\ \bibnamefont
  {Buice}},\ }\bibfield  {title} {\bibinfo {title} {Path {{Integral Methods}}
  for {{Stochastic Differential Equations}}},\ }\href
  {https://doi.org/10.1186/s13408-015-0018-5} {\bibfield  {journal} {\bibinfo
  {journal} {The Journal of Mathematical Neuroscience (JMN)}\ }\textbf
  {\bibinfo {volume} {5}},\ \bibinfo {pages} {1} (\bibinfo {year}
  {2015})}\BibitemShut {NoStop}%
\bibitem [{\citenamefont {Ocker}(2023)}]{ocker_republished_2023}%
  \BibitemOpen
  \bibfield  {author} {\bibinfo {author} {\bibfnamefont {G.~K.}\ \bibnamefont
  {Ocker}},\ }\bibfield  {title} {\bibinfo {title} {Republished: {{Dynamics}}
  of {{Stochastic Integrate-and-Fire Networks}}},\ }\href
  {https://doi.org/10.1103/PhysRevX.13.041047} {\bibfield  {journal} {\bibinfo
  {journal} {Physical Review X}\ }\textbf {\bibinfo {volume} {13}},\ \bibinfo
  {pages} {041047} (\bibinfo {year} {2023})}\BibitemShut {NoStop}%
\bibitem [{\citenamefont {Buice}\ and\ \citenamefont
  {Cowan}(2007)}]{buice_field-theoretic_2007}%
  \BibitemOpen
  \bibfield  {author} {\bibinfo {author} {\bibfnamefont {M.~A.}\ \bibnamefont
  {Buice}}\ and\ \bibinfo {author} {\bibfnamefont {J.~D.}\ \bibnamefont
  {Cowan}},\ }\bibfield  {title} {\bibinfo {title} {Field-theoretic approach to
  fluctuation effects in neural networks},\ }\href
  {https://doi.org/10.1103/PhysRevE.75.051919} {\bibfield  {journal} {\bibinfo
  {journal} {Physical Review E}\ }\textbf {\bibinfo {volume} {75}},\ \bibinfo
  {pages} {051919} (\bibinfo {year} {2007})}\BibitemShut {NoStop}%
\bibitem [{\citenamefont {Buice}\ \emph {et~al.}(2010)\citenamefont {Buice},
  \citenamefont {Cowan},\ and\ \citenamefont {Chow}}]{buice_systematic_2010}%
  \BibitemOpen
  \bibfield  {author} {\bibinfo {author} {\bibfnamefont {M.~A.}\ \bibnamefont
  {Buice}}, \bibinfo {author} {\bibfnamefont {J.~D.}\ \bibnamefont {Cowan}},\
  and\ \bibinfo {author} {\bibfnamefont {C.~C.}\ \bibnamefont {Chow}},\
  }\bibfield  {title} {\bibinfo {title} {Systematic {{Fluctuation Expansion}}
  for {{Neural Network Activity Equations}}},\ }\href
  {https://doi.org/10.1162/neco.2009.02-09-960} {\bibfield  {journal} {\bibinfo
   {journal} {Neural Computation}\ }\textbf {\bibinfo {volume} {22}},\ \bibinfo
  {pages} {377} (\bibinfo {year} {2010})}\BibitemShut {NoStop}%
\bibitem [{\citenamefont {Ocker}\ \emph {et~al.}(2017)\citenamefont {Ocker},
  \citenamefont {Josi{\'c}}, \citenamefont {{Shea-Brown}},\ and\ \citenamefont
  {Buice}}]{ocker_linking_2017}%
  \BibitemOpen
  \bibfield  {author} {\bibinfo {author} {\bibfnamefont {G.~K.}\ \bibnamefont
  {Ocker}}, \bibinfo {author} {\bibfnamefont {K.}~\bibnamefont {Josi{\'c}}},
  \bibinfo {author} {\bibfnamefont {E.}~\bibnamefont {{Shea-Brown}}},\ and\
  \bibinfo {author} {\bibfnamefont {M.~A.}\ \bibnamefont {Buice}},\ }\bibfield
  {title} {\bibinfo {title} {Linking structure and activity in nonlinear
  spiking networks},\ }\href {https://doi.org/10.1371/journal.pcbi.1005583}
  {\bibfield  {journal} {\bibinfo  {journal} {PLOS Computational Biology}\
  }\textbf {\bibinfo {volume} {13}},\ \bibinfo {pages} {e1005583} (\bibinfo
  {year} {2017})}\BibitemShut {NoStop}%
\bibitem [{\citenamefont {Williams}\ and\ \citenamefont
  {Stuart}(1999)}]{williams_mechanisms_1999}%
  \BibitemOpen
  \bibfield  {author} {\bibinfo {author} {\bibfnamefont {S.~R.}\ \bibnamefont
  {Williams}}\ and\ \bibinfo {author} {\bibfnamefont {G.~J.}\ \bibnamefont
  {Stuart}},\ }\bibfield  {title} {\bibinfo {title} {Mechanisms and
  consequences of action potential burst firing in rat neocortical pyramidal
  neurons},\ }\href {https://doi.org/10.1111/j.1469-7793.1999.00467.x}
  {\bibfield  {journal} {\bibinfo  {journal} {The Journal of Physiology}\
  }\textbf {\bibinfo {volume} {521}},\ \bibinfo {pages} {467} (\bibinfo {year}
  {1999})}\BibitemShut {NoStop}%
\bibitem [{\citenamefont {Larkum}\ \emph {et~al.}(2001)\citenamefont {Larkum},
  \citenamefont {Zhu},\ and\ \citenamefont {Sakmann}}]{larkum_dendritic_2001}%
  \BibitemOpen
  \bibfield  {author} {\bibinfo {author} {\bibfnamefont {M.~E.}\ \bibnamefont
  {Larkum}}, \bibinfo {author} {\bibfnamefont {J.~J.}\ \bibnamefont {Zhu}},\
  and\ \bibinfo {author} {\bibfnamefont {B.}~\bibnamefont {Sakmann}},\
  }\bibfield  {title} {\bibinfo {title} {Dendritic mechanisms underlying the
  coupling of the dendritic with the axonal action potential initiation zone of
  adult rat layer 5 pyramidal neurons},\ }\href
  {https://doi.org/10.1111/j.1469-7793.2001.0447a.x} {\bibfield  {journal}
  {\bibinfo  {journal} {The Journal of Physiology}\ }\textbf {\bibinfo {volume}
  {533}},\ \bibinfo {pages} {447} (\bibinfo {year} {2001})}\BibitemShut
  {NoStop}%
\bibitem [{\citenamefont {Schwindt}\ and\ \citenamefont
  {Crill}(1999)}]{schwindt_mechanisms_1999}%
  \BibitemOpen
  \bibfield  {author} {\bibinfo {author} {\bibfnamefont {P.}~\bibnamefont
  {Schwindt}}\ and\ \bibinfo {author} {\bibfnamefont {W.}~\bibnamefont
  {Crill}},\ }\bibfield  {title} {\bibinfo {title} {Mechanisms {{Underlying
  Burst}} and {{Regular Spiking Evoked}} by {{Dendritic Depolarization}} in
  {{Layer}} 5 {{Cortical Pyramidal Neurons}}},\ }\href
  {https://doi.org/10.1152/jn.1999.81.3.1341} {\bibfield  {journal} {\bibinfo
  {journal} {Journal of Neurophysiology}\ }\textbf {\bibinfo {volume} {81}},\
  \bibinfo {pages} {1341} (\bibinfo {year} {1999})}\BibitemShut {NoStop}%
\bibitem [{\citenamefont {Kim}\ and\ \citenamefont
  {Connors}(1993)}]{kim_apical_1993}%
  \BibitemOpen
  \bibfield  {author} {\bibinfo {author} {\bibfnamefont {H.~G.}\ \bibnamefont
  {Kim}}\ and\ \bibinfo {author} {\bibfnamefont {B.~W.}\ \bibnamefont
  {Connors}},\ }\bibfield  {title} {\bibinfo {title} {Apical dendrites of the
  neocortex: Correlation between sodium- and calcium-dependent spiking and
  pyramidal cell morphology},\ }\href
  {https://doi.org/10.1523/JNEUROSCI.13-12-05301.1993} {\bibfield  {journal}
  {\bibinfo  {journal} {Journal of Neuroscience}\ }\textbf {\bibinfo {volume}
  {13}},\ \bibinfo {pages} {5301} (\bibinfo {year} {1993})}\BibitemShut
  {NoStop}%
\bibitem [{\citenamefont {Zhu}(2015)}]{zhu_large_2015}%
  \BibitemOpen
  \bibfield  {author} {\bibinfo {author} {\bibfnamefont {L.}~\bibnamefont
  {Zhu}},\ }\bibfield  {title} {\bibinfo {title} {Large deviations for
  {{Markovian}} nonlinear {{Hawkes}} processes},\ }\href
  {https://doi.org/10.1214/14-AAP1003} {\bibfield  {journal} {\bibinfo
  {journal} {The Annals of Applied Probability}\ }\textbf {\bibinfo {volume}
  {25}},\ \bibinfo {pages} {548} (\bibinfo {year} {2015})}\BibitemShut
  {NoStop}%
\bibitem [{\citenamefont {Delattre}\ \emph {et~al.}(2016)\citenamefont
  {Delattre}, \citenamefont {Fournier},\ and\ \citenamefont
  {Hoffmann}}]{delattre_hawkes_2016}%
  \BibitemOpen
  \bibfield  {author} {\bibinfo {author} {\bibfnamefont {S.}~\bibnamefont
  {Delattre}}, \bibinfo {author} {\bibfnamefont {N.}~\bibnamefont {Fournier}},\
  and\ \bibinfo {author} {\bibfnamefont {M.}~\bibnamefont {Hoffmann}},\
  }\bibfield  {title} {\bibinfo {title} {Hawkes processes on large networks},\
  }\href {https://doi.org/10.1214/14-AAP1089} {\bibfield  {journal} {\bibinfo
  {journal} {The Annals of Applied Probability}\ }\textbf {\bibinfo {volume}
  {26}},\ \bibinfo {pages} {216} (\bibinfo {year} {2016})}\BibitemShut
  {NoStop}%
\bibitem [{\citenamefont {Heesen}\ and\ \citenamefont
  {Stannat}(2021)}]{heesen_fluctuation_2021}%
  \BibitemOpen
  \bibfield  {author} {\bibinfo {author} {\bibfnamefont {S.}~\bibnamefont
  {Heesen}}\ and\ \bibinfo {author} {\bibfnamefont {W.}~\bibnamefont
  {Stannat}},\ }\bibfield  {title} {\bibinfo {title} {Fluctuation limits for
  mean-field interacting nonlinear {{Hawkes}} processes},\ }\href
  {https://doi.org/10.1016/j.spa.2021.05.007} {\bibfield  {journal} {\bibinfo
  {journal} {Stochastic Processes and their Applications}\ }\textbf {\bibinfo
  {volume} {139}},\ \bibinfo {pages} {280} (\bibinfo {year}
  {2021})}\BibitemShut {NoStop}%
\bibitem [{\citenamefont {Pfaffelhuber}\ \emph {et~al.}(2022)\citenamefont
  {Pfaffelhuber}, \citenamefont {Rotter},\ and\ \citenamefont
  {Stiefel}}]{pfaffelhuber_mean-field_2022}%
  \BibitemOpen
  \bibfield  {author} {\bibinfo {author} {\bibfnamefont {P.}~\bibnamefont
  {Pfaffelhuber}}, \bibinfo {author} {\bibfnamefont {S.}~\bibnamefont
  {Rotter}},\ and\ \bibinfo {author} {\bibfnamefont {J.}~\bibnamefont
  {Stiefel}},\ }\bibfield  {title} {\bibinfo {title} {Mean-field limits for
  non-linear {{Hawkes}} processes with excitation and inhibition},\ }\href
  {https://doi.org/10.1016/j.spa.2022.07.006} {\bibfield  {journal} {\bibinfo
  {journal} {Stochastic Processes and their Applications}\ }\textbf {\bibinfo
  {volume} {153}},\ \bibinfo {pages} {57} (\bibinfo {year} {2022})}\BibitemShut
  {NoStop}%
\bibitem [{\citenamefont {Bittner}\ \emph {et~al.}(2015)\citenamefont
  {Bittner}, \citenamefont {Grienberger}, \citenamefont {Vaidya}, \citenamefont
  {Milstein}, \citenamefont {Macklin}, \citenamefont {Suh}, \citenamefont
  {Tonegawa},\ and\ \citenamefont {Magee}}]{bittner_conjunctive_2015}%
  \BibitemOpen
  \bibfield  {author} {\bibinfo {author} {\bibfnamefont {K.~C.}\ \bibnamefont
  {Bittner}}, \bibinfo {author} {\bibfnamefont {C.}~\bibnamefont
  {Grienberger}}, \bibinfo {author} {\bibfnamefont {S.~P.}\ \bibnamefont
  {Vaidya}}, \bibinfo {author} {\bibfnamefont {A.~D.}\ \bibnamefont
  {Milstein}}, \bibinfo {author} {\bibfnamefont {J.~J.}\ \bibnamefont
  {Macklin}}, \bibinfo {author} {\bibfnamefont {J.}~\bibnamefont {Suh}},
  \bibinfo {author} {\bibfnamefont {S.}~\bibnamefont {Tonegawa}},\ and\
  \bibinfo {author} {\bibfnamefont {J.~C.}\ \bibnamefont {Magee}},\ }\bibfield
  {title} {\bibinfo {title} {Conjunctive input processing drives feature
  selectivity in hippocampal {{CA1}} neurons},\ }\href
  {https://doi.org/10.1038/nn.4062} {\bibfield  {journal} {\bibinfo  {journal}
  {Nature Neuroscience}\ }\textbf {\bibinfo {volume} {18}},\ \bibinfo {pages}
  {1133} (\bibinfo {year} {2015})}\BibitemShut {NoStop}%
\bibitem [{\citenamefont {Cichon}\ and\ \citenamefont
  {Gan}(2015)}]{cichon_branch-specific_2015}%
  \BibitemOpen
  \bibfield  {author} {\bibinfo {author} {\bibfnamefont {J.}~\bibnamefont
  {Cichon}}\ and\ \bibinfo {author} {\bibfnamefont {W.-B.}\ \bibnamefont
  {Gan}},\ }\bibfield  {title} {\bibinfo {title} {Branch-specific dendritic
  {{Ca2}}+ spikes cause persistent synaptic plasticity},\ }\href
  {https://doi.org/10.1038/nature14251} {\bibfield  {journal} {\bibinfo
  {journal} {Nature}\ }\textbf {\bibinfo {volume} {520}},\ \bibinfo {pages}
  {180} (\bibinfo {year} {2015})}\BibitemShut {NoStop}%
\bibitem [{\citenamefont {Bittner}\ \emph {et~al.}(2017)\citenamefont
  {Bittner}, \citenamefont {Milstein}, \citenamefont {Grienberger},
  \citenamefont {Romani},\ and\ \citenamefont
  {Magee}}]{bittner_behavioral_2017}%
  \BibitemOpen
  \bibfield  {author} {\bibinfo {author} {\bibfnamefont {K.~C.}\ \bibnamefont
  {Bittner}}, \bibinfo {author} {\bibfnamefont {A.~D.}\ \bibnamefont
  {Milstein}}, \bibinfo {author} {\bibfnamefont {C.}~\bibnamefont
  {Grienberger}}, \bibinfo {author} {\bibfnamefont {S.}~\bibnamefont
  {Romani}},\ and\ \bibinfo {author} {\bibfnamefont {J.~C.}\ \bibnamefont
  {Magee}},\ }\bibfield  {title} {\bibinfo {title} {Behavioral time scale
  synaptic plasticity underlies {{CA1}} place fields},\ }\href
  {https://doi.org/10.1126/science.aan3846} {\bibfield  {journal} {\bibinfo
  {journal} {Science (New York, N.Y.)}\ }\textbf {\bibinfo {volume} {357}},\
  \bibinfo {pages} {1033} (\bibinfo {year} {2017})}\BibitemShut {NoStop}%
\bibitem [{\citenamefont {Martin}\ \emph {et~al.}(1973)\citenamefont {Martin},
  \citenamefont {Siggia},\ and\ \citenamefont
  {Rose}}]{martin_statistical_1973}%
  \BibitemOpen
  \bibfield  {author} {\bibinfo {author} {\bibfnamefont {P.~C.}\ \bibnamefont
  {Martin}}, \bibinfo {author} {\bibfnamefont {E.~D.}\ \bibnamefont {Siggia}},\
  and\ \bibinfo {author} {\bibfnamefont {H.~A.}\ \bibnamefont {Rose}},\
  }\bibfield  {title} {\bibinfo {title} {Statistical {{Dynamics}} of
  {{Classical Systems}}},\ }\href {https://doi.org/10.1103/PhysRevA.8.423}
  {\bibfield  {journal} {\bibinfo  {journal} {Physical Review A}\ }\textbf
  {\bibinfo {volume} {8}},\ \bibinfo {pages} {423} (\bibinfo {year}
  {1973})}\BibitemShut {NoStop}%
\bibitem [{\citenamefont {Dominicis}(1976)}]{dominicis_techniques_1976}%
  \BibitemOpen
  \bibfield  {author} {\bibinfo {author} {\bibfnamefont {C.~D.}\ \bibnamefont
  {Dominicis}},\ }\bibfield  {title} {\bibinfo {title} {{TECHNIQUES DE
  RENORMALISATION DE LA TH{\'E}ORIE DES CHAMPS ET DYNAMIQUE DES
  PH{\'E}NOM{\`E}NES CRITIQUES}},\ }\href
  {https://doi.org/10.1051/jphyscol:1976138} {\bibfield  {journal} {\bibinfo
  {journal} {Le Journal de Physique Colloques}\ }\textbf {\bibinfo {volume}
  {37}},\ \bibinfo {pages} {C1} (\bibinfo {year} {1976})}\BibitemShut {NoStop}%
\bibitem [{\citenamefont {Janssen}(1976)}]{janssen_lagrangean_1976}%
  \BibitemOpen
  \bibfield  {author} {\bibinfo {author} {\bibfnamefont {H.-K.}\ \bibnamefont
  {Janssen}},\ }\bibfield  {title} {\bibinfo {title} {On a {{Lagrangean}} for
  classical field dynamics and renormalization group calculations of dynamical
  critical properties},\ }\href {https://doi.org/10.1007/BF01316547} {\bibfield
   {journal} {\bibinfo  {journal} {Physik der Kondensierten Materie}\ }\textbf
  {\bibinfo {volume} {23}},\ \bibinfo {pages} {377} (\bibinfo {year}
  {1976})}\BibitemShut {NoStop}%
\bibitem [{\citenamefont {Jensen}(1981)}]{jensen_functional_1981}%
  \BibitemOpen
  \bibfield  {author} {\bibinfo {author} {\bibfnamefont {R.~V.}\ \bibnamefont
  {Jensen}},\ }\bibfield  {title} {\bibinfo {title} {Functional integral
  approach to classical statistical dynamics},\ }\href
  {https://doi.org/10.1007/BF01022182} {\bibfield  {journal} {\bibinfo
  {journal} {Journal of Statistical Physics}\ }\textbf {\bibinfo {volume}
  {25}},\ \bibinfo {pages} {183} (\bibinfo {year} {1981})}\BibitemShut
  {NoStop}%
\bibitem [{\citenamefont {Schmutz}\ \emph {et~al.}(2021)\citenamefont
  {Schmutz}, \citenamefont {L{\"o}cherbach},\ and\ \citenamefont
  {Schwalger}}]{schmutz_finite-size_2021}%
  \BibitemOpen
  \bibfield  {author} {\bibinfo {author} {\bibfnamefont {V.}~\bibnamefont
  {Schmutz}}, \bibinfo {author} {\bibfnamefont {E.}~\bibnamefont
  {L{\"o}cherbach}},\ and\ \bibinfo {author} {\bibfnamefont {T.}~\bibnamefont
  {Schwalger}},\ }\bibfield  {title} {\bibinfo {title} {On a finite-size
  neuronal population equation},\ }\href@noop {} {\  (\bibinfo {year}
  {2021})}\BibitemShut {NoStop}%
\bibitem [{\citenamefont {Vinci}\ \emph {et~al.}(2023)\citenamefont {Vinci},
  \citenamefont {Benzi},\ and\ \citenamefont
  {Mattia}}]{vinci_self-consistent_2023}%
  \BibitemOpen
  \bibfield  {author} {\bibinfo {author} {\bibfnamefont {G.~V.}\ \bibnamefont
  {Vinci}}, \bibinfo {author} {\bibfnamefont {R.}~\bibnamefont {Benzi}},\ and\
  \bibinfo {author} {\bibfnamefont {M.}~\bibnamefont {Mattia}},\ }\bibfield
  {title} {\bibinfo {title} {Self-{{Consistent Stochastic Dynamics}} for
  {{Finite-Size Networks}} of {{Spiking Neurons}}},\ }\href
  {https://doi.org/10.1103/PhysRevLett.130.097402} {\bibfield  {journal}
  {\bibinfo  {journal} {Physical Review Letters}\ }\textbf {\bibinfo {volume}
  {130}},\ \bibinfo {pages} {097402} (\bibinfo {year} {2023})}\BibitemShut
  {NoStop}%
\bibitem [{\citenamefont {Fleurantin}\ \emph {et~al.}(2023)\citenamefont
  {Fleurantin}, \citenamefont {Slyman}, \citenamefont {Barker},\ and\
  \citenamefont {Jones}}]{fleurantin_dynamical_2023}%
  \BibitemOpen
  \bibfield  {author} {\bibinfo {author} {\bibfnamefont {E.}~\bibnamefont
  {Fleurantin}}, \bibinfo {author} {\bibfnamefont {K.}~\bibnamefont {Slyman}},
  \bibinfo {author} {\bibfnamefont {B.}~\bibnamefont {Barker}},\ and\ \bibinfo
  {author} {\bibfnamefont {C.~K. R.~T.}\ \bibnamefont {Jones}},\ }\bibfield
  {title} {\bibinfo {title} {A dynamical systems approach for most probable
  escape paths over periodic boundaries},\ }\href
  {https://doi.org/10.1016/j.physd.2023.133860} {\bibfield  {journal} {\bibinfo
   {journal} {Physica D: Nonlinear Phenomena}\ }\textbf {\bibinfo {volume}
  {454}},\ \bibinfo {pages} {133860} (\bibinfo {year} {2023})}\BibitemShut
  {NoStop}%
\bibitem [{\citenamefont {Tsodyks}\ \emph {et~al.}(1997)\citenamefont
  {Tsodyks}, \citenamefont {Skaggs}, \citenamefont {Sejnowski},\ and\
  \citenamefont {McNaughton}}]{tsodyks_paradoxical_1997}%
  \BibitemOpen
  \bibfield  {author} {\bibinfo {author} {\bibfnamefont {M.~V.}\ \bibnamefont
  {Tsodyks}}, \bibinfo {author} {\bibfnamefont {W.~E.}\ \bibnamefont {Skaggs}},
  \bibinfo {author} {\bibfnamefont {T.~J.}\ \bibnamefont {Sejnowski}},\ and\
  \bibinfo {author} {\bibfnamefont {B.~L.}\ \bibnamefont {McNaughton}},\
  }\bibfield  {title} {\bibinfo {title} {Paradoxical {{Effects}} of {{External
  Modulation}} of {{Inhibitory Interneurons}}},\ }\href@noop {} {\bibfield
  {journal} {\bibinfo  {journal} {The Journal of Neuroscience}\ }\textbf
  {\bibinfo {volume} {17}},\ \bibinfo {pages} {4382} (\bibinfo {year}
  {1997})}\BibitemShut {NoStop}%
\bibitem [{\citenamefont {Sanzeni}\ \emph {et~al.}(2020)\citenamefont
  {Sanzeni}, \citenamefont {Akitake}, \citenamefont {Goldbach}, \citenamefont
  {Leedy}, \citenamefont {Brunel},\ and\ \citenamefont
  {Histed}}]{sanzeni_inhibition_2020}%
  \BibitemOpen
  \bibfield  {author} {\bibinfo {author} {\bibfnamefont {A.}~\bibnamefont
  {Sanzeni}}, \bibinfo {author} {\bibfnamefont {B.}~\bibnamefont {Akitake}},
  \bibinfo {author} {\bibfnamefont {H.~C.}\ \bibnamefont {Goldbach}}, \bibinfo
  {author} {\bibfnamefont {C.~E.}\ \bibnamefont {Leedy}}, \bibinfo {author}
  {\bibfnamefont {N.}~\bibnamefont {Brunel}},\ and\ \bibinfo {author}
  {\bibfnamefont {M.~H.}\ \bibnamefont {Histed}},\ }\bibfield  {title}
  {\bibinfo {title} {Inhibition stabilization is a widespread property of
  cortical networks},\ }\href {https://doi.org/10.7554/eLife.54875} {\bibfield
  {journal} {\bibinfo  {journal} {eLife}\ }\textbf {\bibinfo {volume} {9}},\
  \bibinfo {pages} {e54875} (\bibinfo {year} {2020})}\BibitemShut {NoStop}%
\bibitem [{\citenamefont {{Litwin-Kumar}}\ \emph {et~al.}(2016)\citenamefont
  {{Litwin-Kumar}}, \citenamefont {Rosenbaum},\ and\ \citenamefont
  {Doiron}}]{litwin-kumar_inhibitory_2016}%
  \BibitemOpen
  \bibfield  {author} {\bibinfo {author} {\bibfnamefont {A.}~\bibnamefont
  {{Litwin-Kumar}}}, \bibinfo {author} {\bibfnamefont {R.}~\bibnamefont
  {Rosenbaum}},\ and\ \bibinfo {author} {\bibfnamefont {B.}~\bibnamefont
  {Doiron}},\ }\bibfield  {title} {\bibinfo {title} {Inhibitory stabilization
  and visual coding in cortical circuits with multiple interneuron subtypes},\
  }\href {https://doi.org/10.1152/jn.00732.2015} {\bibfield  {journal}
  {\bibinfo  {journal} {Journal of Neurophysiology}\ }\textbf {\bibinfo
  {volume} {115}},\ \bibinfo {pages} {1399} (\bibinfo {year}
  {2016})}\BibitemShut {NoStop}%
\bibitem [{\citenamefont {Palmer}\ \emph {et~al.}(2012)\citenamefont {Palmer},
  \citenamefont {Schulz}, \citenamefont {Murphy}, \citenamefont {Ledergerber},
  \citenamefont {Murayama},\ and\ \citenamefont
  {Larkum}}]{palmer_cellular_2012}%
  \BibitemOpen
  \bibfield  {author} {\bibinfo {author} {\bibfnamefont {L.~M.}\ \bibnamefont
  {Palmer}}, \bibinfo {author} {\bibfnamefont {J.~M.}\ \bibnamefont {Schulz}},
  \bibinfo {author} {\bibfnamefont {S.~C.}\ \bibnamefont {Murphy}}, \bibinfo
  {author} {\bibfnamefont {D.}~\bibnamefont {Ledergerber}}, \bibinfo {author}
  {\bibfnamefont {M.}~\bibnamefont {Murayama}},\ and\ \bibinfo {author}
  {\bibfnamefont {M.~E.}\ \bibnamefont {Larkum}},\ }\bibfield  {title}
  {\bibinfo {title} {The {{Cellular Basis}} of {{GABAB-Mediated
  Interhemispheric Inhibition}}},\ }\href
  {https://doi.org/10.1126/science.1217276} {\bibfield  {journal} {\bibinfo
  {journal} {Science}\ }\textbf {\bibinfo {volume} {335}},\ \bibinfo {pages}
  {989} (\bibinfo {year} {2012})}\BibitemShut {NoStop}%
\bibitem [{\citenamefont {Hay}\ \emph {et~al.}(2011)\citenamefont {Hay},
  \citenamefont {Hill}, \citenamefont {Sch{\"u}rmann}, \citenamefont
  {Markram},\ and\ \citenamefont {Segev}}]{hay_models_2011}%
  \BibitemOpen
  \bibfield  {author} {\bibinfo {author} {\bibfnamefont {E.}~\bibnamefont
  {Hay}}, \bibinfo {author} {\bibfnamefont {S.}~\bibnamefont {Hill}}, \bibinfo
  {author} {\bibfnamefont {F.}~\bibnamefont {Sch{\"u}rmann}}, \bibinfo {author}
  {\bibfnamefont {H.}~\bibnamefont {Markram}},\ and\ \bibinfo {author}
  {\bibfnamefont {I.}~\bibnamefont {Segev}},\ }\bibfield  {title} {\bibinfo
  {title} {Models of {{Neocortical Layer}} 5b {{Pyramidal Cells Capturing}} a
  {{Wide Range}} of {{Dendritic}} and {{Perisomatic Active Properties}}},\
  }\href {https://doi.org/10.1371/journal.pcbi.1002107} {\bibfield  {journal}
  {\bibinfo  {journal} {PLOS Computational Biology}\ }\textbf {\bibinfo
  {volume} {7}},\ \bibinfo {pages} {e1002107} (\bibinfo {year}
  {2011})}\BibitemShut {NoStop}%
\bibitem [{\citenamefont {Markram}\ \emph {et~al.}(1997)\citenamefont
  {Markram}, \citenamefont {L{\"u}bke}, \citenamefont {Frotscher},
  \citenamefont {Roth},\ and\ \citenamefont
  {Sakmann}}]{markram_physiology_1997}%
  \BibitemOpen
  \bibfield  {author} {\bibinfo {author} {\bibfnamefont {H.}~\bibnamefont
  {Markram}}, \bibinfo {author} {\bibfnamefont {J.}~\bibnamefont {L{\"u}bke}},
  \bibinfo {author} {\bibfnamefont {M.}~\bibnamefont {Frotscher}}, \bibinfo
  {author} {\bibfnamefont {A.}~\bibnamefont {Roth}},\ and\ \bibinfo {author}
  {\bibfnamefont {B.}~\bibnamefont {Sakmann}},\ }\bibfield  {title} {\bibinfo
  {title} {Physiology and anatomy of synaptic connections between thick tufted
  pyramidal neurones in the developing rat neocortex.},\ }\href
  {https://doi.org/10.1113/jphysiol.1997.sp022031} {\bibfield  {journal}
  {\bibinfo  {journal} {The Journal of Physiology}\ }\textbf {\bibinfo {volume}
  {500}},\ \bibinfo {pages} {409} (\bibinfo {year} {1997})}\BibitemShut
  {NoStop}%
\bibitem [{\citenamefont {Markram}(1997)}]{markram_network_1997}%
  \BibitemOpen
  \bibfield  {author} {\bibinfo {author} {\bibfnamefont {H.}~\bibnamefont
  {Markram}},\ }\bibfield  {title} {\bibinfo {title} {A network of tufted layer
  5 pyramidal neurons.},\ }\href {https://doi.org/10.1093/cercor/7.6.523}
  {\bibfield  {journal} {\bibinfo  {journal} {Cerebral Cortex}\ }\textbf
  {\bibinfo {volume} {7}},\ \bibinfo {pages} {523} (\bibinfo {year}
  {1997})}\BibitemShut {NoStop}%
\bibitem [{\citenamefont {Bodor}\ \emph {et~al.}(2025)\citenamefont {Bodor},
  \citenamefont {{Schneider-Mizell}}, \citenamefont {Zhang}, \citenamefont
  {Elabbady}, \citenamefont {Mallen}, \citenamefont {Bergeson}, \citenamefont
  {Brittain}, \citenamefont {Buchanan}, \citenamefont {Bumbarger},
  \citenamefont {Dalley}, \citenamefont {Gamlin}, \citenamefont {Joyce},
  \citenamefont {Kapner}, \citenamefont {Kinn}, \citenamefont {Mahalingam},
  \citenamefont {Seshamani}, \citenamefont {Suckow}, \citenamefont {Takeno},
  \citenamefont {Torres}, \citenamefont {Yin}, \citenamefont {Bae},
  \citenamefont {Castro}, \citenamefont {Dorkenwald}, \citenamefont {Halageri},
  \citenamefont {Jia}, \citenamefont {Jordan}, \citenamefont {Kemnitz},
  \citenamefont {Lee}, \citenamefont {Li}, \citenamefont {Lu}, \citenamefont
  {Macrina}, \citenamefont {Mitchell}, \citenamefont {Mondal}, \citenamefont
  {Mu}, \citenamefont {Nehoran}, \citenamefont {Popovych}, \citenamefont
  {Silversmith}, \citenamefont {Turner}, \citenamefont {Yu}, \citenamefont
  {Wong}, \citenamefont {Wu}, \citenamefont {Celii}, \citenamefont
  {Campagnola}, \citenamefont {Seeman}, \citenamefont {Jarsky}, \citenamefont
  {Ren}, \citenamefont {Arkhipov}, \citenamefont {Reimer}, \citenamefont
  {Seung}, \citenamefont {Reid}, \citenamefont {Collman},\ and\ \citenamefont
  {{da Costa}}}]{bodor_synaptic_2025}%
  \BibitemOpen
  \bibfield  {author} {\bibinfo {author} {\bibfnamefont {A.~L.}\ \bibnamefont
  {Bodor}}, \bibinfo {author} {\bibfnamefont {C.~M.}\ \bibnamefont
  {{Schneider-Mizell}}}, \bibinfo {author} {\bibfnamefont {C.}~\bibnamefont
  {Zhang}}, \bibinfo {author} {\bibfnamefont {L.}~\bibnamefont {Elabbady}},
  \bibinfo {author} {\bibfnamefont {A.}~\bibnamefont {Mallen}}, \bibinfo
  {author} {\bibfnamefont {A.}~\bibnamefont {Bergeson}}, \bibinfo {author}
  {\bibfnamefont {D.}~\bibnamefont {Brittain}}, \bibinfo {author}
  {\bibfnamefont {J.}~\bibnamefont {Buchanan}}, \bibinfo {author}
  {\bibfnamefont {D.~J.}\ \bibnamefont {Bumbarger}}, \bibinfo {author}
  {\bibfnamefont {R.}~\bibnamefont {Dalley}}, \bibinfo {author} {\bibfnamefont
  {C.}~\bibnamefont {Gamlin}}, \bibinfo {author} {\bibfnamefont
  {E.}~\bibnamefont {Joyce}}, \bibinfo {author} {\bibfnamefont
  {D.}~\bibnamefont {Kapner}}, \bibinfo {author} {\bibfnamefont
  {S.}~\bibnamefont {Kinn}}, \bibinfo {author} {\bibfnamefont {G.}~\bibnamefont
  {Mahalingam}}, \bibinfo {author} {\bibfnamefont {S.}~\bibnamefont
  {Seshamani}}, \bibinfo {author} {\bibfnamefont {S.}~\bibnamefont {Suckow}},
  \bibinfo {author} {\bibfnamefont {M.}~\bibnamefont {Takeno}}, \bibinfo
  {author} {\bibfnamefont {R.}~\bibnamefont {Torres}}, \bibinfo {author}
  {\bibfnamefont {W.}~\bibnamefont {Yin}}, \bibinfo {author} {\bibfnamefont
  {J.~A.}\ \bibnamefont {Bae}}, \bibinfo {author} {\bibfnamefont {M.~A.}\
  \bibnamefont {Castro}}, \bibinfo {author} {\bibfnamefont {S.}~\bibnamefont
  {Dorkenwald}}, \bibinfo {author} {\bibfnamefont {A.}~\bibnamefont
  {Halageri}}, \bibinfo {author} {\bibfnamefont {Z.}~\bibnamefont {Jia}},
  \bibinfo {author} {\bibfnamefont {C.}~\bibnamefont {Jordan}}, \bibinfo
  {author} {\bibfnamefont {N.}~\bibnamefont {Kemnitz}}, \bibinfo {author}
  {\bibfnamefont {K.}~\bibnamefont {Lee}}, \bibinfo {author} {\bibfnamefont
  {K.}~\bibnamefont {Li}}, \bibinfo {author} {\bibfnamefont {R.}~\bibnamefont
  {Lu}}, \bibinfo {author} {\bibfnamefont {T.}~\bibnamefont {Macrina}},
  \bibinfo {author} {\bibfnamefont {E.}~\bibnamefont {Mitchell}}, \bibinfo
  {author} {\bibfnamefont {S.~S.}\ \bibnamefont {Mondal}}, \bibinfo {author}
  {\bibfnamefont {S.}~\bibnamefont {Mu}}, \bibinfo {author} {\bibfnamefont
  {B.}~\bibnamefont {Nehoran}}, \bibinfo {author} {\bibfnamefont
  {S.}~\bibnamefont {Popovych}}, \bibinfo {author} {\bibfnamefont
  {W.}~\bibnamefont {Silversmith}}, \bibinfo {author} {\bibfnamefont {N.~L.}\
  \bibnamefont {Turner}}, \bibinfo {author} {\bibfnamefont {S.-c.}\
  \bibnamefont {Yu}}, \bibinfo {author} {\bibfnamefont {W.}~\bibnamefont
  {Wong}}, \bibinfo {author} {\bibfnamefont {J.}~\bibnamefont {Wu}}, \bibinfo
  {author} {\bibfnamefont {B.}~\bibnamefont {Celii}}, \bibinfo {author}
  {\bibfnamefont {L.}~\bibnamefont {Campagnola}}, \bibinfo {author}
  {\bibfnamefont {S.~C.}\ \bibnamefont {Seeman}}, \bibinfo {author}
  {\bibfnamefont {T.}~\bibnamefont {Jarsky}}, \bibinfo {author} {\bibfnamefont
  {N.}~\bibnamefont {Ren}}, \bibinfo {author} {\bibfnamefont {A.}~\bibnamefont
  {Arkhipov}}, \bibinfo {author} {\bibfnamefont {J.}~\bibnamefont {Reimer}},
  \bibinfo {author} {\bibfnamefont {H.~S.}\ \bibnamefont {Seung}}, \bibinfo
  {author} {\bibfnamefont {R.~C.}\ \bibnamefont {Reid}}, \bibinfo {author}
  {\bibfnamefont {F.}~\bibnamefont {Collman}},\ and\ \bibinfo {author}
  {\bibfnamefont {N.~M.}\ \bibnamefont {{da Costa}}},\ }\bibfield  {title}
  {\bibinfo {title} {The synaptic architecture of layer 5 thick tufted
  excitatory neurons in mouse visual cortex},\ }\href
  {https://doi.org/10.1038/s41593-025-02004-2} {\bibfield  {journal} {\bibinfo
  {journal} {Nature Neuroscience}\ ,\ \bibinfo {pages} {1}} (\bibinfo {year}
  {2025})}\BibitemShut {NoStop}%
\bibitem [{\citenamefont {Schiller}\ \emph {et~al.}(2000)\citenamefont
  {Schiller}, \citenamefont {Major}, \citenamefont {Koester},\ and\
  \citenamefont {Schiller}}]{schiller_nmda_2000}%
  \BibitemOpen
  \bibfield  {author} {\bibinfo {author} {\bibfnamefont {J.}~\bibnamefont
  {Schiller}}, \bibinfo {author} {\bibfnamefont {G.}~\bibnamefont {Major}},
  \bibinfo {author} {\bibfnamefont {H.~J.}\ \bibnamefont {Koester}},\ and\
  \bibinfo {author} {\bibfnamefont {Y.}~\bibnamefont {Schiller}},\ }\bibfield
  {title} {\bibinfo {title} {{{NMDA}} spikes in basal dendrites of cortical
  pyramidal neurons},\ }\href {https://doi.org/10.1038/35005094} {\bibfield
  {journal} {\bibinfo  {journal} {Nature}\ }\textbf {\bibinfo {volume} {404}},\
  \bibinfo {pages} {285} (\bibinfo {year} {2000})}\BibitemShut {NoStop}%
\bibitem [{\citenamefont {Nevian}\ \emph {et~al.}(2007)\citenamefont {Nevian},
  \citenamefont {Larkum}, \citenamefont {Polsky},\ and\ \citenamefont
  {Schiller}}]{nevian_properties_2007}%
  \BibitemOpen
  \bibfield  {author} {\bibinfo {author} {\bibfnamefont {T.}~\bibnamefont
  {Nevian}}, \bibinfo {author} {\bibfnamefont {M.~E.}\ \bibnamefont {Larkum}},
  \bibinfo {author} {\bibfnamefont {A.}~\bibnamefont {Polsky}},\ and\ \bibinfo
  {author} {\bibfnamefont {J.}~\bibnamefont {Schiller}},\ }\bibfield  {title}
  {\bibinfo {title} {Properties of basal dendrites of layer 5 pyramidal
  neurons: A direct patch-clamp recording study},\ }\href
  {https://doi.org/10.1038/nn1826} {\bibfield  {journal} {\bibinfo  {journal}
  {Nature Neuroscience}\ }\textbf {\bibinfo {volume} {10}},\ \bibinfo {pages}
  {206} (\bibinfo {year} {2007})}\BibitemShut {NoStop}%
\bibitem [{\citenamefont {Larkum}\ \emph {et~al.}(2009)\citenamefont {Larkum},
  \citenamefont {Nevian}, \citenamefont {Sandler}, \citenamefont {Polsky},\
  and\ \citenamefont {Schiller}}]{larkum_synaptic_2009}%
  \BibitemOpen
  \bibfield  {author} {\bibinfo {author} {\bibfnamefont {M.~E.}\ \bibnamefont
  {Larkum}}, \bibinfo {author} {\bibfnamefont {T.}~\bibnamefont {Nevian}},
  \bibinfo {author} {\bibfnamefont {M.}~\bibnamefont {Sandler}}, \bibinfo
  {author} {\bibfnamefont {A.}~\bibnamefont {Polsky}},\ and\ \bibinfo {author}
  {\bibfnamefont {J.}~\bibnamefont {Schiller}},\ }\bibfield  {title} {\bibinfo
  {title} {Synaptic {{Integration}} in {{Tuft Dendrites}} of {{Layer}} 5
  {{Pyramidal Neurons}}: {{A New Unifying Principle}}},\ }\href
  {https://doi.org/10.1126/science.1171958} {\bibfield  {journal} {\bibinfo
  {journal} {Science}\ }\textbf {\bibinfo {volume} {325}},\ \bibinfo {pages}
  {756} (\bibinfo {year} {2009})}\BibitemShut {NoStop}%
\bibitem [{\citenamefont {Rubin}\ \emph {et~al.}(2005)\citenamefont {Rubin},
  \citenamefont {{Gerkin, R.C.}}, \citenamefont {{Bi, G-Q.}},\ and\
  \citenamefont {{Chow, C.}}}]{rubin_calcium_2005}%
  \BibitemOpen
  \bibfield  {author} {\bibinfo {author} {\bibfnamefont {J.~E.}\ \bibnamefont
  {Rubin}}, \bibinfo {author} {\bibnamefont {{Gerkin, R.C.}}}, \bibinfo
  {author} {\bibnamefont {{Bi, G-Q.}}},\ and\ \bibinfo {author} {\bibnamefont
  {{Chow, C.}}},\ }\bibfield  {title} {\bibinfo {title} {Calcium {{Time
  Course}} as a {{Signal}} for {{Spike-Timing-Dependent Plasticity}}},\ }\href
  {https://doi.org/10.1152/jn.00803.2004} {\bibfield  {journal} {\bibinfo
  {journal} {Journal of Neurophysiology}\ }\textbf {\bibinfo {volume} {93}},\
  \bibinfo {pages} {2600} (\bibinfo {year} {2005})}\BibitemShut {NoStop}%
\bibitem [{\citenamefont {Shouval}\ \emph {et~al.}(2010)\citenamefont
  {Shouval}, \citenamefont {Wang},\ and\ \citenamefont
  {Wittenberg}}]{shouval_spike_2010}%
  \BibitemOpen
  \bibfield  {author} {\bibinfo {author} {\bibfnamefont {H.~Z.}\ \bibnamefont
  {Shouval}}, \bibinfo {author} {\bibfnamefont {S.~S.-H.}\ \bibnamefont
  {Wang}},\ and\ \bibinfo {author} {\bibfnamefont {G.~M.}\ \bibnamefont
  {Wittenberg}},\ }\bibfield  {title} {\bibinfo {title} {Spike {{Timing
  Dependent Plasticity}}: {{A Consequence}} of {{More Fundamental Learning
  Rules}}},\ }\bibfield  {journal} {\bibinfo  {journal} {Frontiers in
  Computational Neuroscience}\ }\textbf {\bibinfo {volume} {4}},\ \href
  {https://doi.org/10.3389/fncom.2010.00019} {10.3389/fncom.2010.00019}
  (\bibinfo {year} {2010})\BibitemShut {NoStop}%
\bibitem [{\citenamefont {Graupner}\ and\ \citenamefont
  {Brunel}(2012)}]{graupner_calcium-based_2012}%
  \BibitemOpen
  \bibfield  {author} {\bibinfo {author} {\bibfnamefont {M.}~\bibnamefont
  {Graupner}}\ and\ \bibinfo {author} {\bibfnamefont {N.}~\bibnamefont
  {Brunel}},\ }\bibfield  {title} {\bibinfo {title} {Calcium-based plasticity
  model explains sensitivity of synaptic changes to spike pattern, rate, and
  dendritic location},\ }\href {https://doi.org/10.1073/pnas.1109359109}
  {\bibfield  {journal} {\bibinfo  {journal} {Proceedings of the National
  Academy of Sciences}\ }\textbf {\bibinfo {volume} {109}},\ \bibinfo {pages}
  {3991} (\bibinfo {year} {2012})}\BibitemShut {NoStop}%
\bibitem [{\citenamefont {Helias}\ and\ \citenamefont
  {Dahmen}(2020)}]{helias_statistical_2020}%
  \BibitemOpen
  \bibfield  {author} {\bibinfo {author} {\bibfnamefont {M.}~\bibnamefont
  {Helias}}\ and\ \bibinfo {author} {\bibfnamefont {D.}~\bibnamefont
  {Dahmen}},\ }\href {https://doi.org/10.1007/978-3-030-46444-8} {\emph
  {\bibinfo {title} {Statistical {{Field Theory}} for {{Neural Networks}}}}},\
  Lecture {{Notes}} in {{Physics}}\ (\bibinfo  {publisher} {Springer
  International Publishing},\ \bibinfo {year} {2020})\BibitemShut {NoStop}%
\bibitem [{\citenamefont {{Zinn-Justin}}(2002)}]{zinn-justin_quantum_2002}%
  \BibitemOpen
  \bibfield  {author} {\bibinfo {author} {\bibfnamefont {J.}~\bibnamefont
  {{Zinn-Justin}}},\ }\href@noop {} {\emph {\bibinfo {title} {Quantum {{Field
  Theory}} and {{Critical Phenomena}}}}}\ (\bibinfo  {publisher} {Clarendon
  Press},\ \bibinfo {year} {2002})\BibitemShut {NoStop}%
\bibitem [{\citenamefont {Berges}(2004)}]{berges_introduction_2004}%
  \BibitemOpen
  \bibfield  {author} {\bibinfo {author} {\bibfnamefont {J.}~\bibnamefont
  {Berges}},\ }\bibfield  {title} {\bibinfo {title} {Introduction to
  {{Nonequilibrium Quantum Field Theory}}},\ }in\ \href
  {https://doi.org/10.1063/1.1843591} {\emph {\bibinfo {booktitle} {{{AIP
  Conference Proceedings}}}}},\ Vol.\ \bibinfo {volume} {739}\ (\bibinfo {year}
  {2004})\ pp.\ \bibinfo {pages} {3--62},\ \Eprint
  {https://arxiv.org/abs/hep-ph/0409233} {arXiv:hep-ph/0409233} \BibitemShut
  {NoStop}%
\end{thebibliography}%

\end{document}